\documentclass[usenatbib,twocolumn]{mn2e}

\usepackage{bm}
\usepackage{amsmath}
\usepackage{graphicx}

\def\be{\begin{equation}}
\def\ee{\end{equation}}
\def\ba{\begin{eqnarray}}
\def\ea{\end{eqnarray}}
\def\bal#1\eal{\begin{align}#1\end{align}}

\title[Pair cascades of strongly-magnetized neutron stars]
{Pair cascades in the magnetospheres of strongly-magnetized neutron stars}

\author[Zach Medin and Dong Lai]{Zach Medin$^1$\thanks{Email:
zmedin@physics.mcgill.ca}
and Dong Lai$^2$\thanks{Email: dong@astro.cornell.edu}\\
$^1$Department of Physics, McGill University, 3600 rue University,
Montreal, QC H3A 2T8, Canada\\
$^2$Department of Astronomy, Cornell University, Ithaca, NY 14853, USA}

\begin{document}

\date{}

\pagerange{\pageref{firstpage}--\pageref{lastpage}} \pubyear{2010}

\label{firstpage}

\maketitle

\begin{abstract}
We present numerical simulations of electron-positron pair cascades in
the magnetospheres of magnetic neutron stars for a wide range of
surface fields ($B_p=10^{12}$--$10^{15}$~G), rotation periods
($0.1$--$10$~s), and field geometries. This has been motivated by the
discovery in recent years of a number of radio pulsars with inferred
magnetic fields comparable to those of magnetars. Evolving the cascade
generated by a primary electron or positron after it has been
accelerated in the inner gap of the magnetosphere, we follow the
spatial development of the cascade until the secondary photons and
pairs leave the magnetosphere, and we obtain the pair multiplicity and
the energy spectra of the cascade pairs and photons under various
conditions. Going beyond previous works, which were restricted to
weaker fields ($B\la {\rm a~few}\times 10^{12}$~G), we have
incorporated in our simulations detailed treatments of physical
processes that are potentially important (especially in the high field
regime) but were either neglected or crudely treated before, including
photon splitting with the correct selection rules for photon
polarization modes, one-photon pair production into low Landau levels
for the $e^\pm$, and resonant inverse Compton scattering from polar
cap hot spots. We find that even for $B\gg B_Q=4\times 10^{13}$~G,
photon splitting has a small effect on the multiplicity of the cascade
since a majority of the photons in the cascade cannot split.
One-photon decay into $e^+e^-$ pairs at low-Landau levels, however,
becomes the dominant pair production channel when $B\ga 3\times
10^{12}$~G; this tends to suppress synchrotron radiation so that the
cascade can develop only at a larger distance from the stellar
surface. Nevertheless, we find that the total number of pairs and
their energy spectrum produced in the cascade depend mainly on the
polar cap voltage $B_p P^{-2}$, and are weakly dependent on $B_p$ (and
$P$) alone.  We discuss the implications of our results for the radio
pulsar death line and for the hard X-ray emission from magnetized
neutron stars.
\end{abstract}



\section{Introduction}
\label{sec:intro}

The pair cascade in the magnetosphere of a pulsar has long been
considered an essential ingredient for the pulsar's nonthermal
emission, from radio to gamma rays (e.g.,
\citealt{sturrock71,ruderman75,melrose04,thompson04}). More recently
it has been suggested that the pair cascade is also necessary for
nonthermal emission from magnetars (e.g.,
\citealt{beloborodov07,thompson08a,thompson08b}; see \citealt{woods06}
for a review of magnetars). The basic pair cascade involves several
steps: (i) acceleration of primary particles by an electric field
parallel to the magnetic field; (ii) gamma ray emission by the
accelerated particles moving along the magnetic field lines (either by
curvature radiation or inverse Compton upscattering of surface
photons); (iii) field-assisted photon decay into electron-positron
pairs as the angle between the photon and the magnetic field line
becomes sufficiently large, or pair production by two-photon
annihilation in weak-field regimes; (iv) gamma ray emission by the
newly-created particles as they lose their transverse energy through
synchrotron emission; (v) further pair production and gamma ray
emission via steps (iii) and (iv). The dense, relativistic (Lorentz
factors $\gamma \ga 100$) electron-positron plasma generated by this
cascade is a required input in many models for the pulsar radio
emission (e.g.,
\citealt*{melrose95,melrose04,beskin99,melikidze00,lyubarsky02,lyubarsky08,lyutikov07}),
while the high-energy photons emitted in pair cascade models can
reproduce the observed pulse profiles and phase-resolved spectra of
gamma-ray pulsars once the three-dimensional emission geometry is
taken into account (see, e.g.,
\citealt*{romani95,cheng00,dyks03,harding08,bai09}). We note in
passing that the dense pair plasma generated by this cascade also
plays an important role in models of pulsar wind nebulae (see
\citealt{arons07} for a review).

The behavior of the pair cascade in the superstrong field regime
(magnetic field strengths $B \ga B_Q\equiv4.414\times 10^{13}$~G) and
its effect on emission from pulsars and magnetars is somewhat
puzzling. For example, of the dozen-or-so observed magnetars, only two
show pulsed radio emission, and it is of a completely different nature
than the emission from ``standard'' radio pulsars (e.g., the radio
pulsations are transient and appear to be correlated with strong X-ray
outbursts from the magnetars; see \citealt{camilo07,camilo08}). In
contrast, several radio pulsars with inferred surface field strengths
similar to those of magnetars have been discovered (e.g.,
\citealt*{kaspi05,vranevsevic07}). Why the standard mechanism for
pulsed radio emission turns off for magnetars but not for these
pulsars is unknown.

There have been only a few publications devoted to numerical
simulations of the pair cascade in pulsar magnetospheres. For
moderate-strength magnetic fields ($B\la 5\times10^{12}$~G),
significant progress has been made. \citet{daugherty82} present
simulations of the cascade initiated by a single electron injected
from the neutron star surface, emitting photons through curvature
radiation, for (polar) surface field strengths $B_p$ up to
$5\times10^{12}$~G and rotation periods $P=0.033$--$1$~s. In a later
paper (\citealt{daugherty96}) they consider gamma ray emission from
the entire open-field-line region of the magnetosphere, using a
simplified acceleration model and for Vela-like pulsar parameters
($B_p=3\times10^{12}$~G and $P=0.089$~s). \citet*{sturner95} present a
similar simulation to that of Daugherty \& Harding, but for cascades
initiated by electrons upscattering photons through the inverse
Compton process (again for Vela-like parameters). \citet{hibschman01b}
develop a semi-analytic model of the inner gap cascade, both for
curvature radiation-initiated and inverse Compton scattering-initiated
cascades, applicable for $B\la 3\times10^{12}$~G (see also
\citealt*{zhang00}). Cascades occurring in the outer magnetosphere
have also been simulated, by \citet{romani96} for Vela- and Crab-like
($B_p=4\times10^{12}$~G and $P=0.033$~s) parameters (see also
\citealt*{cheng86a,cheng86b,cheng00}).

However, for superstrong magnetic fields ($B \ga B_Q\equiv4.414\times
10^{13}$~G) only limited aspects of the full cascades have been
studied. For example, \citet{arendt02} simulate the cascade for $B_p
\le 10^{13}$~G and $P=0.033$~s (for both a pure dipole and a more
complex field geometry), but with the simplification that all photons
radiated by the primary particle are emitted from the
surface. \citet{baring01} (see also \citealt*{harding97}) use this
same simplification to study the effects of photon splitting on the
cascade for field strengths up to $B=2\times10^{14}$~G (however, they
assumed that both photon modes can split, and thus overestimated the
effect of photon splitting; see Section~\ref{sub:photon}).
\citet{baring07} model the process of resonant inverse Compton
scattering of photons from the neutron star surface (with the
blackbody temperature $T=6\times10^6$~K) in the same field range, but
only for single scattering events (see also \citealt{dermer90}). The
magnetosphere acceleration zone in the superstrong, twisted field
regime of magnetars is investigated analytically by
\citet{beloborodov07} for cascades occurring in the closed field line
region of the magnetosphere and by \citet{thompson08a,thompson08b} in
the open field line region.

In this paper we present numerical simulations of the pair cascade
from onset to completion. Motivated by the lack of full cascade
results for the superstrong field regime, and in light of the
unexplained differences between the observed emission properties of
high-field radio pulsars and magnetars, we run our simulations in
magnetospheres with field strengths up to $10^{15}$~G\@. We consider
several important factors that affect high-field cascades, including
photon splitting, pair creation in low Landau levels, photon
polarization modes ($\perp$ or $\parallel$ to the magnetic field
direction), and resonant inverse Compton scattering. We use our
simulations to generate spectra of the high-energy photons and the
electron-positron plasma produced by the cascade. Additionally, we use
our simulations to comment on the conditions for when the radio
emission mechanism no longer operates in the neutron star
magnetosphere, the so-called ``pulsar death line'' (e.g.,
\citealt*{ruderman75,chen93,hibschman01a,harding02,hardingetal02,
medin07b}). While the
results of our simulation are most applicable to cascades occurring in
the open field line region of the magnetosphere (since the primary
particles are injected into the magnetosphere along open field lines),
some of our results are also relevant to cascades occurring in the
closed field line region for magnetars, e.g., the products of a
cascade initiated by a photon injected into a non-dipole magnetosphere.

A necessary component of any pair cascade simulation is a model of the
magnetosphere acceleration zone, or ``gap'', where the cascade
originates. In real magnetospheres of pulsars and magnetars, the
acceleration of primary particles is coupled to the rest of the
cascade (e.g., charged particles produced in the cascade can screen
out the acceleration potential). However, there is significant
uncertainty about the precise nature of the acceleration gap. A
number of models have been proposed for the location of the gap, from
inner magnetosphere accelerators (both ``vacuum'' and
``space-charge-limited flow'' types; see, e.g.,
\citealt{ruderman75,arons79,muslimov92,hibschman01a,medin07b,thompson08a,thompson08b}),
to outer magnetosphere accelerators (e.g.,
\citealt{cheng86a,cheng86b,romani96,cheng00,takata06}), to hybrid
inner-outer magnetosphere accelerators (``slot'' gaps and extended
outer gaps; e.g., \citealt{arons83,muslimov03,muslimov04,hirotani06}).
Non-steady (oscillatory) inner gaps have also been discussed recently
(e.g., \citealt{sakai03,levinson05,beloborodov08,luo08}). Numerical
simulations of force-free global magnetospheres including
magnetic-field twisting near the light cylinder have been performed
(e.g., \citealt*{contopoulos99,gruzinov05,
spitkovsky06,timokhin06,komissarov06,kalapotharakos09}), but they do
not yet include any particle acceleration, or pair creation
self-consistently. Therefore, in this paper we decouple particle
acceleration from the rest of the cascade and focus on the cascade
produced by a primary electron\footnote{Although the primary particle
could also be a positron (or even an ion), we assume here for
simplicity that the pulsar is oriented such that electrons are
accelerated away from the star.} injected into the magnetosphere with
a given initial Lorentz factor $\gamma_0$
(cf. \citealt{daugherty82}). We also consider the cascade produced by
a single ``primary'' photon emitted by the primary electron, in the
case where photon emission within the acceleration gap is important
(i.e., for cascades where the dominant mechanism for high-energy
photon production is inverse Compton scattering).

The outline of the paper is as follows. In Section~\ref{sec:accel} we
summarize our method for estimating the initial parameters (e.g.,
$\gamma_0$) of the primary cascade particles, for use in our
simulations. In Section~\ref{sec:model} we describe the details of the
numerical simulations, both for cascades with photon production
dominated by curvature radiation and by resonant inverse Compton
scattering (resonant ICS, or RICS). In Section~\ref{sec:results} we
present our results (e.g., photon and pair plasma spectra) for a wide
range of parameters: surface magnetic fields $B=10^{12}$--$10^{15}$~G,
rotation periods $0.1$--$10$~s, surface temperatures
$T=(0.3$--$3)\times10^6$~K, and pure dipole and more complex field
geometries. In Section~\ref{sec:discuss} we summarize our findings and
discuss their implications for the radio emission and high-energy
(hard X-ray and gamma-ray) emission from pulsars and magnetars.  Some
technical details (on our treatment of inverse Compton scattering, on
our treatment of attenuation coefficients and $e^+e^-$ energy levels
for pair production, and on deriving semi-analytic fits to our
numerical results) are given in the appendix.

\section{Estimating the initial parameters for the primary particles}
\label{sec:accel}

\subsection{Primary electrons}
\label{sub:eparam}

In our cascade simulation (described in Section~\ref{sec:model}) we do
not include an actual acceleration region, since we wish in this work
to remain as model-independent as possible. Instead, we model the
effect of this region on the cascade by giving the primary electron an
initial energy $\gamma_0 m_ec^2$ equivalent to the energy it would
reach upon traversing the entire gap, and injecting it into the
magnetosphere at the neutron star surface (cf. \citealt{daugherty82}).
Obviously, this approximation excludes a proper treatment of the slot
gap and outer gap acceleration models. However, in most parts of the
polar cap region (i.e., except for the boundary region adjacent to the
open field lines), the main voltage drop occurs near the stellar
surface, regardless of the nature of the ``gap'' (vacuum gap or space
charge limited flow). Such inner gap models and other global models
with near-surface acceleration are allowed in our analysis.

For a dipole magnetic field geometry, most active pulsars with inner
gap accelerators have gap voltage drops in the range $\Phi \sim
(1$--$2)\times10^{13}$~V, regardless of the acceleration model (e.g.,
\citealt{hibschman01a,medin07b}, hereafter ML07). For the surface
field strengths we are considering, $B \ge 10^{12}$~G, the primary
electrons are not radiation-reaction limited within these gaps (ML07;
cf. the millisecond pulsar models of \citealt*{harding05}), so we can
set $\gamma_0=e\Phi/m_ec^2$. We therefore restrict $\gamma_0$ to the
range $(2$--$4)\times10^7$ for dipole fields. Note that these large
voltage drops do not occur in pulsars where the gap electric field is
fully screened due to inverse Compton scattering by the primary
electron. We discuss this case in Section~\ref{sub:phparam}.

\begin{figure*}
\begin{center}
\begin{tabular}{cc}
\resizebox{0.45\textwidth}{!}{\includegraphics{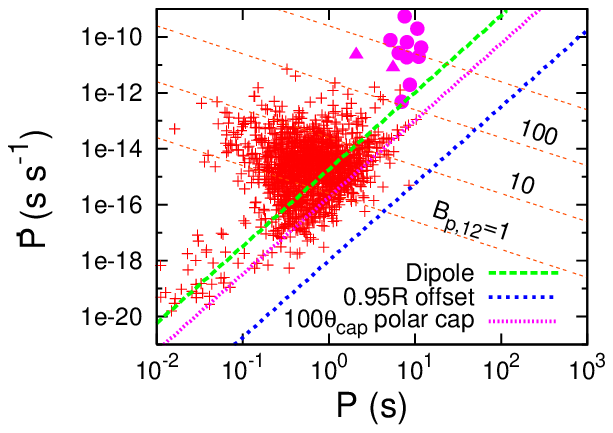}} &
\resizebox{0.45\textwidth}{!}{\includegraphics{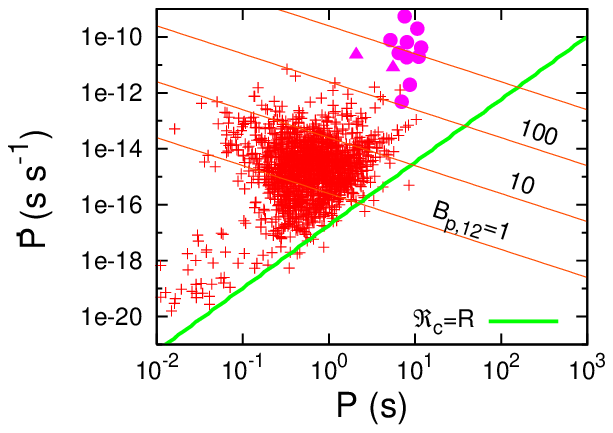}} \\
\end{tabular}
\end{center}
\caption[Pulsar death lines]
{Pulsar death lines. Death lines are shown for pulsars with dipole
magnetic fields, dipole fields offset from the center of the star by
$\Delta r=0.95R$, and magnetic fields with extended polar caps $100$
times larger than the dipole value $\theta_{\rm cap}=\sqrt{\Omega
R/c}$ (left panel); and for pulsars with magnetic field curvatures
${\cal R}_c=R$ at the surface (right panel). Note that these death
lines do not apply for the millisecond pulsar population in the lower
left corner of the diagram, as their short periods and low magnetic
field strengths cause the primary electron to be radiation reaction
limited.  In each panel, rotation-powered pulsars (ATNF catalog,
http://www.atnf.csiro.au/research/pulsar/psrcat) are labeled by
crosses, while magnetars (McGill catalog,
http://www.physics.mcgill.ca/$\sim$pulsar/magnetar/main.html) are
labeled by solid circles and the two radio magnetars are labeled by
solid triangles.  }
\label{fig:death}
\end{figure*}

The voltage drop across the gap can be no larger than the voltage drop
across the entire polar cap of the neutron star (e.g.,
\citealt{ruderman75}):
\be
\Phi_{\rm cap} \simeq \frac{\Omega B_p}{2c}\frac{R^3\Omega}{c} = 7\times10^{12} B_{p,12} P_0^{-2}~{\rm V} \,,
\label{eq:phicap}
\ee
where $R$ is the radius of the star (assumed in this paper to be
10~km), $P_0$ is the spin period in units of 1~s, and
$B_p=10^{12}B_{p,12}$~G is the polar surface magnetic field
strength. If the voltage drop, $\Phi$, required to initiate pair
cascades is not available, i.e., $\Phi > \Phi_{\rm cap}$, the
magnetosphere should not produce pulsed radio emission; the locus of
points where $\Phi = \Phi_{\rm cap}$ defines the pulsar death
line.\footnote{This applies to the vacuum gap model. In the
space-charge-limited-flow model the condition is $\Phi =
\kappa_g\Phi_{\rm cap}$, with $\kappa_g \simeq 0.15$ (e.g.,
\citealt{hibschman01a}).} A typical death line for an inner gap model,
plotted in $P$-$\dot{P}$ space, is shown on the left panel of
Fig.~\ref{fig:death}.  The line was made using three assumptions: (i)
The magnetosphere field geometry is dipolar. (ii) The pair cascade
occurs primarily above the gap, (through curvature radiation) once the
primary electron has reached a large Lorentz factor $\gamma_0 \sim
10^7$. (iii) The spindown power of the pulsar, given by
\be
\dot{E} = -I\Omega\dot{\Omega} = \frac{4\pi^2 I\dot{P}}{P^3} \,,
\label{eq:Edot}
\ee
is approximately equal to the spindown power of a magnetic dipole with
its magnetic field and rotational axes orthogonal to each other:
\be
\dot{E} \simeq \frac{B_p^2\Omega^4 R^6}{6c^3} = \frac{2\Phi_{\rm cap}^2c}{3} \,.
\label{eq:orthoEdot}
\ee
The polar magnetic field strength inferred from this frequently-used
approximation is
\be
B_{p,12} \simeq 2.0\sqrt{P_0 \dot{P}_{-15}} \,,
\ee
where $\dot{P}_{-15}$ is the period derivative in units of
$10^{-15}$~s/s and $I = 10^{45}$~g-cm$^2$ is assumed.

A well-known problem with the death line made using
these assumptions is that it cuts right through the middle of the main
group of pulsars
(\citealt{ruderman75,hibschman01a,harding02,medin07b}); i.e., the model 
incorrectly predicts that there will be no radio emission from
many neutron stars that are observed to be active pulsars.

Several authors have proposed models of the neutron star magnetosphere
that shift the theoretical death line closer to the observed death
line by altering one or more of the assumptions made above. In some
models the magnetosphere geometry is not a centered dipole, but
instead is an offset dipole (\citealt{arons98}), or a twisted dipole
(e.g., \citealt*{thompson02}), or contains quadrupole or higher
multipole components (e.g., \citealt{pavan09}). Numerical simulations
suggest that due to twisting of the field lines near the light
cylinder the polar cap is slightly larger than in the pure dipole case
(e.g., \citealt{contopoulos99,spitkovsky06}); the polar cap could be
significantly larger if, as the pulsar spins down, field line
reconnection (conversion of open field lines into closed field lines)
is too slow to keep pace with the expanding light cylinder
(\citealt{contopoulos05}). All of these models increase the size of
the theoretically allowed $P$-$\dot{P}$ space for pulsars by
decreasing the radius of curvature of the magnetic field lines ${\cal
R}_c$ in the region where the pair cascade occurs, i.e., above the
polar cap. Since pair creation is more efficient along tightly-curved
field lines, a smaller radius of curvature allows the cascade to occur
at a lower $\gamma_0$. However, the magnetosphere must be highly
non-dipolar near the polar cap in order for the models to include all
pulsars on the active side of the death line. For example, the offset
dipole model can fully match observation only if the dipole is offset
by $0.95R$ or more, while an expanded polar cap (due to a twisted
field or delayed field reconnection) must be hundreds of times larger
than the pure dipole cap; see Fig.~\ref{fig:death}. It is unclear
whether such a strongly non-dipolar field is stable
(cf. \citealt{beloborodov07}, where a field twist $\Delta\phi \ga
1$~radian is unstable even in magnetars). In addition, observations of
pulsar radio emission suggest that this emission is coming from a
purely dipole region of the magnetosphere (e.g., \citealt{rankin03}),
so any non-dipole structure at the surface must give way to a dipole
configuration at the altitude where radio pulses are generated
(typically about $1\%$ of the light cylinder radius; e.g.,
\citealt*{gangadhara01,dyks04}). In our simulation, we model the
effects of a non-dipole magnetosphere by giving the magnetic field a
large curvature (${\cal R}_c = R$; cf. \citealt{ruderman75}) at the
surface and a dipole curvature at higher altitudes ($r>2R$); the
implementation of this non-dipole model into our simulation is
discussed in more detail in Section~\ref{sec:model}. The death line in
this approximation is shown on the right panel of
Fig.~\ref{fig:death}. Because of the increased efficiency of the
cascades in these highly-curved magnetic fields, the Lorentz factor of
the primary electron upon emerging from the gap is a factor of
$\sim10$ lower than in the dipole case, in the range $\gamma_0 \simeq
(2$--$4)\times10^6$ (\citealt{hibschman01a,medin07b}).

In some models the pair cascade occurs primarily within the gap, due
to efficient inverse Compton scattering by the primary electron,
rather than above the gap. These cascades occur at much lower energies
of the primary electron ($\gamma \sim 10^3$--$10^4$), such that all
observed pulsars can provide the voltage necessary to initiate this
type of cascade. However, ICS cascades are generally very weak,
producing $\la 10$ electron-positron pairs per primary electron, while
most radio emission models assume a secondary particle density
100-1000 times that of the primary electron beam (e.g., for the
development of a two-stream instability; see \citealt{usov02}). The
inclusion of this type of cascade into our simulations is discussed
below, in Section~\ref{sub:phparam}.

In some models the spindown power of the neutron star differs from
that of an orthogonal magnetic dipole, Eq.~(\ref{eq:orthoEdot}). For
example, \citet{contopoulos06} show that for nearly-aligned pulsars
(with the angle between the magnetic field and rotation axes $\alpha
\la 30^\circ$), the spindown power can be approximated
by
\be
\dot{E} \simeq \frac{2\Phi_{\rm cap}^2c}{3}\left(1-\frac{\Phi}{\Phi_{\rm cap}}\right) \,.
\ee
This equation is nearly the same as Eq.~(\ref{eq:orthoEdot}) for young
pulsars, where $\Phi \ll \Phi_{\rm cap}$. However, for pulsars near
death, with $\Phi \la \Phi_{\rm cap}$, the spindown power is much
lower for a given polar cap voltage. Conversely, for a given $P$ and
$\dot{P}$ [which determines the observed spindown power,
Eq.~(\ref{eq:Edot})] the polar cap voltage is much larger than would
be assumed by using Eq.~(\ref{eq:orthoEdot}). Depending on alignment
and how close the gap potential drop $\Phi$ is to $\Phi_{\rm cap}$,
many pulsars which were predicted to be dead may actually have a
potential drop large enough to generate pair cascades ($\Phi \simeq
10^7$~V). According to Contopoulos \& Spitkovsky, the standard death
line shown in Fig.~\ref{fig:death} is consistent with the observed
$P$-$\dot{P}$ values for all pulsars if the magnetic inclination
angles $\alpha$ of nearly-dead pulsars are weighted towards
$\alpha=0$. If this is the case, we can use $\gamma_0 \simeq
(2$--$4)\times10^7$ for all pulsars and do not need to invoke a
strongly-curved magnetic field geometry (${\cal R}_c \simeq R$) or
efficient inverse Compton scattering by the primary electron in order
to reproduce the observed pulsar death line.

\subsection{Primary photons}
\label{sub:phparam}

As the primary electron traverses the acceleration region it can pass
through two distinct regimes (see, e.g.,
\citealt{hibschman01a}). First, for Lorentz factors $\gamma \sim
10^2$--$10^4$, the electron efficiently upscatters photons through the
inverse Compton process. Second, for Lorentz factors $\gamma \ga
10^6$, the electron efficiently emits curvature radiation. The above
treatment of the acceleration zone (Section~\ref{sub:eparam}) is best
suited for pulsars where the electrons reach the second regime. In
that case we can safely ignore the contributions to the cascade made
by photons emitted before the primary electron reaches full energy
($\gamma_0$), since the number and energy of photons emitted through
curvature radiation increase strongly with $\gamma$ (i.e.,
$\dot{N}_{\rm CR} \propto \gamma$, $E_{\rm CR} \propto \gamma^3$). The
approximation is poor, however, if inverse Compton scattering and
subsequent pair production within the gap is efficient enough to
screen the accelerating potential before the electrons can reach the
second regime. In that case the photons produced in the gap are
critical to the cascade, while the photons produced above the gap have
a negligible effect on the cascade (the upscattered photons must
travel a finite distance before pair production in order to screen the
gap; in that distance the primary electron is accelerated to above
resonance and exits the gap with a Lorentz factor between the first
and second regimes of efficient photon production).

Because ICS is strongly peaked at resonance, primary electrons
traveling through this second type of gap will emit a large number of
photons at a characteristic ``resonance'' energy and very few at other
energies. The effect of this type of gap on the cascade is better
modeled by $N_0$ photons of energy $\epsilon_0$ emitted from the
surface (cf. \citealt{arendt02}), rather than one electron of energy
$\gamma_0 m_ec^2$.  We therefore run a second version of the
simulation, this time tracking the cascade initiated by a ``primary''
photon. The quantitative results of this simulation can be multiplied
by $N_0$ to obtain the full cascade results (e.g., the number of
electron-positron pairs produced per primary electron).

For a primary electron resonantly upscattering primary photons, we
estimate the value of $\epsilon_0$ as follows. When the primary
electron reaches a Lorentz factor $\gamma$, it upscatters photons to a
mean energy (e.g., \citealt{beloborodov07})
\be
\epsilon = \gamma\left(1-\frac{1}{\sqrt{1+2\beta_Q}}\right)m_ec^2 \,,
\ee
where $\beta_Q=B/B_Q$ is the ratio of the magnetic field strength to
the critical quantum field strength, $B_Q=4.414\times 10^{13}$~G.
The primary electron is most efficient at scattering photons when
\be
\gamma = \gamma_{\rm crit} \simeq \epsilon_c/kT \,,
\label{eq:gammac}
\ee
where $T$ is the surface temperature of the star and $\epsilon_c=\hbar
eB/m_ec$ is the electron cyclotron energy.\footnote{Note that the
actual resonance condition is $\epsilon_i\gamma(1-\beta\cos\psi)
=\epsilon_c$, where $\epsilon_i\sim kT$ is the initial (before
scattering) photon energy, $\beta=\sqrt{1-1/\gamma^2}$ is the ratio of
the electron speed to the speed of light and $\psi$ is the incident
angle of the photon with respect to the electron's trajectory.
However, because the scattering rate depends inversely on $\gamma$
(see Appendix \ref{sec:rics}), photons with $\cos\psi \ll 1$ are far
more likely to scatter off the electron than photons with $\cos\psi
\la 1$.} Therefore, for photons scattered from near the surface, where
$B=B_p$, the typical energy of a scattered photon is
\be
\epsilon_{\rm RICS} \simeq 70 B_{p,12} T_6^{-1} f\left(\beta_Q\right)~{\rm MeV} \,,
\label{eq:rics0}
\ee
where $T_6$ is the surface temperature in units of $10^6$~K and
$f\left(\beta_Q\right) = 1-1/\sqrt{1+2\beta_Q}$ is evaluated at the
surface. Setting $\epsilon_0=\epsilon_{\rm RICS}$ and assuming a $T_6$
range of $0.3$--$3$, we obtain $\epsilon_0$ in the range $1$--$10$~MeV
at $B_{p,12}=1$ up to $(0.4$--$4)\times 10^5$~MeV at $B_{p,12}=1000$.

The number of resonant ICS photons scattered by the primary electron
is more difficult to estimate, since it depends on the acceleration
model. Inner gaps with space-charge-limited flows have (e.g.,
\citealt{hibschman01a,thesis})
\be
N_0 \simeq 10 B_{p,12}^{-1} P_0^{3/4} T_6^{5/2} \,.
\label{eq:N0}
\ee
Inner vacuum gaps, with accelerating electric fields on the order of 6
times larger, have $N_0$ values at least 20-100 times smaller (the
primary electron is more rapidly accelerated out of resonance; see,
e.g., ML07).

Note that we can also use this second cascade simulation as a
diagnostic tool for the main simulation. For example, we can study the
partial cascade initiated by a single curvature radiation photon
emitted at some altitude in the magnetosphere to understand how the
strength of the local magnetic field affects the cascade. The
characteristic energy of curvature photons is
\be
\epsilon_{\rm CR} = \frac{3\gamma^3\hbar c}{2{\cal R}_c} \,.
\label{eq:ecr}
\ee
For dipole fields the typical curvature photon has an energy
$\epsilon_{\rm CR} \la 10^3$--$10^4$~MeV (for $\gamma \le \gamma_0 \sim
10^{7.5}$), while for fields with ${\cal R}_c=R$ we have
$\epsilon_{\rm CR} \la 10^2$--$10^3$~MeV (for $\gamma \le \gamma_0 \sim
10^{6.5}$).

\section{Numerical simulation of pair cascades: Physics ingredients and methods}
\label{sec:model}

\begin{figure}
\includegraphics[width=\columnwidth]{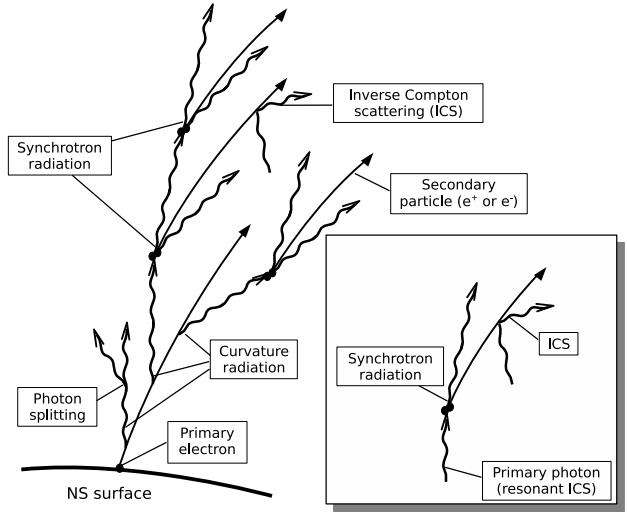}
\caption[The magnetosphere pair cascade]
{A schematic diagram showing the magnetosphere pair cascade, from
initiation by a high-energy electron to completion. Photon splitting
is also shown. The inset shows the beginning of a cascade
initiated by a photon upscattered through the inverse Compton
process.}
\label{fig:cascade}
\end{figure}

The general picture of the pair cascade as modeled by our numerical
simulation is sketched in Fig.~\ref{fig:cascade}. At the start of the
simulation, an electron with initial Lorentz factor $\gamma_0 \sim
10^{6.5}$--$10^{7.5}$ (Section~\ref{sub:eparam}) travels outward from
the stellar surface along the last open field line. As it travels it
emits high-energy photons through curvature radiation or inverse
Compton upscattering. The simulation tracks these photons as they
propagate from the point of emission through the magnetosphere, until
they decay into electron-positron pairs through magnetic pair
production or escape to infinity. In the superstrong field regime, the
photon (if it has the correct polarization; see
Section~\ref{sub:photon}) also has a finite probability of splitting
into two photons before pair production, in which case we follow the
two child photons in a similar way. The electrons and positrons
created by these photons are tracked as they radiate away their
transverse momenta through synchrotron radiation and then gradually
lose their forward momenta through inverse Compton
scattering. Subsequent generations of photons and electrons/positrons
are also tracked, in a recursive manner, and the total numbers and
energies of photons and electrons $+$ positrons that escape the
magnetosphere are recorded. We track each particle until it is
destroyed or reaches a height comparable to the light cylinder
radius. Tracking particles out to the light cylinder is most important
for the primary electron, since although in general there is no
discernible pair production above the radius $r \sim 10R$, curvature
radiation continues up to very high altitudes (albeit very weakly,
with photon energies $\epsilon \la 10$~MeV near the light cylinder).

In the second version of our cascade simulation, a photon with energy
$\epsilon_0 \sim 10^3$--$10^5$~MeV (see Section~\ref{sub:phparam})
travels outward from the last open field line at the stellar surface,
in the direction tangent to the magnetic field at that point. While in
theory the photon should be emitted at an angle to the field line of
$\Delta\Theta_{\rm ph} \la 1/\gamma_e$, where $\gamma_e$ is the final
Lorentz factor of the electron after emitting the photon, in practice
this does not matter, as there is no change in the final products of
the cascade whether we use $\Delta\Theta_{\rm ph} = 1/\gamma_e$ or
$\Delta\Theta_{\rm ph} = 0$. (This is true even for photons
upscattered by resonant ICS at $B \ga B_Q$; see
Section~\ref{sub:rics}). The primary photon and subsequent
generations of electrons/positrons and photons are tracked in the same
way as in the main cascade simulation. The cascade as modeled by the
second simulation is sketched in the inset of Fig.~\ref{fig:cascade}.

The input parameters for our simulation are the initial energy of the
electron ($\gamma_0m_ec^2$) or photon ($\epsilon_0$), its initial
position (in most cases, the intersection of the last open field line
with the stellar surface), the general pulsar parameters (surface
magnetic field strength $B_p=10^{12}$--$10^{15}$~G, rotation period
$P=0.33$--$5$~s, and surface temperature $T=10^6$~K or
$5\times10^6$~K), and the geometry of the magnetic field.  In each run
of the simulation, the magnetic field structure is given by one of two
topologies: (i) a pure dipole field geometry; or (ii) a more complex
field geometry near the stellar surface which gradually reverts to
dipole at higher altitudes (a non-dipole, or ``multipole'' field
geometry). Modeling the dipole field geometry is straightforward (see,
e.g., \citealt*{hoensbroech98}), but there is no obviously correct way
to model the geometry for the multipole field case (see
Section~\ref{sub:eparam}). Two features of a multipole field geometry
have a strong effect on the pair cascade dynamics and must be
incorporated into our model: First, the radius of curvature ${\cal
R}_c$ is much smaller than dipole (we choose ${\cal R}_c=R$, the
stellar radius) near the surface of the star. This leads to a much
larger number and peak energy of photons emitted through curvature
radiation than in the dipole field case. Second, as a photon
propagates through the magnetosphere the angle between the photon and
the field, which scales like $\Delta \Theta_{\rm ph} \sim s_{\rm
ph}/{\cal R}_c$, where $s_{\rm ph}$ is the distance traveled by the
photon from the point of emission, grows much faster than dipole. This
leads to a much more rapid decay of photons into pairs than in the
dipole case. The integration of these two features into our model is
discussed in the relevant subsections below (Section~\ref{sub:primary}
and Section~\ref{sub:photon}, respectively). Note that
\citet{arendt02} consider the first aspect of a multipole field
geometry in their model (that ${\cal R}_c=R$) but ignore the
second. In all of the simulation runs we assume that the local
magnetic field strength varies as in the dipole case,
\be
B(r,\theta,\phi) = B_p \left(\frac{R}{r}\right)^3 \frac{\sqrt{3\cos^2\theta+1}}{2} \,,
\label{eq:Blocal}
\ee
where $(r,\theta,\phi)$ are the spherical coordinates (with the
magnetic north pole at $r=R$ and $\theta=0$). Our approximation
therefore ignores any amplification of the field strength near the
surface caused by the complex topology.

For simplicity we consider a ``two-dimensional'' cascade model in
which all photons are emitted and travel in the plane defined by the
local magnetic field line. Both the photons and the
electrons/positrons are tracked in the ``corotating'' frame (the frame
rotating with the star), and any bending of the photon path due to
rotation is ignored -- this is expected to be valid since the cascade
takes place far inside the light cylinder.  Thus we shall also call
this corotating frame the ``lab'' frame for the remainder of the
paper. With this approximation the particle positions and trajectories
are defined only in terms of $r$ and $\theta$ in our simulation. We
justify this approximation below (Sections~\ref{sub:primary} and
\ref{sub:photon}). As an additional simplification we ignore any
effects of general relativity on the photon/particle trajectory.

The cascade simulation can naturally be divided into three parts: (i)
the propagation and photon emission of the primary electron; (ii)
photon propagation, pair production, and splitting; and (iii) the
propagation and photon emission of the secondary\footnote{In this
paper we use the term ``secondary'' to refer to any cascade particle
except the primary electron, positron, or photon that initiates the
cascade. The fourth generation of electrons and positrons, e.g., are
all ``secondary'' particles.} electrons and positrons. Each of these
aspects of the simulation is described in a separate subsection
below. At the end of this section, cascades initiated by primary
photons are discussed.

\subsection{Propagation and photon emission of the primary electron}
\label{sub:primary}

In our cascade simulation, the primary electron starts at the position
$(r_0,\theta_0)=(R,\theta_0)$ (i.e., at some angle $\theta_0$
from the magnetic pole
on the neutron star surface) with the initial energy $\gamma_0
m_ec^2$, and moves outward along the local magnetic field line. The
initial position of the primary electron is chosen so that it moves
along the last open field line, whose location at the surface is given
by the polar cap angle: $\theta_0 = \theta_{\rm
cap}\equiv\sqrt{R/r_{\rm LC}}$, where $r_{\rm LC}=c/\Omega$ is the
light cylinder radius.\footnote{Since the last open field line is also
the most-tightly curved, this choice for the primary electron's
location gives us the largest possible cascade in our simplified
model. However, in a real magnetosphere there is no cascade at all
along the last open field line and weak cascades for field lines very
near the last ($\theta \la \theta_{\rm cap}$), since the potential
drop is zero on the boundary of the open field region. The strongest
cascades occurs on field lines neither at the edge of the open field
region nor at the center (where ${\cal R}_c \rightarrow \infty$).}

The primary electron moves outward along the field line in a stepwise
fashion. The lengths of the steps $\Delta s(r)$ are chosen so that a
uniform amount of energy $\Delta \gamma$ (we choose $\sim
0.001\gamma_0$) is lost by the electron in each step ($\gamma
\rightarrow \gamma-\Delta\gamma$):
\be
\Delta s(r) \simeq -\frac{\Delta\gamma}{d\gamma/ds} \,.
\ee
For an electron emitting curvature radiation,
\be
\frac{d\gamma}{ds} = -\frac{2}{3}\gamma^4 \frac{\alpha_f^2 a_0}{{\cal R}_c^2} \,, 
\ee
where $\alpha_f=e^2/(\hbar c)$ is the fine structure constant and $a_0$
is the Bohr radius. For a dipole field the radius of curvature is given by
\be
{\cal R}_c = \frac{r}{\sin\theta}\frac{\left(1+3\cos^2\theta\right)^{3/2}}{3+3\cos^2\theta} \,,
\label{eq:rcdipole}
\ee
while for a near-surface multipole field we use ${\cal R}_c=R$. As
discussed in Section~\ref{sub:eparam}, we do not consider photon
emission due to inverse Compton scattering here, since this process is
very inefficient once the primary electron has reached the energy
$\gamma_0 m_ec^2$. We do, however, consider in our simulation the
photon emission due to ICS by the secondary electrons and positrons
(see Section~\ref{sub:secondary}) which typically have $\gamma \ll
\gamma_0$. Note that we also indirectly include ICS in our second
cascade simulation (described in Section~\ref{sub:phprimary}), which
models photon-initiated cascades, by choosing photon energies
$\epsilon_0$ that are typical of ICS photons.

As the electron moves a distance $\Delta s$ along the field it emits
photons with energies divided into discrete bins (our simulation uses
$\sim50$ bins). The energy in each bin, $\epsilon$, is a constant
multiple of the characteristic energy of curvature photons
$\epsilon_{\rm CR} = 3\gamma^3\hbar c/(2{\cal R}_c)$, with
$\epsilon/\epsilon_{\rm CR}$ in the range $10^{-4}$--$10$. The number
of photons in a given energy bin emitted in one step is given by the
classical spectrum of curvature radiation (e.g., \citealt{jackson98}),
\be
\Delta N_\epsilon \simeq \Delta\epsilon \frac{dN}{d\epsilon} \simeq
\frac{\sqrt{3}}{2\pi}\frac{\alpha_f\Delta s}{{\cal R}_c}\frac{\gamma\Delta\epsilon}{\epsilon} F\left(\frac{\epsilon}{\epsilon_{\rm CR}}\right) \,,
\label{eq:curvature}
\ee
where $\Delta\epsilon$ is the spacing between energy bins and the
values of ${\cal R}_c$ and $\gamma$ used are averages over the
interval $\Delta s$. Here, $F(x)=x \int_x^\infty K_{5/3}(t) dt$ and
$K_{5/3}(x)$ is the $n=5/3$ Bessel function of the second kind. Note that  
$F(x) \propto x^{1/3}$ for $x \ll 1$, and 
$F(x) \propto \sqrt{x}e^{-x}$ for $x \gg 1$ (e.g., \citealt{erber66}).

\begin{figure}
\includegraphics[width=\columnwidth]{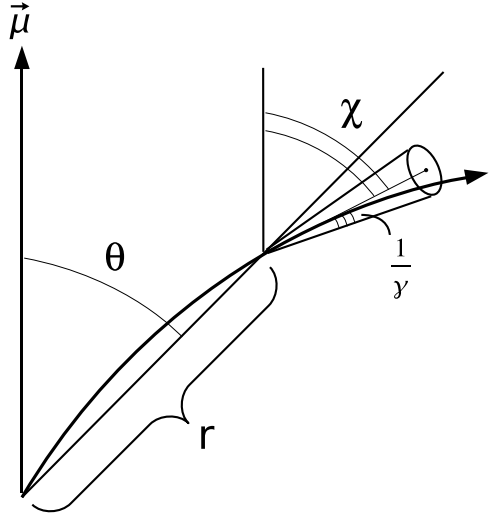}
\caption[The photon emission angle]
{A schematic diagram illustrating the photon emission angle. The
direction of the magnetic dipole axis is given by $\vec{\mu}$. The electron
(positron) follows the curved field line to the point $(r,\theta)$,
then emits a photon in a cone of width $1/\gamma$, inclined with respect
to the magnetic axis by an angle $\chi$.}
\label{fig:chiangle}
\end{figure}

The photons are emitted in the direction nearly tangent to the field
line at the current location of the electron $(r,\theta)$. For a dipole
field geometry the angle between the local magnetic field and the
magnetic dipole axis is given by
\be
\chi(\theta) = \theta+\arctan\left(\frac{\tan\theta}{2}\right) \,;
\label{eq:chiangle}
\ee
see Fig.~\ref{fig:chiangle}. There is an additional contribution to
the emission angle of $\sim1/\gamma$, due to relativistic beaming. In
reality this beaming angle is in a random direction; however, for our
two-dimensional approximation it can only be in the plane of the
magnetic field. The photon emission angle is given by the (projected)
sum of these two angles:
\be
\Theta_{\rm ph} = \chi + \frac{1}{\gamma}\cos\Pi \,,
\label{eq:thph}
\ee
where $\Pi$ is a random angle between 0 and $2\pi$. Note that ignoring
the three-dimensional aspect of the photon emission introduces an
error in the emission angle of order $1/\gamma$. This affects the
location at which the photon decays (into pairs) 
in our simulation, since photon decay
depends strongly on the intersection angle between the photon and the
magnetic field (see Section~\ref{sub:photon} below). However, as the
photon propagates through the magnetosphere these errors (which are on
the order of $1/\gamma \sim 10^{-7}$ for curvature photons and
$10^{-3}$ for resonant ICS photons) quickly become negligible in
comparison to the photon-magnetic field intersection angle, which
grows like $s_{\rm ph}/{\cal R}_c$ (and so reaches the angle
$1/\gamma$ by $s_{\rm ph} \sim 10^{-5}R$ for curvature radiation and
$\sim 0.1R$ for RICS).

We also use Eq.~(\ref{eq:thph}) for simulation runs with a multipole
field geometry. This is obviously a simplification, but we have found
that in practice the photon propagation direction has little effect on
the overall cascade product (as long as it points generally outward).
Far more important for the cascade is how the angle between the photon
and the magnetic field changes as the photon travels. As is discussed
in Section~\ref{sub:photon}, we artificially force this angle to
change more rapidly with distance than in the dipole case, to account
for the effect of the stronger field line curvature.

The total energy lost over each step is
\be
\sum_\epsilon \epsilon \Delta N_\epsilon \simeq \Delta\gamma m_ec^2 \,.
\ee
Only one photon is tracked for each energy bin $\epsilon$ at each step
$\Delta s$, so the photon is given a weighting factor $\Delta
N_{\epsilon}$. In addition to its initial position (the position of
the electron at the point of emission $r,\theta$) and propagation
direction ($\Theta_{\rm ph}$), the photon has a polarization
direction. For curvature radiation the polarization fraction is
between 50\% and 100\% polarized parallel to the magnetic field
curvature, depending on photon frequency (\citealt{jackson98}; see
also \citealt{rybicki79}). We therefore randomly assign the photon a
polarization in the ratio of one $\perp$ (perpendicular to the
field) to every seven $\parallel$ (parallel to the field) photons,
corresponding to 75\% averaged parallel polarization.

\subsection{Photon propagation, pair production, and splitting}
\label{sub:photon}

In our simulation, the photon is emitted/scattered from the point
$(r_{0,\rm ph},\theta_{0,\rm ph})$ with energy $\epsilon$, polarization
$\parallel$ or $\perp$, and weighting factor $\Delta N_\epsilon$ (to
represent multiple photons; Section~\ref{sub:primary}). It has an
optical depth to pair production, $\tau$, and to photon splitting,
$\tau_{\rm sp}$, both of which are set to zero at the moment of the
photon's creation.  The photon propagates in a straight line from the
point of emission, at an angle $\Theta_{\rm ph}$ with respect to the
magnetic dipole axis.  Note that in the corotating frame (which is the
frame we are working in for most of our simulation; but see
Section~\ref{sub:secondary}) the path of the photon is in reality
curved, with the angular deviation from a straight line growing
approximately as $s_{\rm ph} \Omega/c = s_{\rm ph}/r_{\rm LC}$
(cf. \citealt*{harding78}). Like the beaming angle
(Section~\ref{sub:primary}), this curved path modifies the growth of
the photon-magnetic field intersection angle and the location
of photon decay in our simulation. However, the total intersection angle grows
much faster with photon distance $s_{\rm ph}$ than the deviation does
($\sim s_{\rm ph}/{\cal R}_c$ versus $s_{\rm ph}/r_{\rm LC}$, or a
factor of $r_{\rm LC}/{\cal R}_c \simeq 100 P_0^{1/2}$ larger for
dipole fields), so we can safely ignore this deviation.

\begin{figure}
\includegraphics[width=\columnwidth]{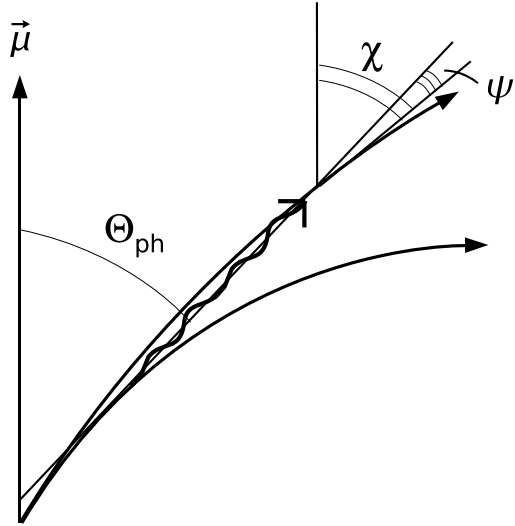}
\caption[The angle between the photon and the magnetic field]
{A schematic diagram for deriving the angle between the photon and the
magnetic field, $\psi$. The direction of the magnetic dipole axis is
given by $\vec{\mu}$. The photon propagates through the magnetosphere
with angle $\Theta_{\rm ph}$ with respect to the magnetic axis [see
Eq.~(\ref{eq:thph})]. The local magnetic field makes an angle $\chi$
with respect to the magnetic axis [Eq.~(\ref{eq:chiangle})].}
\label{fig:psiangle}
\end{figure}

In each step the photon travels a short distance through the
magnetosphere, $\Delta s_{\rm ph} < 0.05r_{\rm ph}$, where $(r_{\rm
ph},\theta_{\rm ph})$ refers to the current position of the photon;
our method for choosing the value of $\Delta s_{\rm ph}$ for a given
photon is discussed at the end of this section. At the new position
the change in the optical depth for pair production, $\Delta \tau$,
and for photon splitting, $\Delta \tau_{\rm sp}$, are calculated:
\be
\Delta \tau \simeq \Delta s_{\rm ph} R_{\parallel,\perp} \,,
\ee
\be
\Delta \tau_{\rm sp} \simeq \Delta s_{\rm ph} R^{\,\rm sp}_{\parallel,\perp} \,,
\ee
where $R_{\parallel,\perp}=R'_{\parallel,\perp} \sin\psi$ is the
attenuation coefficient for the $\parallel$ or $\perp$ polarized photons,
$\psi$ is the angle of intersection between the photon and the local
magnetic field, and $R'$ is the attenuation coefficient in the
``perpendicular'' frame (i.e., the frame where the photon 
propagates perpendicular to the local magnetic field). For a dipole
field geometry the intersection angle is given by
\be
\psi = \chi(\theta_{\rm ph}) - \Theta_{\rm ph} \,,
\label{eq:psi}
\ee
where $\Theta_{\rm ph}$ is given by Eq.~(\ref{eq:thph}) and
$\chi(\theta_{\rm ph})$ is the angle between the magnetic axis and the
magnetic field at the current location of the photon
[Eq.~(\ref{eq:chiangle})]; see Fig.~\ref{fig:psiangle} for a
sketch. For the near-surface multipole field geometry we set
\be
\tan\psi = \frac{s_{\rm ph}}{{\cal R}_c} = \frac{s_{\rm ph}}{R} \,.
\ee
This approximation has the advantage of accounting for the effect of a
strong field curvature on the photon propagation without requiring
knowledge of the actual field topology.

The total attenuation coefficient (in the perpendicular frame) for
pair production is given by (suppressing the subscripts
$\parallel,\perp$) $R'= \Sigma_{jk} R'_{jk}$, where $R'_{jk}$ is the
attenuation coefficient for the channel in which the photon produces
an electron in Landau level $j$ and a positron in Landau level $k$,
and the sum is taken over all possible states for the
electron-positron pair. Since pair production is symmetric with
respect to the electron and the positron, $R'_{jk} = R'_{kj}$; for
simplicity we hereafter use $R'_{jk}$ to represent the combined
probability of creating the pair in either the state $(jk)$ or $(kj)$
(i.e., $R'^{\,\rm new}_{jk} = R'^{\,\rm old}_{jk}+R'^{\,\rm
old}_{kj}$). For a given channel $(jk)$, the threshold condition for
pair production is
\be
\epsilon'>E_j+E_K,
\label{eq:thresh}
\ee
where $\epsilon'=\epsilon \sin\psi$ is the photon energy in the
perpendicular frame and $E_n=m_ec^2\sqrt{1+2\beta_Q n}$ is the minimum
energy of an electron/positron in Landau level $n$ (the energy of an
electron/positron with the momentum along the magnetic field
$p_\parallel=0$). In dimensionless form, the condition
[Eq.~(\ref{eq:thresh})] can be written as
\bal
x = {}& \frac{\epsilon'}{2m_ec^2} = \frac{\epsilon}{2m_ec^2}\sin\psi \nonumber\\
 > {}& x_{jk} \equiv \frac{1}{2}\left[\sqrt{1+2\beta_Q j} + \sqrt{1+2\beta_Q k}\right] \,.
\label{eq:xjk}
\eal
Note that $x_{jk}$ satisfies
\be
x_{00} < x_{01} < x_{02} < \!\!
\begin{array}{ll}
x_{11} < x_{03} < \cdots \,, & \beta_Q < 4 \,; \\
x_{03} < x_{11} < \cdots \,, & \beta_Q > 4 \,.
\end{array}
\label{eq:xcond}
\ee
The first three attenuation coefficients (corresponding to the three
lowest threshold levels $x_{00},x_{01},x_{02}$) for both $\parallel$
and $\perp$ polarizations are given in Appendix~\ref{sec:pair},
Eqs.~(\ref{par00eq})-(\ref{perp02eq}); see also
\citet{daugherty83}. Note that $R'_{\perp,00}=0$, and thus the first
non-zero attenuation coefficient for $\perp$ polarized photons is
actually $R'_{\perp,01}$, not $R'_{\perp,00}$.

\begin{figure}
\includegraphics[width=\columnwidth]{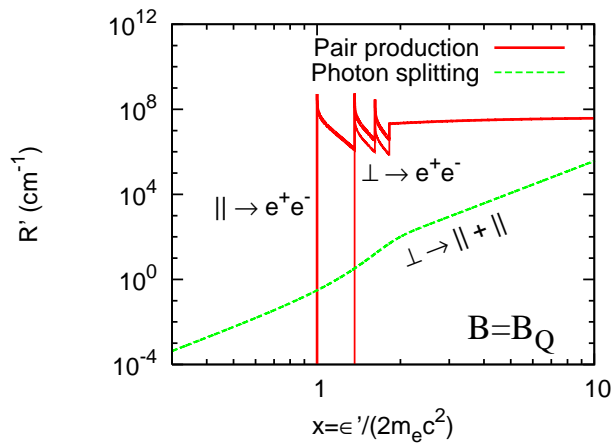}
\caption[Attenuation coefficients for both photon splitting and pair production]
{Attenuation coefficients in the perpendicular frame (the frame where
the photon is traveling perpendicular to the magnetic field), for both
photon splitting, labeled by $\perp \rightarrow \parallel +
\parallel$, and pair production, labeled by $\parallel \rightarrow e^+
e^-$ and $\perp \rightarrow e^+ e^-$. The local magnetic field
strength is $B=B_Q\equiv4.414\times10^{13}$~G\@.}
\label{fig:atten}
\end{figure}

In our simulation a photon is typically created with $x$ below the
first threshold ($x_{00}$ or $x_{01}$, depending on the photon
polarization). As long as $x$ remains below the first threshold,
$R'=0$ and the optical depth to pair production remains zero.  As the
photon propagates into the magnetosphere and crosses the first
threshold, $R'>0$, $\Delta\tau>0$, and $\tau$ begins to grow. As it
continues to travel outward, both $\tau$ and the number of Landau
levels available for pair production $j_{\rm max}$ and $k_{\rm max}$
increase. Depending on the local magnetic field strength
[Eq.~(\ref{eq:Blocal})], the photon may reach a large enough optical
depth ($\tau\sim 1$) for pair production after crossing only a few
thresholds (so that $j_{\rm max}$ and $k_{\rm max}$ are small) or
after crossing many thresholds (so that $j_{\rm max}$ and $k_{\rm
max}$ are very large). For ``weak'' magnetic fields ($\beta_Q \la
0.1$) the optical depth increases slowly with $s_{\rm ph}$ and it is
valid to use the $j_{\rm max},k_{\rm max} \gg 1$ asymptotic
attenuation coefficient for pair production (e.g., \citealt{erber66}),
\be
R'_{\parallel,\perp} \simeq \frac{0.23}{a_0}\beta_Q\exp\left(-\frac{4}{3x\beta_Q}\right) \,,
\label{eq:asymp}
\ee
which applies for both polarizations. For stronger fields, however,
pairs are produced in low Landau levels, and the more accurate
coefficients of Daugherty \& Harding must be used. In
Appendix~\ref{sub:atten} we find that the critical magnetic field
strength separating these two regimes is 
\be 
B_{\rm crit} \simeq 3\times10^{12}~{\rm G}
\label{eq:Bcrit}
\ee
[see Eq.~(\ref{eq:critatten})]. 
We also find that the
boundary between the two regimes is very sharp: pairs are either
created at the first few Landau levels ($n\le 2$) for $B \ga B_{\rm
crit}$ or in very high Landau levels for $B \la B_{\rm crit}$, with
very few electrons/positrons created in intermediate Landau
levels. Therefore, in our simulation we only consider the first three
attenuation coefficients for $\parallel$-polarized photons
($R'_{\parallel,00},R'_{\parallel,01},R'_{\parallel,02}$) and the
first two non-zero attenuation coefficients for $\perp$-polarized
photons ($R'_{\perp,01},R'_{\perp,02}$). If the photon reaches the
threshold for the $(03)$ or $(11)$ channel [whichever is reached
first; see Eq.~(\ref{eq:xcond})], we use the asymptotic formula,
Eq.~(\ref{eq:asymp}). The total attenuation coefficient for pair
production (as given by this approximation) is plotted in
Fig.~\ref{fig:atten} for both $\parallel$ and $\perp$ polarizations at
$\beta_Q=1$.

We include photon splitting in our simulations. Based on the kinetic
selection rule (\citealt{adler71,usov02}, but see \citealt{baring01}),
only the process $\perp\rightarrow\parallel\,\parallel$ is allowed.
Therefore, for $\parallel$-polarized photons, the attenuation
coefficient for photon splitting is zero ($R'^{\,\rm sp}_{\parallel} =
0$). For $\perp$-polarized photons we use the following formula,
adapted from the numerical calculation of \citet{baring97}:
\be
R'^{\,\rm sp}_{\perp\rightarrow\parallel\,\parallel} \simeq \frac{\frac{\alpha_f^2}{60\pi^2a_0}\left(\frac{26}{315}\right)^2 (2x)^5 \beta_Q^6}{\left[g(\beta_Q,x)+0.05\right]\left[0.25g(\beta_Q,x)+20\right]} \,,
\label{eq:split}
\ee
where $g(\beta_Q,x) = \beta_Q^3 \exp(-0.6x^3)$. For $x \le 1$, this
expression reproduces the results of Baring \& Harding to better than
$10\%$ at both $\beta_Q \le 0.5$ and $\beta_Q \gg 1$, while
underestimating the results at $\beta_Q=1$ by less than $30\%$.  The
$\perp\rightarrow\parallel\,\parallel$ attenuation coefficient for
photon splitting is plotted in Fig.~\ref{fig:atten} at
$B=B_Q$. Because the attenuation coefficient $R'^{\,\rm sp}$ drops
rapidly with field strength for $\beta_Q < 1$, photon splitting is
unimportant for $\beta_Q \la 0.5$ (e.g., \citealt{baring01}). However,
for $\perp$-polarized photons propagating in superstrong fields
$\beta_Q \ga 0.5$, photon splitting is the dominant attenuation
process: even though above the first threshold ($x \ge x_{01}$ for
$\perp$ photons) the attenuation coefficient for photon splitting is
much smaller than that for pair production, in superstrong fields the
photon splits before reaching the first threshold (see
Fig.~\ref{fig:atten}).

In the simulation, whenever $\tau \ge 1$ or $\tau_{\rm sp} \ge 1$ the
photon is destroyed (i.e., turned into a pair or two photons). More
precisely, the photon should only be destroyed with probability
$1-\exp(-\tau)$. But in practice we find that such a refinement has a
negligible effect on the cascade result. If $\tau_{\rm sp} \ge 1$ the
photon splits into two. As a simplification we assume that each photon
takes half of the energy of the parent photon
(cf. \citealt{baring97}); therefore, at the point of photon splitting
a new photon is created with an energy $0.5\epsilon_0$ and a weighting
factor $2\Delta N_\epsilon$ (i.e, the simulation photon represents two
actual photons). The new photon is $\parallel$-polarized (for the
$\perp\rightarrow\parallel\,\parallel$ process) and is assumed to be
traveling in the same direction as the parent photon, $\Theta_{\rm
ph}$. If $\tau \ge 1$ the photon creates an electron-positron pair.
For $B\la B_{\rm crit}\sim 3\times 10^{12}$~G, the pairs are created
in high Landau levels (see above), and we assume that the electron and
positron each shares half of the photon energy and travels in the same
direction as the photon: thus $\gamma m_ec^2=\epsilon/2$ and the
electron/positron's magnetic pitch angle is $\Psi=\psi$.  This
approximation is valid as long as $x \beta_Q \la 0.1$ (see
\citealt{daugherty83}), which according to ML07 is satisfied for $B\la
B_{\rm crit}$.  When $B\ga B_{\rm crit}$, the electron and positron
are created in low Landau levels (we choose the maximum allowed
values, $j_{\rm max},k_{\rm max}$, since this channel dominates the
total attenuation coefficient), with energies given by
Eq.~(\ref{eq:Ee}) of Appendix~\ref{sub:levels}.

In the simulation we try to find the photon-magnetic field
intersection angle at which pair creation occurs, $\psi_{\rm pair}$,
to an error of less than $10\%$. If the error in $\psi_{\rm pair}$ is
too large, the electron and positron will be created in the wrong
Landau level and will emit too many or too few synchrotron photons
(see Section~\ref{sub:secondary}). To accurately determine $\psi_{\rm
pair}$ we use the following procedure in our simulation: The photon's
first full step, $s_0$, should be small enough that the probability of
pair production is negligible at $s_0$ but large enough that the
probability grows rapidly with subsequent steps. At high fields a good
choice for $s_0$ is the location of the first non-zero threshold
($x=x_{00}$ for $\parallel$ polarization or $x=x_{01}$ for $\perp$
polarization), since the attenuation coefficients are large enough to
allow pair production in a distance much shorter than $1$~cm. At low
fields a good choice is the point where $x\beta_Q = 1/20$. At this
point the mean free path for pair production is much larger than the
gap height while for $x\beta_Q = 1/10$, e.g., the mean free path is
much smaller than the gap height. Therefore, $s_0$ is chosen such
that it solves
\be
x =
\left\{
\begin{array}{ll}
x_{00} \,, & \beta_Q > 1/20 \,\mbox{ and $\parallel$ polarization;} \\
x_{01} \,, & \beta_Q > 1/20 \,\mbox{ and $\perp$ polarization;} \\
1/(20\beta_Q) \,, & \beta_Q < 1/20 \,.
\end{array}
\right.
\label{eq:s0}
\ee
Note that both $x$ and $\beta_Q$ depend on $s_0$, so the value of
$s_0$ must be found numerically.  Since $\sin\psi \simeq s_{\rm
ph}/{\cal R}_c$ (for small angles), this distance is approximately
given by
\be
s_0 \simeq {\cal R}_c \frac{2m_ec^2}{\epsilon}\left(1+\frac{1}{20\beta_Q}\right).
\ee
In our simulation, for $\parallel$ polarizations (no photon
splitting), the photon moves directly to $s_0$ in one step.  For
$\perp$ polarizations, the photon moves to $s_0$ in 10 steps (with
step sizes $0.1s_0$), allowing for the possibility of photon splitting
before reaching this point. In either case, once the photon reaches
$s_0$ it steps outward in the manner described at the beginning of
this section. At high fields ($B\ga B_{\rm crit}$) we choose the step
size to be $\Delta s_{\rm ph} = 0.1s_0(x_{01}-x_{00})/x_{00}$ for
$\parallel$ polarizations or $0.1s_0(x_{02}-x_{01})/x_{02}$ for
$\perp$ polarizations.  At low fields ($B\la B_{\rm crit}$) we choose
$\Delta s_{\rm ph} = 0.1s_0$.

Sometimes the photon does not pair produce (or split) before exiting
the magnetosphere. Conveniently, we do not have to track the photons
out to the light cylinder to know whether pair production will
occur. Once a photon reaches the $x_{03}$ or $x_{11}$ threshold, such
that the asymptotic expression for pair production
Eq.~(\ref{eq:asymp}) can be used (i.e., when $B \la B_{\rm crit}$),
then the growth in optical depth depends ``exponentially'' on $x
\beta_Q$ [since $\Delta \tau \propto \exp\{-1/(x\beta_Q)\}$]. Because
$x \beta_Q \propto s_{\rm ph}(r_{0,\rm ph}+s_{\rm ph})^{-7/2}$ reaches
a maximum at $s_{\rm ph} \simeq 0.4r_{0,\rm ph}$ and then rapidly
decreases (cf. \citealt{hibschman01a}), we assume in our simulation
that if the photon does not pair produce by
\be
s_{\rm ph,max} = 0.5r_{0,\rm ph} \,,
\label{eq:smax}
\ee
it will never pair produce and instead escapes the magnetosphere. Here
$r_{0,\rm ph}$ is the altitude of the photon at the emission point.
Note that Eq.~(\ref{eq:smax}) is also approximately valid for our
treatment of non-dipole fields (${\cal R}_c=R$ near the stellar
surface), since once the photon has traveled a distance $s_{\rm ph}
\simeq s_{\rm ph,max}$ it is in the dipole regime ($r > 2R$).

\subsection{Propagation and photon emission of the secondary electrons and positrons}
\label{sub:secondary}

\subsubsection{Synchrotron radiation}
\label{sub:synch}

In the corotating frame (the ``lab'' frame) the secondary electron (or
positron) is created with energy $\gamma m_ec^2$, pitch angle $\Psi$
[with the corresponding Landau level $n$; see Eq.~(\ref{eq:gmp})
below] and weighting factor $\Delta N_\epsilon$
(Section~\ref{sub:primary}). For the purpose of tracking the
synchrotron emission from the electron it is easier to work in the
``circular'' frame, the frame in which the electron has no momentum
along the magnetic field direction and only moves transverse to the
field in a circular motion. Note that this frame is in general
different from the perpendicular frame (defined in
Section~\ref{sub:photon}) of the progenitor photon; only if the
electron-positron pair is created exactly at threshold [$x=x_{jk}$;
see Eq.~(\ref{eq:xjk})] are the two frames the same. The energy of the
electron in the circular frame, $E_\perp=\gamma_\perp m_ec^2$, is
related to that in the lab frame by
\be
\gamma_\perp =
\sqrt{\gamma^2\sin^2\Psi+\cos^2\Psi} = \sqrt{1+2\beta_Q n} \,.
\label{eq:gmp}
\ee
Note this expression also gives a relation between $\gamma_\perp$ and
$n$; we shall use $\gamma_\perp$ and $n$ interchangeably to refer to
the electron's energy in the circular frame.

In the circular frame $E_\perp$ is radiated away through synchrotron
emission on the timescale
\bal
t_{\rm synch} \simeq {}& \left|\frac{E_\perp}{P_{\rm synch}}\right| = \frac{\gamma_\perp m_ec^2}{\frac{2e^2}{3c^3}(\gamma_\perp^2-1) c^2 \omega_c^2} \nonumber\\
 \simeq {}& 5\times10^{-16} B_{12}^{-2} \gamma_\perp^{-1}~{\rm s} \,,
\eal
where $\omega_c=eB/m_ec$ is the electron cyclotron frequency and
$B_{12}$ is the local magnetic field strength $B$ in units of
$10^{12}$~G. This decay time is much shorter than other relevant
cascade timescales (e.g., the timescale for $B$ or ${\cal R}_c$ to
change significantly, which is of order $r/c \ga 10^{-4}$~s, or the
timescale for the emission of resonant ICS photons, discussed later in
this section). Therefore, in the simulation the electron is assumed to
lose all of its perpendicular momentum $p_\perp$ ``instantaneously''
due to synchrotron radiation, before moving from its initial position
(cf. \citealt{daugherty82}). The final energy of the electron once it
reaches the ground Landau level (or $p_\perp=0$) is given by
\be
\gamma_\parallel = (1-\beta^2\cos^2\Psi)^{-1/2} = \gamma/\gamma_\perp \,,
\label{eq:gpar}
\ee
where $\beta=\sqrt{1-1/\gamma^2}$ is the electron velocity.

Since the synchrotron photon may carry an energy comparable to
$E_\perp$ of the parent electron, it is necessary to track the
electron energy after each photon is emitted in order to obtain
accurate synchrotron spectrum (this is in contrast to the case of
curvature radiation discussed in Section~\ref{sub:primary}, where a
large number of curvature photons can be emitted without significantly
affecting the energy of the parent electron). As a simplification, in
the circular frame the synchrotron photons are assumed to be emitted
isotropically in the plane of motion, such that no velocity kick is
imparted to the electron; thus the frame corresponding to circular
motion of the electron does not change over the course of the
synchrotron emission process. In other words, as the electron loses
its $p_\perp$, the Lorentz factors $\gamma$ and $\gamma_\perp$
decrease but $\gamma_\parallel$ is constant, and Eq.~(\ref{eq:gpar})
remains valid during the entire synchrotron emission process.

We adopt the following procedure in our simulation: In the circular
frame, the electron Lorentz factor $\gamma_\perp$ drops from its
initial value to $\gamma_\perp=1$ in a series of steps; when
$\gamma_\perp=1$ (i.e., $n=0$) synchrotron emission stops. In each
step one synchrotron photon is emitted, with an energy
$\epsilon_\perp$ that depends strongly on the ``current'' value of
$\gamma_\perp$. After the photon is emitted the energy of the electron
is reduced by the amount $\Delta
\gamma_\perp=\epsilon_\perp/m_ec^2$. In the next step another photon
is emitted with a new value of $\epsilon_\perp$, and so on.

In the simulation the photon energy $\epsilon_\perp$ of the
synchrotron radiation is chosen in one of three ways, depending on the
Landau level number $n$ of the electron.  (i) If the electron is
created in a high Landau level ($n\ge 3$), the energy of the photon is
chosen randomly, but with a weighting based on the asymptotic
synchrotron spectrum\footnote{This expression differs from the
classical synchrotron spectrum (e.g., \citealt{rybicki79}) in two
ways: First, a factor of $f=1-\epsilon_\perp/(\gamma_\perp m_ec^2)$
appears in several places throughout Eq.~(\ref{eq:synch}); when the
photon energy is equal to the electron energy
($\epsilon_\perp=\gamma_\perp m_ec^2$ or $f=0$) the asymptotic
expression goes to zero. Second, a term containing the function $G(x)$
appears in Eq.~(\ref{eq:synch}). While such a term appears in the
classical expressions for the radiation spectra of both $\parallel$-
and $\perp$-polarized photons, in the classical expression for the
total radiation spectra these terms cancel out. However, when the
quantum effect of the electron spin is considered there is an
asymmetry between the perpendicular and parallel polarizations such
that the $G(x)$ term remains.} (e.g., \citealt{sokolov68,harding87})
\bal
\frac{d^2 N}{dt\,d\epsilon_\perp} {}& = \frac{\sqrt{3}}{2\pi}\frac{\alpha_f \omega_c}{\epsilon_\perp} \nonumber\\
 & \times \left[f F\left(\frac{\epsilon_\perp}{f \epsilon_{\rm SR}}\right) + \left(\frac{\epsilon_\perp}{\gamma_\perp m_ec^2}\right)^2 G\left(\frac{\epsilon_\perp}{f \epsilon_{\rm SR}}\right)\right] \,,
\label{eq:synch}
\eal
where
\be
\epsilon_{\rm SR} = \frac{3}{2}\gamma_\perp^2 \hbar\omega_c
\label{eq:esr}
\ee
is the characteristic energy of the synchrotron photons,
$f=1-\epsilon_\perp/(\gamma_\perp m_ec^2)$ is the fraction of the
electron's energy remaining after photon emission,
$F(x)=x\int_x^\infty K_{5/3}(t)\,dt$, and $G(x)=x K_{2/3}(x)$ [cf.
Eq.~(\ref{eq:curvature})]. (ii) If $n=2$, the energy of the photon is
either that required to lower the electron to its ground state
($n=2\rightarrow0$) or the first excited state ($n=2\rightarrow1$),
with a probability that depends on the local magnetic field
strength. We do not use the exact transition rates for the $n=2$
state here. Instead, we use the following simplified prescription,
based on the results of \citet*{herold82} (see also
\citealt{harding87}): If $\beta_Q < 1$ the energy of the photon is
chosen to be that required to lower the electron to the first excited
state, $\epsilon_\perp =
m_ec^2\left(\sqrt{1+4\beta_Q}-\sqrt{1+2\beta_Q}\right)$. If $\beta_Q
\ga 1$ the energy of the photon is randomly chosen to be that required
to lower the electron to either the ground state [$\epsilon_\perp =
m_ec^2\left(\sqrt{1+4\beta_Q}-1\right)$], 50\% of the time, or the
first excited state, 50\% of the time.  (iii) If $n = 1$, the energy of the
photon is that required to lower the electron to its ground state,
$\epsilon_\perp = m_ec^2\left(\sqrt{1+2\beta_Q}-1\right)$. If the
electron is not in the ground state after emission of the synchrotron
photon (which could happen for the $n=2$ and $n \ge 3$ cases discussed
above, but not for the $n=1$ case), $\gamma_\perp$ is recalculated and
a new photon energy is chosen.

The energy of the photon is transformed from the circular frame into
the ``lab'' frame using
\be
\epsilon = \gamma_\parallel \epsilon_\perp \,.
\ee
The photon carries with it the same weighting factor $\Delta
N_\epsilon$ as the secondary particle that emitted it. Because the
photon is emitted in a random direction perpendicular to the magnetic
field in the circular frame, in the lab frame the angle of emission
(relative to the dipole axis) is approximately given by
\be
\Theta_{\rm ph}\simeq \chi + \Psi\cos\Pi \,,
\ee
where $\Pi$ is a random angle between 0 and $2\pi$, $\chi$ is
the angle between the local magnetic field and the dipole axis
and is given by Eq.~(\ref{eq:chiangle}), and the pitch angle is given by
Eqs.~(\ref{eq:gmp}) and (\ref{eq:gpar}):
\be
\Psi = \arcsin\left(\sqrt{\frac{\gamma_\perp^2-1}{\gamma_\perp^2\gamma_\parallel^2-1}}\right) \,.
\ee
For synchrotron radiation the polarization fraction is between 50\%
and 100\% polarized perpendicular to the magnetic field (which
is the exact opposite of the curvature radiation case; see
\citealt{rybicki79}). Therefore we randomly assign the photon a
polarization in the ratio of one $\parallel$ to every seven
$\perp$ photons (corresponding to a 75\% perpendicular polarization).

\subsubsection{Resonant inverse Compton scattering}
\label{sub:rics}

Once the electron loses all of its perpendicular momentum, it moves
along the magnetic field line in a stepwise fashion while upscattering
surface thermal photons through RICS.  The step size $\Delta s$ is
related to $\Delta N_{\rm RICS}$, the number of photons scattered in
each step, by
\be
\Delta s \simeq \frac{c \Delta N_{\rm RICS}}{dN_{\rm RICS}/dt} \,.
\ee
In our simulation we choose $\Delta N_{\rm RICS}$ to be
\be
\Delta N_{\rm RICS} = \min\left(1\,,\,0.1R\frac{dN_{\rm RICS}/dt}{c}\right) \,.
\ee
In other words, $\Delta N_{\rm RICS}=1$ if the RICS process is
efficient enough to produce at least one resonant photon within a
distance of $0.1R$; otherwise $\Delta N_{\rm RICS}$ is chosen so that
the electron step size is $\Delta s=0.1R$.
Using Eq.~(\ref{eq:dNdt}) from Appendix~\ref{sec:rics}, we have
\be
\Delta s \simeq \frac{\Delta N_{\rm RICS}}{ \left[\frac{\beta_Q}{\gamma_\parallel^2\beta_\parallel a_0} \left(\frac{kT}{m_ec^2}\right) \ln\frac{1-e^{-\epsilon_c/[\gamma_\parallel(1-\beta_\parallel)kT]}}{1-e^{-\epsilon_c/[\gamma_\parallel(1-\beta_\parallel\cos\psi_{\rm crit})kT]}}\right]} \,,
\ee
where $\beta_\parallel = \sqrt{1-1/\gamma_\parallel^2}$ is the speed
of the electron after it has completed synchrotron emission (so that
$p_\perp=0$), and $\psi_{\rm crit}$ is the incidence angle with
respect to the electron's trajectory of photons coming from the edge
of the surface ``hot spot'' (see Fig.~\ref{fig:ICS}). The mean energy
of the scattered photons is (e.g., \citealt{beloborodov07})
\be
\epsilon = \gamma_\parallel \left(1-\frac{1}{\sqrt{1+2\beta_Q}}\right) m_ec^2 \,,
\label{eq:Erics}
\ee
and the energy loss of the electron in each step is given by
\be
\Delta \gamma_\parallel m_ec^2 = -\epsilon \Delta N_{\rm RICS} \,.
\ee

In the lab frame the photon's angle of emission is approximately given
by
\be
\Theta_{\rm ph}\simeq \chi + \frac{1}{\gamma_e}\cos\Pi \,,
\label{eq:anglerics}
\ee
where $\Pi$ is a random angle between 0 and $2\pi$ and $\chi$ is the
angle between the local magnetic field and the dipole axis
[Eq.~(\ref{eq:chiangle}]. Here, $\gamma_e$ is the final Lorentz factor
of the electron after emitting the photon; from Eq.~(\ref{eq:Erics}),
its value is approximately
\be
\gamma_e \simeq \frac{\gamma_\parallel}{\sqrt{1+2\beta_Q}} \,.
\ee
While we include the $1/\gamma_e\cos\Pi$ term in
Eq.~(\ref{eq:anglerics}) for completeness, we find that it is not
important for our simulation. This is true even at $B \ga B_Q$, where
$\gamma_e$ is much smaller than the initial Lorentz factor of the
electron and pair production can occur almost immediately after the
photon is scattered (\citealt{beloborodov07}). The extra distance
traveled by the photons in order to pair produce when the photons are
upscattered tangent to the local magnetic field (i.e., when
$\Theta_{\rm ph} = \chi$ is assumed) has a negligible effect on the
overall cascade.

In the superstrong field regime, the final polarization state of a
photon upscattered through RICS is given by the results of
\citet{gonthier00}. For $B \la B_Q$, both below and above resonance
more $\perp$ photons are produced than $\parallel$ photons, at a ratio
of $\simeq 3:1$. The same situation occurs for $B \ga B_Q$ below
resonance; above resonance, however, the situation reverses and more
$\parallel$ photons are produced than $\perp$ photons. We therefore
assign the photons a polarization in the ratio of one $\parallel$ to
every three $\perp$ photons for $B < B_Q$, and a polarization in the
ratio of one $\parallel$ to every $\perp$ photon for $B \ge B_Q$
(based on the assumption that approximately $50\%$ of the photons are
slightly below resonance and $50\%$ are slightly above). In practice,
however, we find that the cascade does not depend sensitively on the
initial photon polarization. At low fields ($B \la 3\times10^{12}$~G)
the polarization has no effect on the cascade, since the asymptotic
attenuation coefficient for pair production is used; at high fields a
$\perp$ photon is split into two $\parallel$ photons before it can
pair produce, and the resulting cascade is not much different from the
cascade of a single $\parallel$ photon with twice the energy.

In our simulation we consider thermal photon emission from three types
of surface ``hot spots'' (see \citealt{pons07} for a review of neutron
star surface temperatures in strong magnetic fields): a large cool
spot, $T_6=0.3$ and $\theta_{\rm spot}=\pi/2$, representing emission
from the entire surface of a neutron star; a mid-sized warm spot,
$T_6=1.0$ and $\theta_{\rm spot}=0.3$; and a small hot spot, $T_6=3.0$
and $\theta_{\rm spot}=0.1$, representing emission from a heated polar
region 1~km across (which, though small, is still significantly larger
than the polar cap region, unless $P \le 0.01$~s). We also consider
the case where ICS has no effect on the cascade, which we find occurs
for $T_6 \la 0.1$ (neutron stars too cold), $\theta_{\rm spot} \la
0.01$ (hot spots too small), or $r_0 \ga R(1+2\theta_{\rm spot})$
(particles injected too far away from the surface; this is most
relevant for photon-initiated cascades discussed in
Section~\ref{sub:phprimary} below).

\subsection{Cascades initiated by a primary photon}
\label{sub:phprimary}

In the second version of the simulation, a photon is created with
energy $\epsilon_0$ at the position $(r_{0,\rm ph},\theta_{0,\rm
ph})$.  We typically choose $r_{0,\rm ph}=R$ and $\theta_{0,\rm
ph}=\theta_{\rm cap}$ (cf. Section~\ref{sub:primary}), since the
resonant ICS photon density is largest at $r \simeq R$; however, we
are also interested in photons emitted at a higher altitude (e.g., for
surface field strengths $B_p \ga 10^{13}$~G the behavior of the
cascade with $r_{0,\rm ph}=R$ and with $r_{0,\rm ph}=3R$ are very
different; see Section~\ref{sec:results}). We set $\Delta
N_\epsilon=1$, such that each photon in the simulation represents
exactly one photon in reality; we can later multiply the simulation
results by $N_0$ of Eq.~(\ref{eq:N0}) if we wish to compare cascades
dominated by RICS and by curvature radiation (see
Section~\ref{sub:phparam}). The photon is injected tangent to the
magnetic field [$\Theta_{\rm ph} = \chi$; cf. Eq.~(\ref{eq:thph})],
since we find almost no difference in the final photon or pair spectra
if we add a beaming angle $1/\gamma \sim 10^{-7}$--$10^{-3}$. As was
discussed in Section~\ref{sub:rics}, resonant ICS photons have a
polarization ratio $\parallel$ to $\perp$ of approximately 1:3 for $B
< B_Q$ and approximately 1:1 for $B \ge B_Q$. In
Section~\ref{sub:photi}, results for the photon-initiated cascades, we
choose the initial photon to be polarized perpendicular to the
magnetic field, to create a cascade with particle multiplicities as
large as possible. Our results therefore represent upper bounds on the
actual cascade multiplicities. The actual cascade should not differ
greatly from that presented in Section~\ref{sub:photi}, however, as
cascades initiated by photons polarized parallel to the magnetic field
are only slightly lower in particle multiplicity and are qualitatively
similar in spectral shape. Once the initial parameters of the photon
have been chosen, the simulation proceeds in the exact same way as
described in Sections~\ref{sub:photon}--\ref{sub:secondary}: the
photon steps outward in a straight line from the point of emission
until its optical depth is large enough to pair produce or split, etc.

Note that due to the discrete, random nature of the synchrotron
emission and the small number of particles involved in the cascade,
photon-initiated cascades will have photon and pair spectra that are
coarse and that vary between simulation runs. In order to
smooth/average the spectra to some extent, we modify the synchrotron
emission procedure of Section~\ref{sub:secondary} for secondary
particles in high Landau levels ($n \ge 3$). In every step 10 photons
are emitted, each with a weighting factor of $0.1\Delta N_\epsilon$
(rather than one photon with a weighting factor of $\Delta
N_\epsilon$, as before). Each photon has a different energy
$\epsilon_{\perp,i}$ [chosen randomly according to
Eq.~(\ref{eq:synch})], so that the total energy lost by the secondary
particle becomes $\gamma_\perp m_ec^2 = 0.1\sum_{i=1}^{10}
\epsilon_{\perp,i}$. We do not apply this procedure to the synchrotron
emission from secondary particles in Landau levels $n=1$ or $n=2$, as
it would not gain anything; each of the 10 photons emitted would have
the same value of $\epsilon_{\perp,i}$.

\section{Results}
\label{sec:results}

In this section we present the results of our simulations of photon-
and electron-initiated cascades (Sections~\ref{sub:photi} and
\ref{sub:eleci}, respectively), for a variety of different surface
field strengths, rotation periods, field geometries, and initial
energies of the primary particle. For each type of cascades, we
present the ``final'' spectra of the cascade photons and pairs as they
cross the light cylinder and escape from the magnetosphere. For the
electron-initiated cascades we also show the spectra at several
intermediate stages (i.e., the spectra of all photons and pairs that
cross the height $r=1.2R,~2R,~5R$, etc.). The photon spectra are
plotted over the energy range 10~keV-1~TeV, since for energies
$\la1$~keV the thermal photons dominate the spectra while above
$\sim1$~TeV fewer than one photon is produced per primary electron.
We are particularly interested in the pair multiplicities, i.e., the
total number of cascade electrons + positrons produced per primary
particle. We use $n_E$ to denote the number of electrons and positrons
per ``primary'' photon and $N_E$ to denote the number per primary
electron; the two multiplicities are related by
\be
N_E = N_0 \times n_E \,,
\ee
where $N_0$ is the number of photons produced by the primary electron
(see Section~\ref{sub:phparam}). From our numerical results we infer
various empirical relations for each cascade; quantitative arguments
for the validity of several of these relations are given in
Appendix~\ref{sec:empirical}. 

We first present our results for photon-initiated cascades (see
Section~\ref{sub:phprimary}), as they are simpler and aid us in our
discussion of the results for the full cascade (initiated by a primary
electron).

\subsection{Results: photon-initiated cascades}
\label{sub:photi}

\begin{figure*}
\begin{center}
\resizebox{0.3816\textwidth}{!}{\includegraphics*[viewport=0 21 169 215]{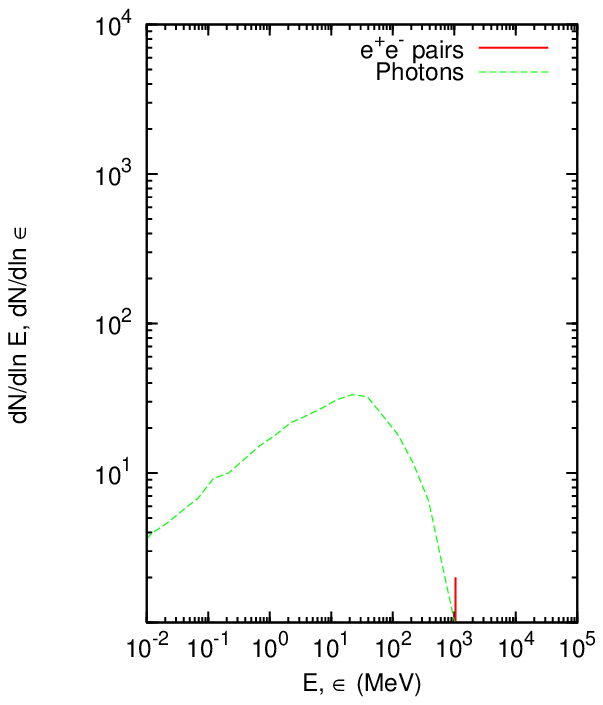}} $\!\!\!\!$
\resizebox{0.2913\textwidth}{!}{\includegraphics*[viewport=45 21 174 215]{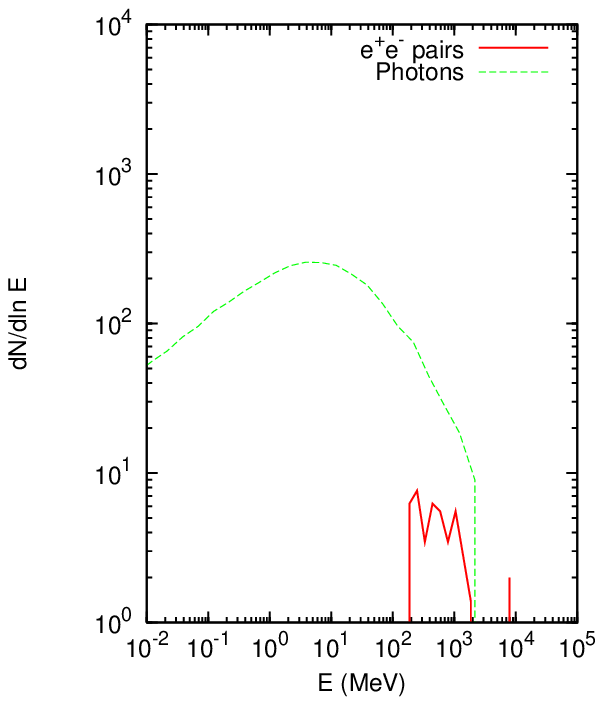}} \\
\vspace{-0.6ex}
\resizebox{0.3816\textwidth}{!}{\includegraphics*[viewport=0 0 169 194]{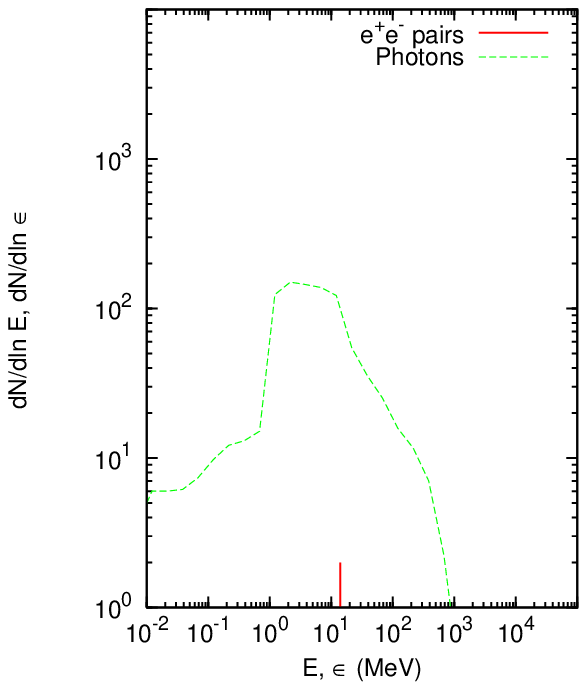}} $\!\!\!\!$
\resizebox{0.2913\textwidth}{!}{\includegraphics*[viewport=45 0 174 194]{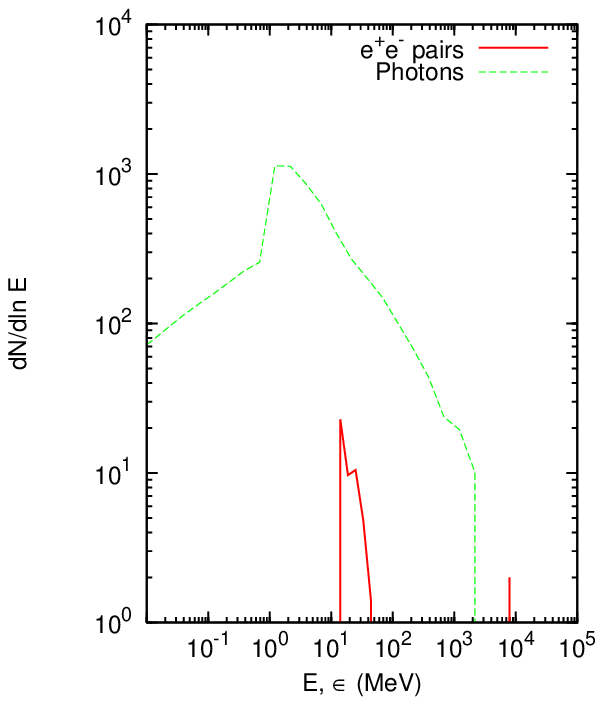}} \\
\end{center}
\caption[The final photon and pair spectra of photon-initiated cascades for $B_{p,12}=1$]
{The final photon and pair spectra of photon-initiated cascades for
surface magnetic fields $B_{p,12}=1$. The NS spin period is $P_0=1$
and a dipole field geometry is adopted. In the upper panels, ICS is
assumed to be inactive, while in the lower panels, ICS from a hot spot
with $T_6=1$, $\theta_{\rm spot}=0.3$ is included in the
simulation. The primary photon is injected from the surface and has an
energy of $10^4$~MeV (left panels) or $10^5$~MeV (right panels); for
photons with energy $10^3$~MeV, no cascade is initiated. The spike in
the pair spectra of each panel represents the electron-positron pair
produced by the primary photon. The spectra in the top panels (where
ICS is inactive) are nearly identical to the spectra generated, e.g.,
by a photon injected at $r_{0,\rm ph}=3R$ above a star with surface
field $B_{p,12}=3^3=27$, such that the local field strength at the
injection point is $B=10^{12}$~G [Eq.~(\ref{eq:Blocal})].}
\label{fig:photiLowB}
\end{figure*}

\begin{figure*}
\begin{center}
\resizebox{0.3816\textwidth}{!}{\includegraphics*[viewport=0 21 169 215]{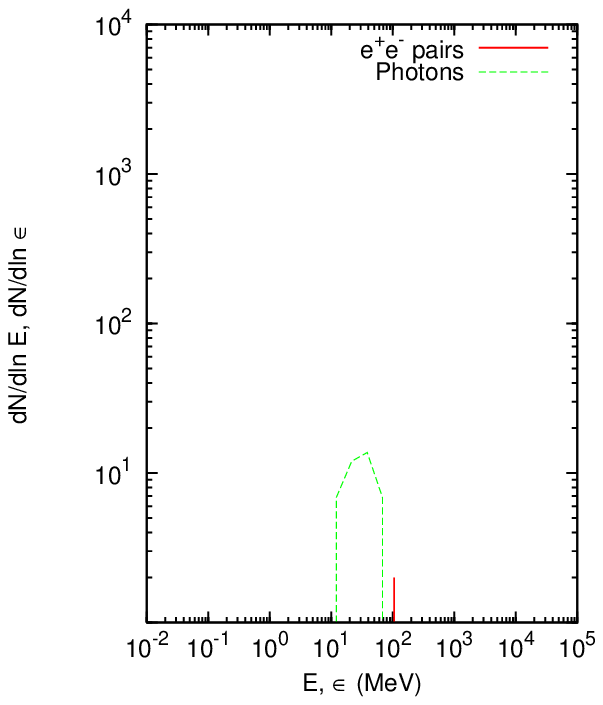}} $\!\!\!\!$
\resizebox{0.28\textwidth}{!}{\includegraphics*[viewport=45 21 169 215]{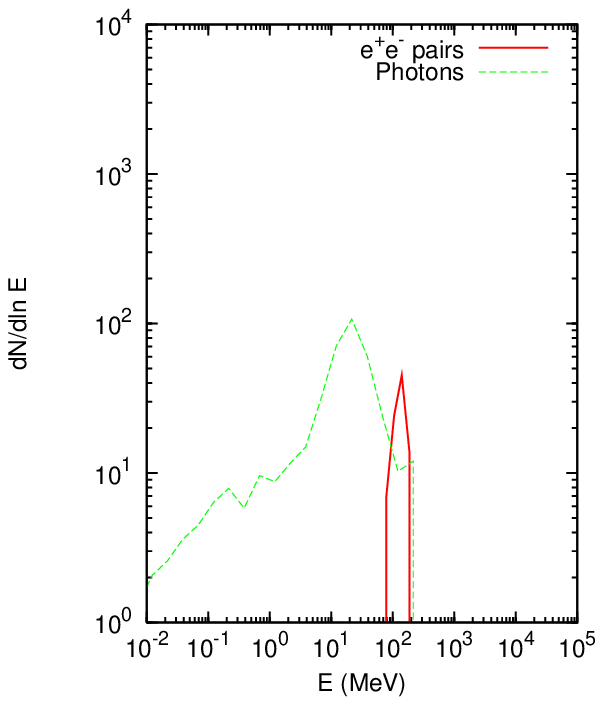}} $\!\!\!\!$
\resizebox{0.2913\textwidth}{!}{\includegraphics*[viewport=45 21 174 215]{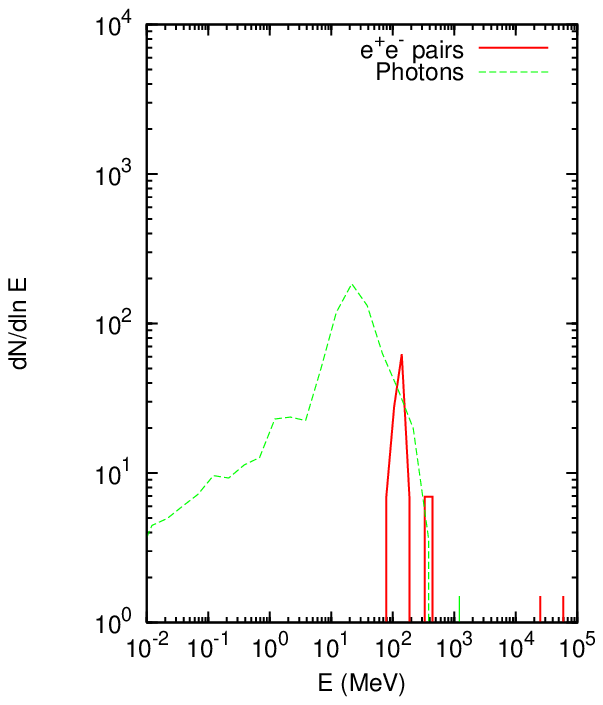}} \\
\vspace{-0.6ex}
\resizebox{0.3816\textwidth}{!}{\includegraphics*[viewport=0 0 169 194]{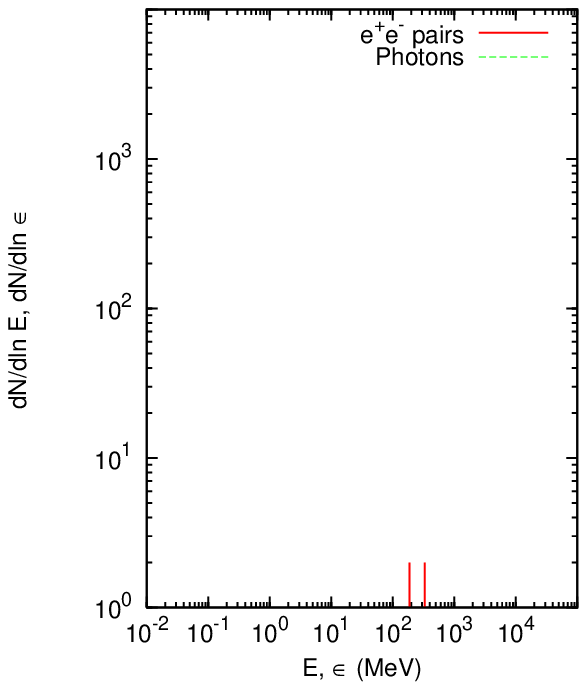}} $\!\!\!\!$
\resizebox{0.28\textwidth}{!}{\includegraphics*[viewport=45 0 169 194]{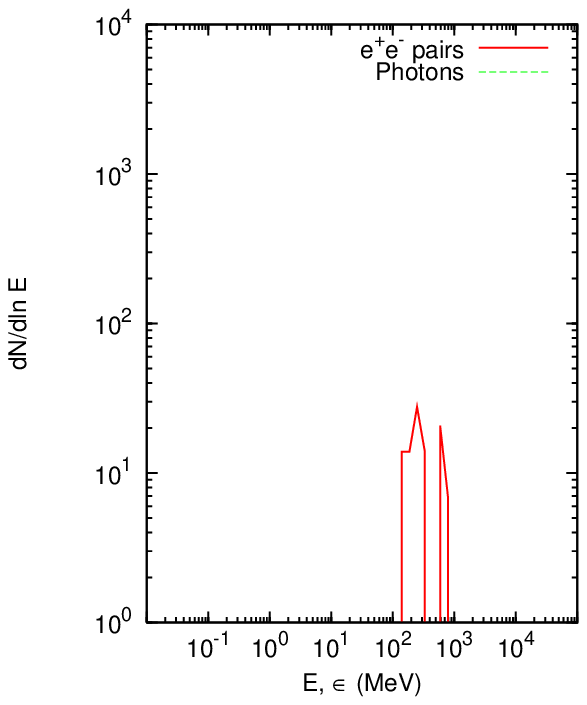}} $\!\!\!\!$
\resizebox{0.2913\textwidth}{!}{\includegraphics*[viewport=45 0 174 194]{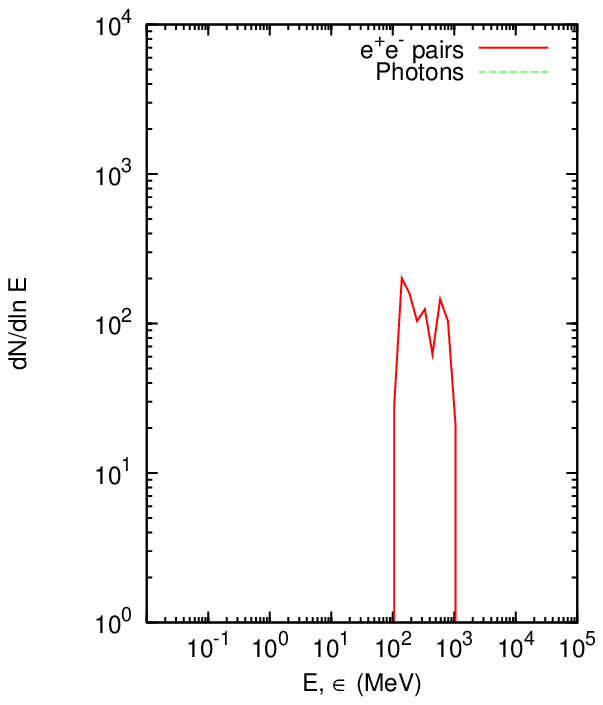}} \\
\end{center}
\caption[The final photon and pair spectra of photon-initiated cascades with ICS from a $T_6=1$ hot spot, for $B_{p,12}=10,100$]
{The final photon and pair spectra of photon-initiated cascades with
active ICS, for surface magnetic fields $B_{p,12}=10$ (upper panels)
and $100$ (lower panels). The pulsar spin period is $P_0=1$, a dipole
field geometry is adopted, photon splitting
$\perp\rightarrow\parallel\,\parallel$ is active, and thermal photons
are emitted from a hot surface spot with $T_6=1$, $\theta_{\rm
spot}=0.3$. The primary photon has an energy of $10^3$~MeV (left
column of panels), $10^4$~MeV (middle column), or $10^5$~MeV (right
column). The spikes in the pair spectra of several panels represent
the electron-positron pair produced by the primary photon. Note that
there are actually two such pairs in the bottom left panel
($B_{p,12}=100,\epsilon_0=10^3$~MeV), as the primary photon has split
in that case; while in the upper right panel
($B_{p,12}=10,\epsilon_0=10^5$~MeV) the electron and positron have
significantly different energies from each other and so are
represented by two shorter spikes.  Also note that photon spectra do
not appear in the lower panels. This is due to a combination of weak
synchrotron emission and efficient pair production near the surface of
a $B_{p,12} \ge 100$ neutron star; very few secondary photons are
created, and none of them survive to escape the magnetosphere.}
\label{fig:photiHighB}
\end{figure*}

\begin{figure*}
\begin{center}
\resizebox{0.3816\textwidth}{!}{\includegraphics*[viewport=0 0 169 201]{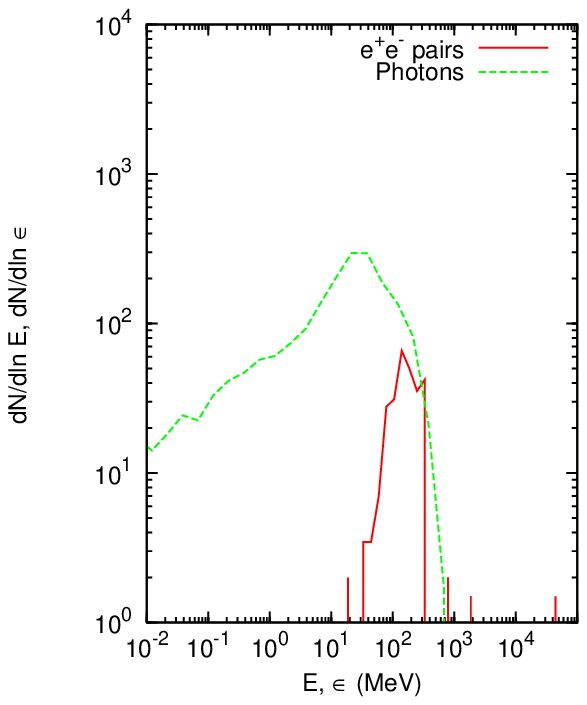}} $\!\!\!\!$
\resizebox{0.28\textwidth}{!}{\includegraphics*[viewport=45 0 169 201]{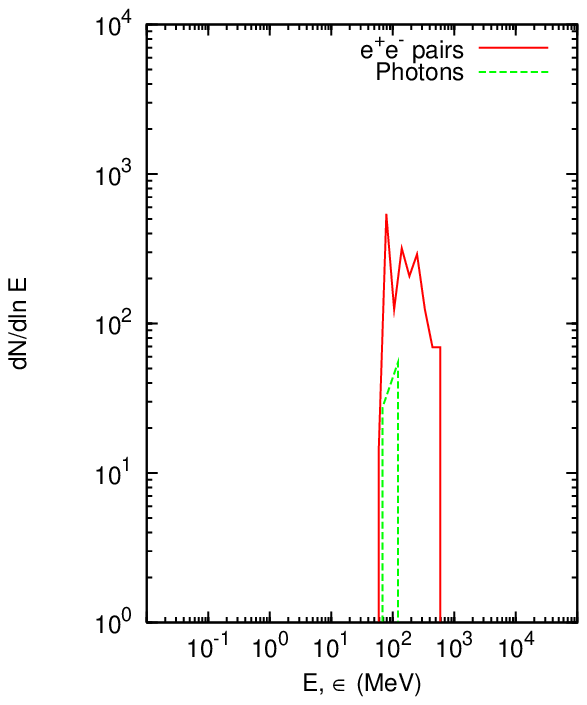}} $\!\!\!\!$
\resizebox{0.2913\textwidth}{!}{\includegraphics*[viewport=45 0 174 201]{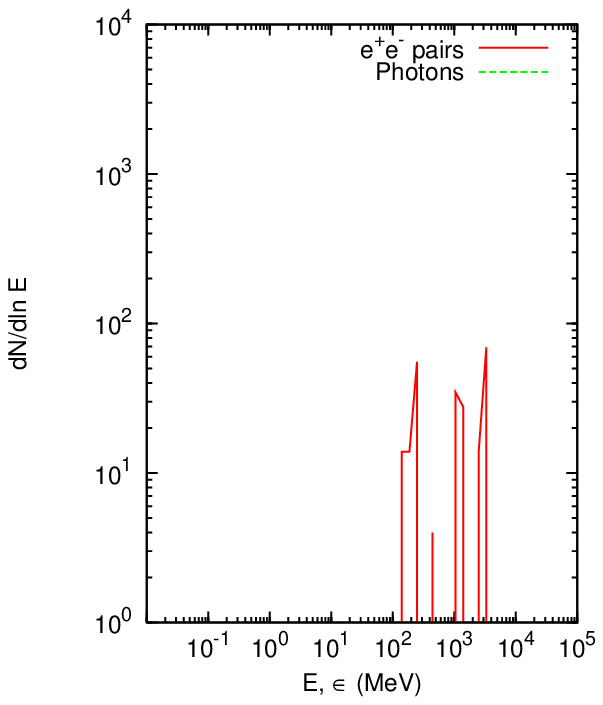}} \\
\end{center}
\caption[The final photon and pair spectra of photon-initiated cascades with ICS from a $T_6=3$ hot spot, for $B_{p,12}=10,100,1000$]
{The final photon and pair spectra of photon-initiated cascades with
active ICS, for surface magnetic fields $B_{p,12}=10$ (left panel),
$100$ (middle panel), and $1000$ (right panel). The NS period is
$P_0=1$, a dipole field geometry is adopted, and thermal photons are
emitted from a hot spot with $T_6=3$, $\theta_{\rm spot}=0.1$. The
primary photon has $\epsilon_0=10^5$~MeV. The spikes in the pair
spectra of the left and right panels represent, from shortest to
tallest, one, two, or four electrons/positrons.}
\label{fig:photiSPOT}
\end{figure*}

\begin{figure*}
\begin{center}
\resizebox{0.3816\textwidth}{!}{\includegraphics*[viewport=0 0 169 201]{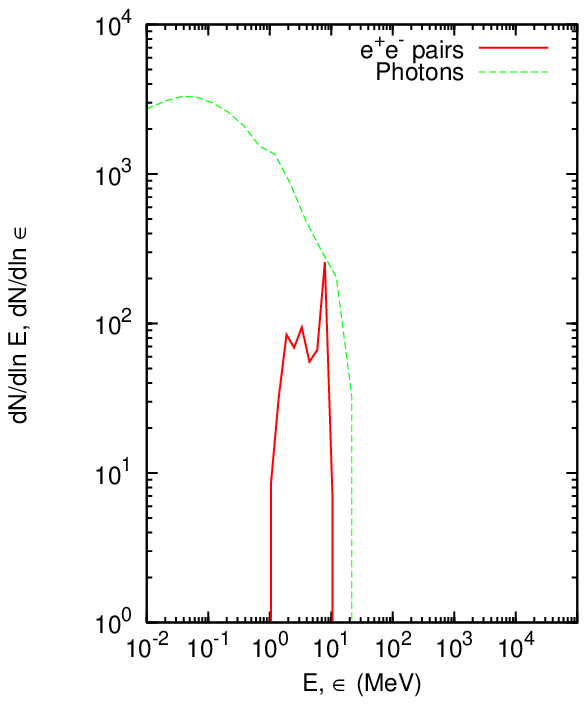}} $\!\!\!\!$
\resizebox{0.28\textwidth}{!}{\includegraphics*[viewport=45 0 169 201]{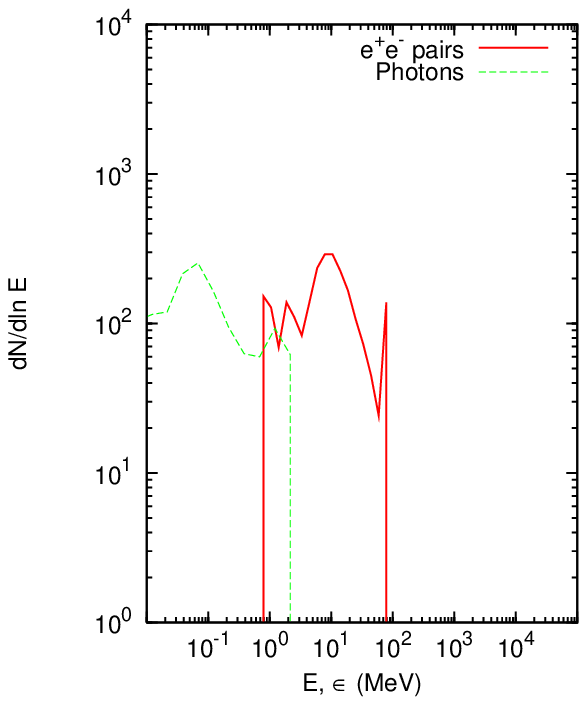}} $\!\!\!\!$
\resizebox{0.2913\textwidth}{!}{\includegraphics*[viewport=45 0 174 201]{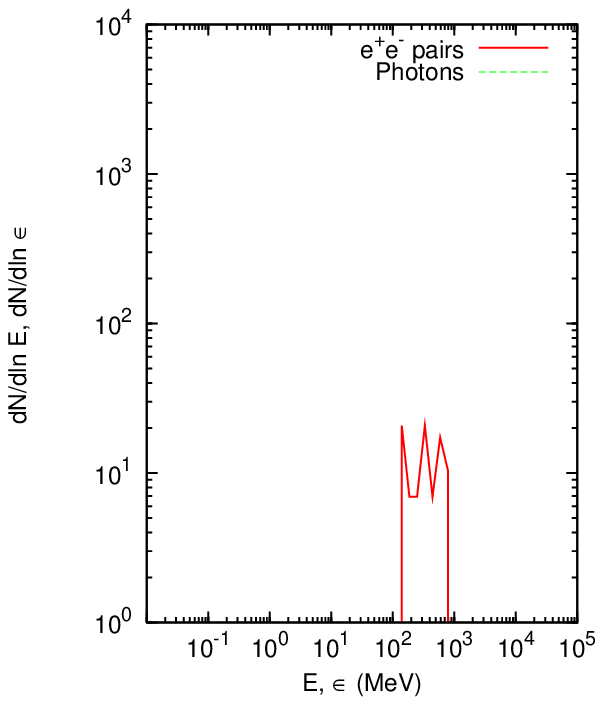}} \\
\end{center}
\caption[The final photon and pair spectra of photon-initiated cascades for ${\cal R}_c=R$ and $B_{p,12}=1,10,100$]
{The final photon and pair spectra of photon-initiated cascades for a
non-dipole magnetosphere with local radius of curvature ${\cal R}_c=R$, 
for surface magnetic fields $B_{p,12}=1$ (left panel), $10$
(middle panel), and $100$ (right panel). Here, the primary photon has
$\epsilon_0=10^4$~MeV and ICS is active from a hot spot with $T_6=1$,
$\theta_{\rm spot}=0.3$.}
\label{fig:photiMULT}
\end{figure*}

Our results for photon-initiated cascades are presented in
Figs.~\ref{fig:photiLowB}--\ref{fig:photiMULT}.  We consider primary
photons with energies in the range of $10^3$--$10^5$~MeV; for
$B_{p,12}=1$--$1000$, the primary electron should emit very few
photons (via either resonant ICS or curvature radiation) above this
energy range (see Section~\ref{sec:accel}).  Unless otherwise stated,
the primary photon is emitted from near the surface, in the direction
tangent to the last open field line. Thus the radius of curvature near
the point of emission is ${\cal R}_c \simeq 9\times10^7 P_0^{1/2}$~cm
for dipole fields [Eq.~(\ref{eq:rcdipole})].

We find significant differences in the behavior of the cascades at
magnetic field strengths below and above $B_{\rm crit} \simeq
3\times10^{12}$~G [Eq.~(\ref{eq:critatten})]. At low fields $B\la
B_{\rm crit}$, the primary photon can pair produce if
[Eq.~(\ref{eq:phmin2}); see also \citealt{hibschman01a}]
\be
\epsilon_0 > \epsilon_{\rm min} \sim 3000 B_{p,12}^{-1} {\cal R}_8~{\rm MeV} \,,
\label{eq:phmin}
\ee
where ${\cal R}_8$ is the radius of curvature ${\cal R}_c$ in units of
$10^8$~cm, evaluated at the surface along the last open field
line. Strong cascades, where more than one electron-positron pair is
produced, typically occur at energies $\sim10$ times $\epsilon_{\rm
min}$. For $\epsilon_0$ in the range from $\epsilon_{\rm min}$ to
$\sim10^5$~MeV, we find that the multiplicities of photons and
$e^+e^-$ particles produced in the cascade are
\be
n_\epsilon \sim \frac{\epsilon_0}{500~\rm MeV} {\cal R}_8^{-1}
\label{eq:Nph}
\ee
and
\be
n_E \sim \frac{\epsilon_0}{10^4~\rm MeV} B_{p,12} {\cal R}_8^{-1} \,,
\label{eq:Ne}
\ee
respectively. These results are (largely) independent of the hot spot
model used. When ICS is inactive, the cascade electron/positron has
final energy (after it has finished radiating synchrotron photons)
extending from [Eq.~(\ref{eq:Emax2})]
\be
E_{\rm max} \sim 0.1 B_{p,12} \epsilon_0
\label{eq:Emax}
\ee
(for the first pair produced) down to $\sim0.1B_{p,12}\epsilon_{\rm
min}$ for the lowest-energy pairs, and the total energy of the pairs
is [Eq.~(\ref{eq:Etot2})]
\be
{\cal E}_{\rm tot} \sim 2E_{\rm max} + 0.1B_{p,12}\epsilon_{\rm min} n_E \ln \left(\frac{0.075\epsilon_0}{\epsilon_{\rm min}}\right) \,.
\label{eq:Etot}
\ee
When ICS is active from a hot spot (Section~\ref{sub:secondary}), the
number of pairs produced does not change, since the photons produced
through ICS at these field strengths have energies $\sim B_{p,12}^2
T_6^{-1}$~MeV [Eq.~(\ref{eq:rics0})] and can not pair produce. The
total pair energy ${\cal E}_{\rm tot}$ decreases, however, since the
ICS process transfers energy from the pairs to photons. Although
resonant ICS is most important for electrons and positrons at
$\gamma_{\rm crit} \simeq \epsilon_c/kT$ [Eq.~(\ref{eq:gammac}); see
Section~\ref{sub:phparam}], we find in these cascades that all
electrons and positrons with energies in the range of
\bal
E_{\rm RICS} \sim {}& (0.3-30) \gamma_{\rm crit}m_ec^2 \nonumber\\
 \simeq {}& (20-2000) B_{p,12} T_6^{-1}~{\rm MeV}
\label{eq:Eics}
\eal
are strongly affected. Thus hot surface spots with higher $T$ tend to
lower ${\cal E}_{\rm tot}$ more. As expected, we find that photon
splitting does not affect the cascade at these field strengths (see
Section~\ref{sub:photon}). The photon and pair cascade spectra for
$B_{p,12}=1$ are shown in Fig.~\ref{fig:photiLowB}, both when ICS is
inactive and when ICS is active from a ``warm spot'' ($T_6=1$,
$\theta_{\rm spot}=0.3$).

At high fields ($B\ga B_{\rm crit}$), a primary photon injected from the
surface will pair produce when
\be
\epsilon_0 > \epsilon_{\rm min} \sim 200\,{\cal R}_8~{\rm MeV} \,,
\label{eq:phminHighB}
\ee
largely independent of field strength. When ICS is inactive, almost
all of the cascade energy resides
in the pairs; i.e., ${\cal E}_{\rm tot} \simeq
\epsilon_0$. The pair cascade will be very weak regardless of photon
energy, with $n_E < 10$ and $n_\epsilon=0$ or $1$ (i.e., at most one photon
escapes the magnetosphere without pair production). This is because the
$e^\pm$ pairs are produced exclusively through the
$(jk)=(00)$ or $(01)$ channel (see Section~\ref{sub:photon}), so that
at most one synchrotron photon is emitted per pair. For $B_{p,12} \ga
20$, photon splitting causes all pairs to be produced with
$(jk)=(00)$, such that the cascades are even weaker: $n_E\le 4$ and
$n_\epsilon=0$. When ICS is active, both $n_\epsilon$ and $n_E$ can be
larger, though not as large as would be predicted by an extrapolation
of Eqs.~(\ref{eq:Nph}) and (\ref{eq:Ne}) to high fields. In order for
ICS to affect the cascade, however, the primary photon must have an energy
\be
\epsilon_0 \ga 70 B_{p,12} T_6^{-1}~{\rm MeV} \,;
\label{eq:eminICS}
\ee
the energy of the electron/positron produced by this photon, $E_{\rm
max} \sim (0.1$--$0.5)\epsilon_0$ [Eq.~(\ref{eq:EmaxHighB})], must be
larger than the minimum energy at which ICS is effective,
$\sim0.3\gamma_{\rm crit} m_ec^2$ (see
Appendix~\ref{sub:photemper}). Note that this energy is approximately
equal to $\epsilon_{\rm RICS}$, the energy of a typical ICS photon
upscattered by the primary electron at high fields
[Eq.~(\ref{eq:rics0})]; therefore, at high fields a typical ICS photon
is able to initiate a weak cascade. We find that for $B_p \la 0.5B_Q$
a significant fraction ($\sim20\%$--$60\%$) of the total cascade energy
$\epsilon_0$ resides in the photons; as in the low field case
[Eq.~(\ref{eq:Etot})], this fraction decreases as either $\epsilon_0$
or $B_{p,12}$ increases. For $B_p \ga 0.5 B_Q$, even ignoring photon
splitting, the total photon energy fraction is very low, $<10\%$; but
with photon splitting included, it is almost negligible, $<1\%$. The
photon and pair cascade spectra for $B_{p,12}=10$ and $100$ when ICS
is active from a ``warm spot'' ($T_6=1$, $\theta_{\rm spot}=0.3$) are
shown in Fig.~\ref{fig:photiHighB}.

As discussed in Section~\ref{sub:secondary}, we consider three hot
surface spot models for active ICS: a ``cool'' $T_6=0.3$, $\theta_{\rm
spot}=\pi/2$ spot; a ``warm'' $T_6=1$, $\theta_{\rm spot}=0.3$ spot;
and ``hot'' $T_6=3$, $\theta_{\rm spot}=0.1$ spot. At low fields we
find that the only effect the various hot spot models have is to lower
${\cal E}_{\rm tot}$ relative to the total cascade energy $\epsilon_0$
(see above). At high fields, the cascades due to warm and hot spots
are similar in multiplicities $n_\epsilon$ and $n_E$ for energies
$\epsilon_0 \la 10^4$~MeV; but for energies $\epsilon_0 \ga 10^5$~MeV,
the hot spots give multiplicities $\sim3$ times larger than warm
spots; see Fig.~\ref{fig:photiSPOT} [from Eq.~(\ref{eq:eminICS}), at
$B_{p,12}=1000$ a strong cascade requires $T_6 \ge 3$]. Cool spots
give much smaller multiplicities (factors of $>10$ smaller) than warm
or hot spots regardless of the primary photon energy.

At low fields, the multiplicities of photons and $e^+e^-$ particles
produced per primary photon, $n_\epsilon$ and $n_E$, depend on ${\cal
R}_c$ to the $(-1)$ power [Eqs.~(\ref{eq:Nph}) and (\ref{eq:Ne})],
such that cascades in non-dipole magnetospheres can be much larger
than in dipole magnetospheres. At high fields, we find that this
dependence on curvature radius is much weaker: $n_E\propto {\cal
R}_c^{-1/2}$ at $B_p \simeq 0.5B_Q$ and $n_E\propto {\cal R}_c^{-1/4}$
power at larger fields. Thus for high fields, the cascades in
non-dipole and dipole magnetospheres are of similar sizes. The photon
and pair cascade spectra for $B_{p,12}=1$, $10$, and $100$ and a
${\cal R}_c=R$ non-dipole magnetosphere are shown in
Fig.~\ref{fig:photiMULT}.

Two aspects of the cascade depend strongly on the altitude 
($r_{0,\rm ph}$) at which the primary photon is injected: the local magnetic
field strength, $B\propto r^{-3}$, and the effectiveness of ICS, which
is completely negligible for $r_{0,\rm ph}\ge3R$ (regardless of the temperature
and size of the hot spot). The other cascade parameters have a much
weaker dependence on altitude (e.g., radius of curvature ${\cal R}_c
\propto \sqrt{r_{0,\rm ph}}$). We find that the photon and pair
spectra for primary photons injected at $r_{0,\rm ph}>R$ are very
similar to the spectra for photons injected at the surface, as long as
ICS is inactive and the local magnetic field strengths are the same in
both cases. For example, the spectra for $r_{0,\rm ph}=3R$ and
$B_{p,12}=27$ [such that the local field strength at the point of
injection is $B_{12}=1$; Eq.~(\ref{eq:Blocal})] is nearly identical
to the spectra given in the first row of Fig.~\ref{fig:photiLowB},
where $r_{0,\rm ph}=R$, $B_{p,12}=1$, and ICS is inactive.

We can use the pair multiplicity per primary photon,
$n_E\left(\epsilon_0\right)$, obtained from our numerical simulations
to estimate the pair multiplicity per primary electron, $N_E = N_0
n_E$, when ICS is the dominant cascade emission process. For
$\epsilon_0$ and $N_0$, we use the expressions for the typical energy
$\epsilon_{\rm RICS}$ and total number $\simeq 10 B_{p,12}^{-1}
P_0^{3/4} T_6^{5/2}$ of resonant ICS photons upscattered by the
primary electron [Eqs.~(\ref{eq:rics0}) and (\ref{eq:N0}); see
Section~\ref{sub:phparam}]. The results are shown in
Table~\ref{tab:Neics}. Note that although $\epsilon_{\rm RICS}$ is
independent of the acceleration model used, the number of upscattered
photons, $N_0$, as given by Eq.~(\ref{eq:N0}) is applicable for only
inner gap accelerators with space-charge-limited flow; inner vacuum
gap accelerators, for example, would yield $N_0$ about $20$--$100$
times lower. Also note that, for a given acceleration model, an
accurate determination of $N_E$ requires that rather than just setting
$\epsilon_0=\epsilon_{\rm RICS}$, a distribution of energies be used
which takes into account resonant (and possibly non-resonant)
scattering away from the thermal peak (i.e., $\gamma \ne \gamma_{\rm
crit}$): although fewer photons are upscattered at energies greater
than $\epsilon_{\rm RICS}$, these photons can have an important effect
on the total multiplicity of pairs produced, since $n_E$ grows
approximately linearly with photon energy $\epsilon_0$; the
$\epsilon_0>\epsilon_{\rm RICS}$ photons are especially important for
cascades where $n_E\left(\epsilon_{\rm RICS}\right)=0$ (e.g., cascades
with $B_{p,12}=10$ and $P_0=1$).

\begin{table*}
\caption[Electron/positron multiplicities for RICS-initiated cascades]
{Pair multiplicity $N_E$ when resonant ICS is the dominant photon
emission process of the primary electron. For simplicity, all photons
are assumed to be upscattered from the thermal peak ($\gamma_{\rm
crit}$; see text). Each entry in the table gives $N_E = N_0 n_E$ for a
different surface magnetic field strength $B_p$, radius of curvature
(either ${\cal R}_c=R$ or a dipole field curvature with the pulsar
spin period $P_0$ is specified), and hot spot temperature $T$ and size
$\theta_{\rm spot}$. The pair multiplicity is zero for $B_{p,12}=1$
and the hot spot models used here, even when ${\cal R}_c=R$; we
therefore omitted these entries from the table.}
\centering
\begin{tabular}{c c c c c c c}
\hline\hline
 & \multicolumn{3}{c}{$T_6=1$, $\theta_{\rm spot}=0.3$} & \multicolumn{3}{c}{$T_6=3$, $\theta_{\rm spot}=0.1$} \\
 & $P_0=10$ & $P_0=1$ & ${\cal R}_c=R$ & $P_0=10$ & $P_0=1$ & ${\cal R}_c=R$ \\
\hline
$B_{p,12}=10$ & 0 & 0 & 14 & 0 & 0 & 220 \\
\hline
$B_{p,12}=100$ & 2.2 & 1.2 & 1.2 & 35 & 6.2 & 31 \\
\hline
$B_{p,12}=1000$ & 0.7 & 0.1 & 0.1 & 11 & 1.9 & 4.4 \\
\hline\hline
\end{tabular}
\label{tab:Neics}
\end{table*}

\subsection{Results: electron-initiated cascades}
\label{sub:eleci}

\begin{figure*}
\begin{center}
\begin{tabular}{cc}
\resizebox{0.45\textwidth}{!}{\includegraphics{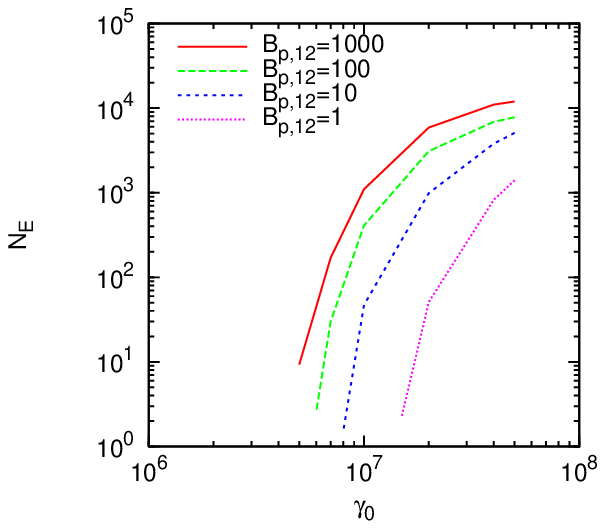}} &
\resizebox{0.45\textwidth}{!}{\includegraphics{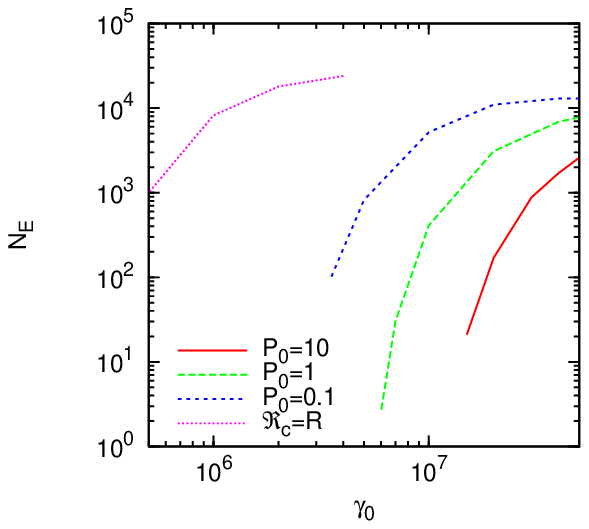}} \\
\end{tabular}
\end{center}
\caption[Secondary electron/positron multiplicities as a function of $\gamma_0$]
{Secondary electron/positron multiplicities $N_E$ as a function of the
initial Lorentz factor of the primary electron $\gamma_0$, for $P_0=1$
and a dipole geometry at several different surface field strengths
(left panel), and for $B_{p,12}=100$ at several different periods/field
geometries (right panel).}
\label{fig:multiplicity}
\end{figure*}

\begin{figure*}
\begin{center}
\resizebox{0.3816\textwidth}{!}{\includegraphics*[viewport=0 0 169 155]{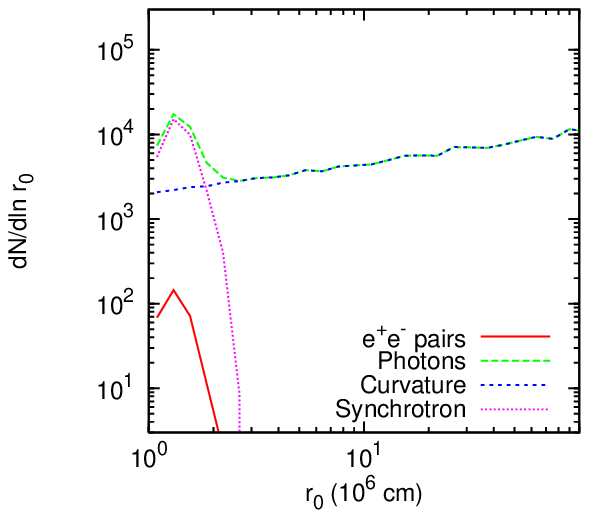}} $\!\!\!\!$
\resizebox{0.28\textwidth}{!}{\includegraphics*[viewport=45 0 169 155]{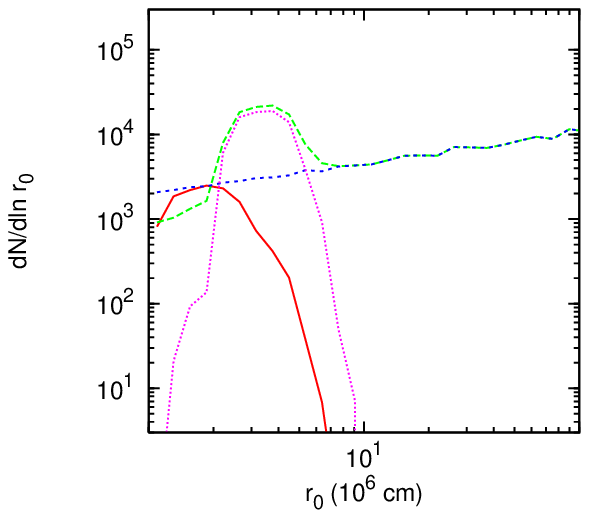}} $\!\!\!\!$
\resizebox{0.2913\textwidth}{!}{\includegraphics*[viewport=45 0 174 155]{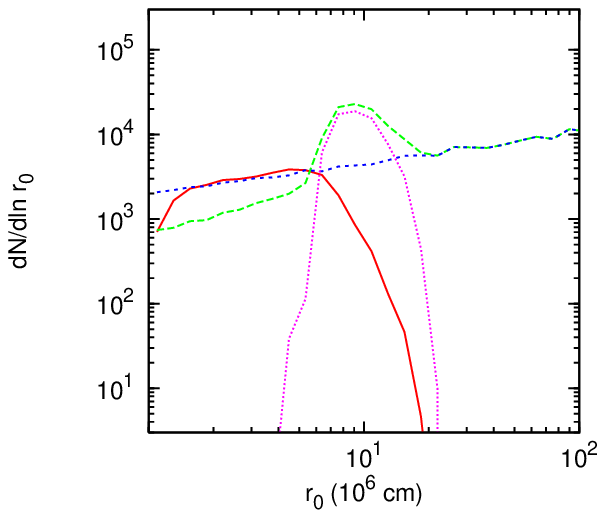}} \\
\end{center}
\caption[The number of photons and pairs as a function of radius]
{The number of photons and electrons $+$ positrons as a function of
the radius where they are created, $r_{0,\rm ph}$ or $r_0$, for
$B_{p,12}=1$ (left panel), $44.14$ (middle panel), and $1000$ (right
panel). The neutron star spin period is assumed to be $P_0=1$ and a
dipole field geometry is used. The
$\perp\rightarrow\parallel\,\parallel$ photon splitting mode is
active, and the primary electron has $\gamma_0=2\times10^7$; the
effects of ICS are not included. The curve labeled ``$e^+e^-$ pairs''
shows where the secondary electrons and positrons are created,
``Photons'' shows where the photons that escape the magnetosphere
(i.e., that do not split or pair produce) are created, ``Curvature''
shows where the curvature photons are emitted by the primary electron
(which continues in a similar manner beyond the graph out to the light
cylinder), and ``Synchrotron'' shows where the synchrotron photons are
emitted by the secondary pairs.}
\label{fig:dndlnr}
\end{figure*}

\begin{figure*}
\begin{center}
\resizebox{0.35\textwidth}{!}{\includegraphics*[viewport=0 0 168 155]{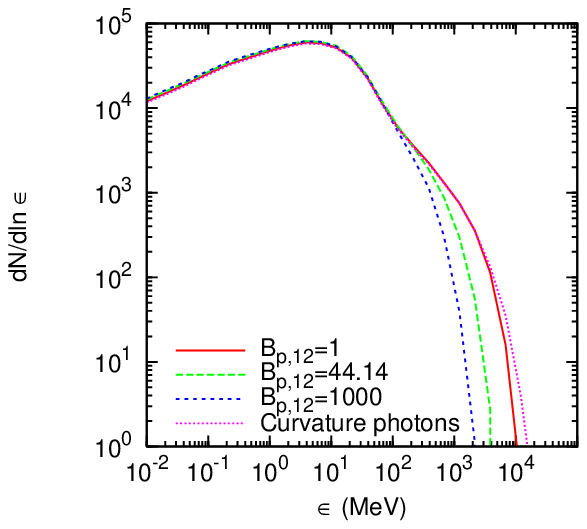}} $\!\!\!\!$
\resizebox{0.35\textwidth}{!}{\includegraphics*[viewport=44 0 212 155]{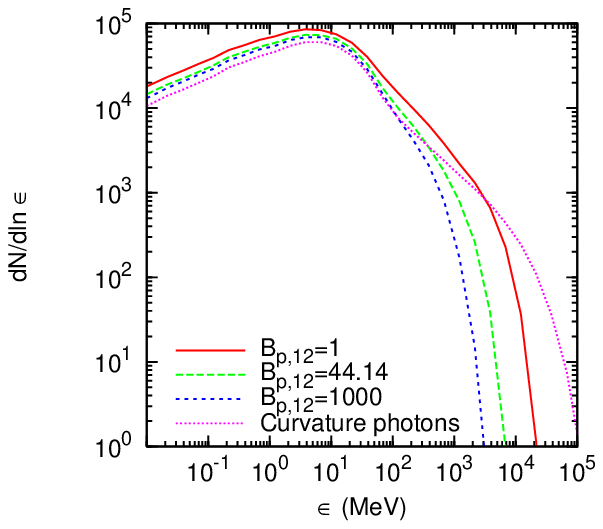}} \\
\resizebox{0.35\textwidth}{!}{\includegraphics*[viewport=0 0 168 155]{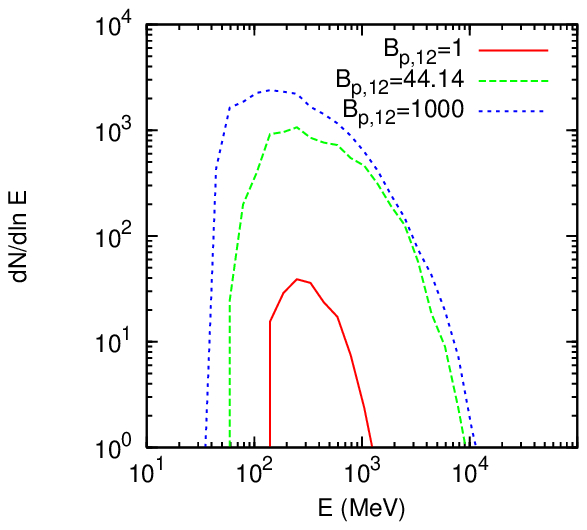}} $\!\!\!\!$
\resizebox{0.35\textwidth}{!}{\includegraphics*[viewport=44 0 212 155]{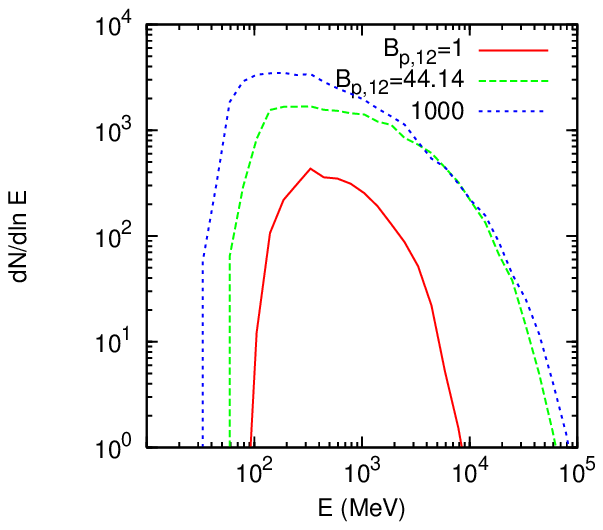}} \\
\end{center}
\caption[The final photon and pair spectra when $\gamma_0=2\times10^7$ and $4\times10^7$]
{The final photon and pair spectra of electron-initiated cascades.
The primary electron has Lorentz factor $\gamma_0=2\times10^7$ (the
left two panels) and $4\times10^7$ (right two panels), and the surface
magnetic field strengths are $B_{p,12}=1$, $44.14$, and $1000$. The
neutron star spin period is $P_0=1$ and a dipole field geometry is
adopted. The secondary photon spectra are shown in the top two
panels; the pair spectra are shown in the bottom two panels. The curve
labeled ``Curvature photons'' in each of the top two panels (the
dot-dashed line) shows the spectrum of curvature radiation emitted by
the primary electron, which is the same for all field strengths.}
\label{fig:BComp}
\end{figure*}

\begin{figure*}
\begin{center}
\resizebox{0.3544\textwidth}{!}{\includegraphics*[viewport=0 0 169 155]{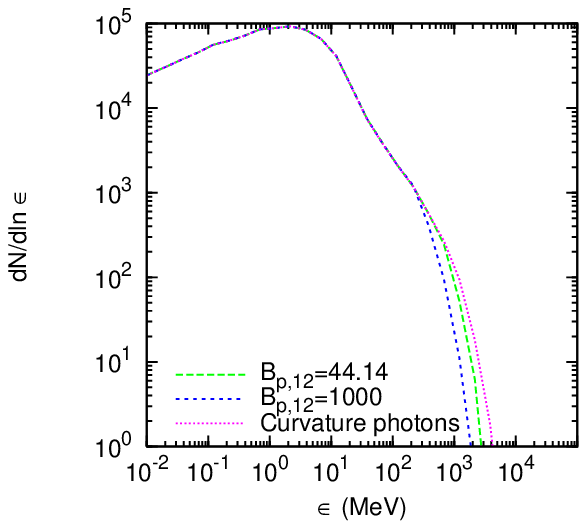}} $\!\!\!\!$
\resizebox{0.2705\textwidth}{!}{\includegraphics*[viewport=45 0 174 155]{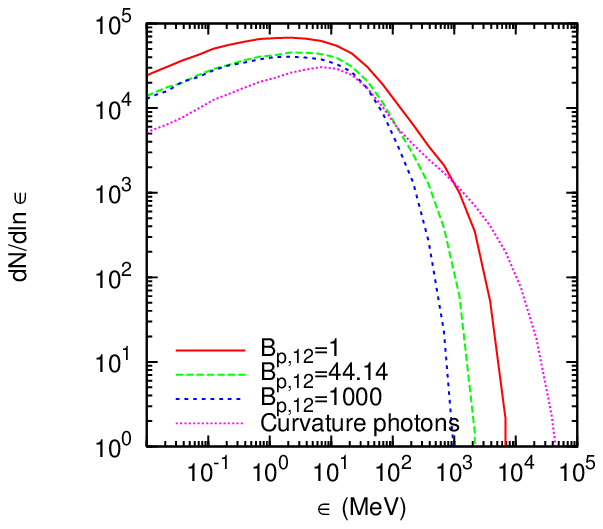}} $\!\!\!\!$
\resizebox{0.3753\textwidth}{!}{\includegraphics*[viewport=-10 0 169 155]{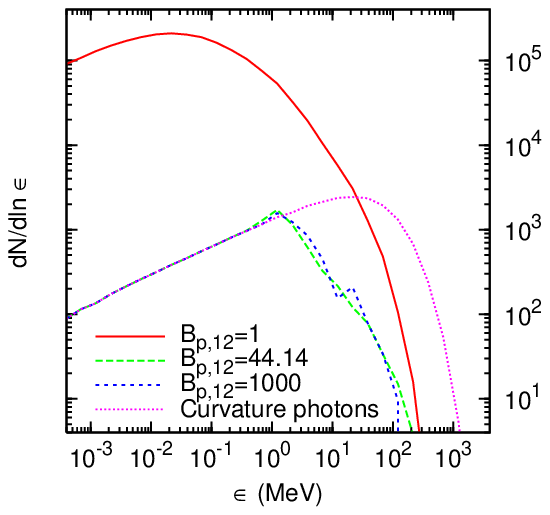}} \\
\resizebox{0.3544\textwidth}{!}{\includegraphics*[viewport=0 0 169 155]{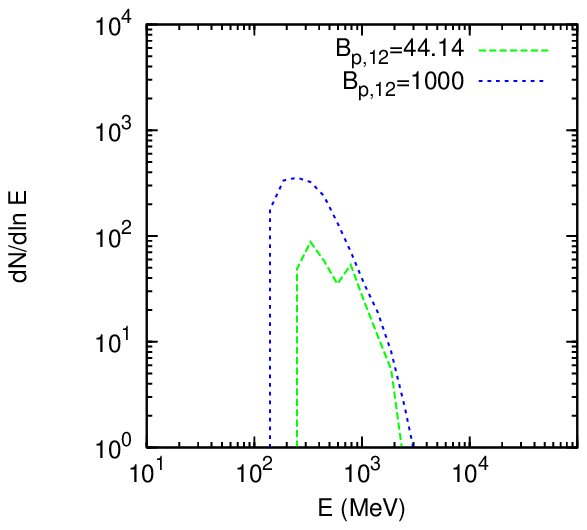}} $\!\!\!\!$
\resizebox{0.26\textwidth}{!}{\includegraphics*[viewport=45 0 169 155]{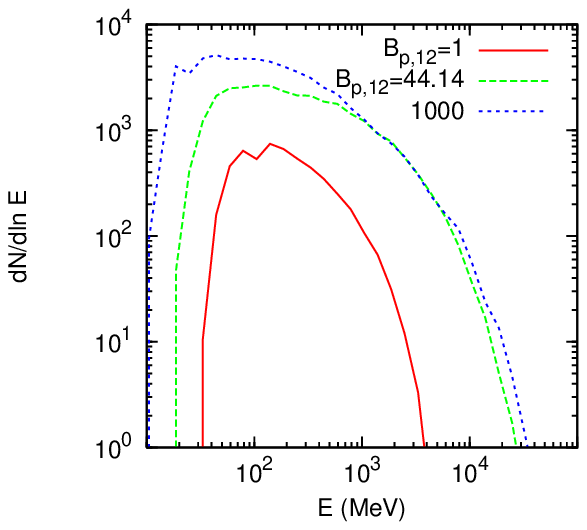}} $\!\!\!\!$
\resizebox{0.26\textwidth}{!}{\includegraphics*[viewport=45 0 169 155]{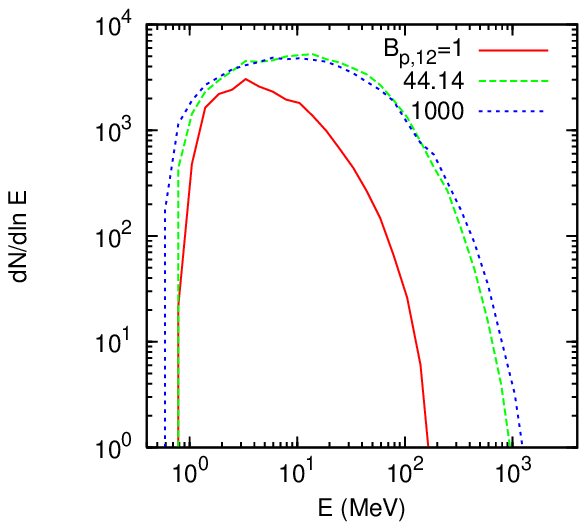}} \\
\end{center}
\caption[The final photon and pair spectra for dipole magnetospheres with $P_0=10$ and $0.1$ and a non-dipole magnetosphere with ${\cal R}_c=R$]
{The final photon and pair spectra of electron-initiated cascades for
dipole magnetospheres with $P_0=10$ (left panels) and $0.1$ (middle
panels), and non-dipole magnetospheres with ${\cal R}_c=R$ (right
panels). The surface field strengths are $B_{p,12}=1$, $44.14$, and
$1000$. The Lorentz factor of the primary electron is
$\gamma_0=2\times10^7$ for the dipole magnetospheres (left and middle
columns) and $\gamma_0=2\times10^6$ for the non-dipole magnetosphere
(right column; see Section~\ref{sec:accel}). The photon spectra are
shown in the top three panels; the pair spectra are shown in the
bottom three panels. The curve labeled ``Curvature photons'' in each
of the top three panels (the dot-dashed line) shows the spectrum of
curvature radiation emitted by the primary electron, which is the same
for all field strengths. Note that the top right panel, corresponding
to the photon spectra when ${\cal R}_c=R$, has a different vertical
(and horizontal) scale than the top left and top middle panels.}
\label{fig:PComp}
\end{figure*}

\begin{figure*}
\begin{center}
\resizebox{0.3816\textwidth}{!}{\includegraphics*[viewport=0 0 169 155]{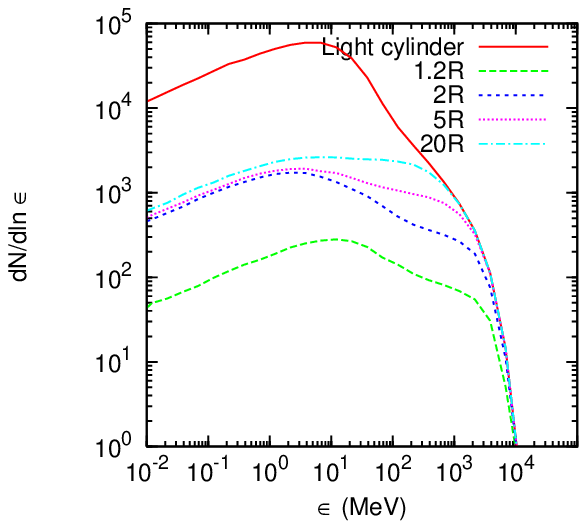}} $\!\!\!\!$
\resizebox{0.28\textwidth}{!}{\includegraphics*[viewport=45 0 169 155]{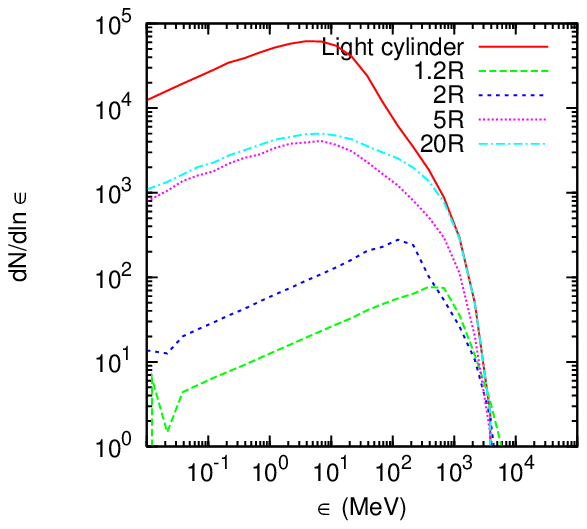}} $\!\!\!\!$
\resizebox{0.2913\textwidth}{!}{\includegraphics*[viewport=45 0 174 155]{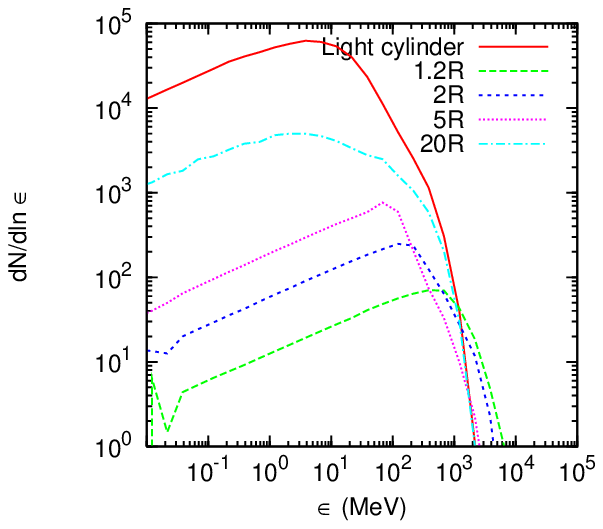}} \\
\resizebox{0.3816\textwidth}{!}{\includegraphics*[viewport=0 0 169 155]{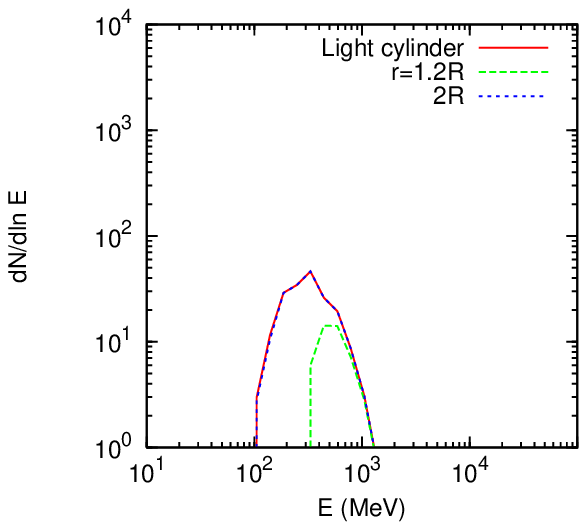}} $\!\!\!\!$
\resizebox{0.28\textwidth}{!}{\includegraphics*[viewport=45 0 169 155]{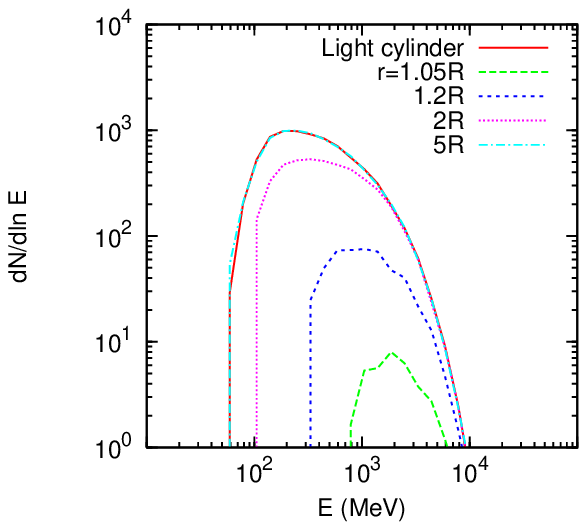}} $\!\!\!\!$
\resizebox{0.2913\textwidth}{!}{\includegraphics*[viewport=45 0 174 155]{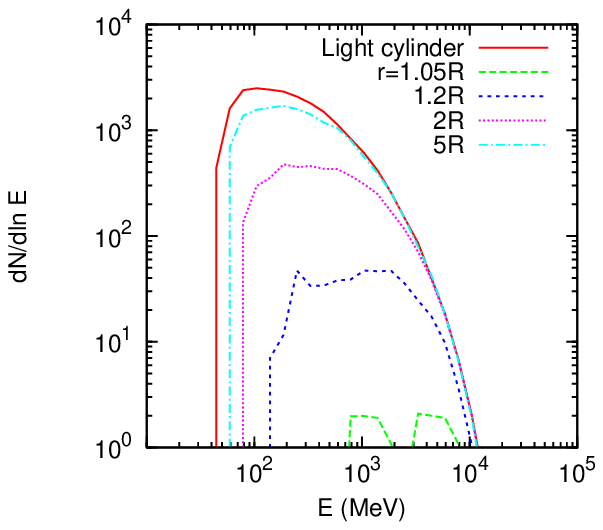}} \\
\end{center}
\caption[Cumulative photon and pair spectra at $r=1.05R,1.2R,2R,5R,20R$]
{The cumulative photon and pair spectra at magnetosphere radii
$r=1.05R$, $1.2R$, $2R$, $5R$, $20R$, and $r_{\rm LC}$ (denoted
``Light cylinder''), for $B_{p,12}=1$ (left two panels), $44.14$
(middle panels), and $1000$ (right panels).  The NS spin period is
$P_0=1$ and a dipole field geometry is used, and
$\gamma_0=2\times10^7$. The photon spectra are shown in the top three
panels; the pair spectra are shown in the bottom three panels. For a
given magnetosphere radius $r$, every cascade particle that is created
below $r$ and survives until reaching $r$ is recorded in the
spectrum.}
\label{fig:BCompR}
\end{figure*}

\begin{figure*}
\begin{center}
\begin{tabular}{cc}
\resizebox{0.4\textwidth}{!}{\includegraphics{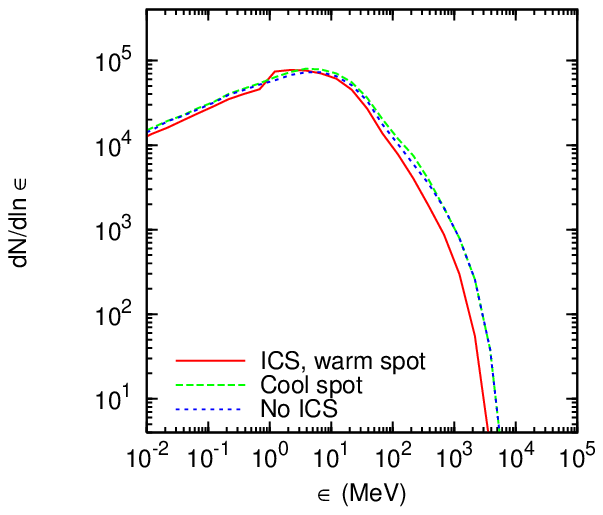}} &
\resizebox{0.4\textwidth}{!}{\includegraphics{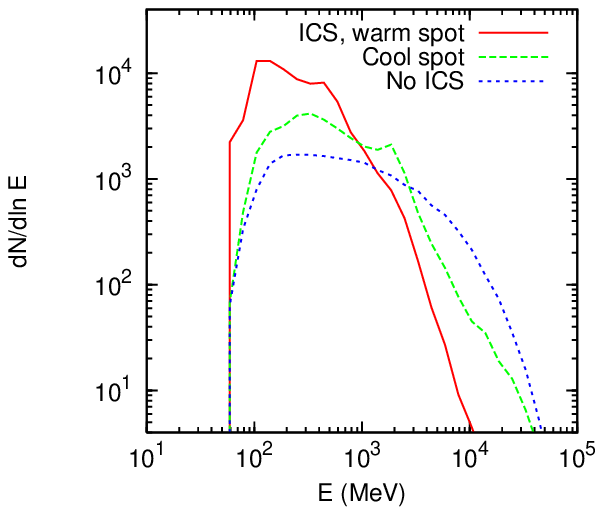}} \\
\end{tabular}
\end{center}
\caption[The effect of resonant inverse Compton scattering on the final photon and pair spectra]
{The effect of resonant inverse Compton scattering on the final photon
and pair spectra. Two surface hot spot models are considered:
a ``warm'' spot ($T_6=1$, $\theta_{\rm spot}=0.3$) and 
``cool'' spot ($T_6=0.3$, $\theta_{\rm spot}=\pi/2$). The case where 
ICS is inactive is also shown. The other cascade parameters are:
$B_{p,12}=44.14$, $P_0=1$ (a dipole field geometry is
used), and $\gamma_0=4\times10^7$. The photon spectra are shown in the
left panel and the pair spectra are shown in the right panel.}
\label{fig:ICScomp}
\end{figure*}

Our results for electron-initiated cascades are presented in
Table~\ref{tab:energies} and
Figs.~\ref{fig:multiplicity}--\ref{fig:ICScomp}. In our simulation,
the primary electron is emitted from the surface along the last open
field line ($\theta_0=\theta_{\rm cap}$). We consider the cases of
$\gamma_0=2\times10^7$ and $4\times10^7$ in dipole magnetospheres, as
well as the case of $\gamma_0=2\times10^6$ in non-dipole ${\cal
R}_c=R$ magnetospheres, as discussed in Section~\ref{sub:eparam}.
Although a larger initial primary energy $\gamma_0$ gives rise to more
cascade particles with a larger total energy, the behavior of the
cascade at $\gamma_0 > 4\times10^7$ is qualitatively similar to that
at $\gamma_0=(2$--$4)\times10^7$ (or $\gamma_0=2\times10^6$ and ${\cal
R}_c=R$) and key quantities such as cascade multiplicities and
energies can be extrapolated from our results. For most of this
section we ``turn off'' ICS in our simulation and allow photon
splitting only through the $\perp\rightarrow\parallel\,\parallel$
mode; at the end of this section we discuss how changing these
simulation parameters affects the cascade.

In Table~\ref{tab:energies} we list some key quantitative results of
our simulations: For each cascade (characterized by the spin period
for dipole fields or the curvature radius for multipole fields, the
surface field strength $B_p$ and the primary electron energy $\gamma_0
m_ec^2$), we give $\gamma_f m_ec^2$ (the final energy of the primary
electron when it escapes the light cylinder), $\varepsilon_{\rm tot}$
(the total energy of the cascade photons), ${\cal E}_{\rm tot}$ (the
total energy of the secondary $e^+e^-$ pairs), and $N_E$ (the
multiplicity of $e^+e^-$ pairs produced per primary electron). Note
that the total cascade energy must satisfy $\gamma_0 m_ec^2 = \gamma_f
m_ec^2+{\cal E}_{\rm tot} +\varepsilon_{\rm tot}$. We find that $N_E$
is largest for cascades with strong surface fields, short rotation
periods, multipole geometries, or large initial energies for the
primary electron, but that regardless of cascade parameters the
particle multiplicity saturates at $N_E \sim 10^4$. Increasing $B_p$
or $\gamma_0$, or decreasing $P$ or ${\cal R}_c$, tends to increase
the ratio $f_E={\cal E}_{\rm tot}/(\varepsilon_{tot}+{\cal E}_{\rm
tot})$; i.e., under these conditions a larger fraction of the
``secondary'' energy, $\gamma_0 m_ec^2-\gamma_f m_ec^2$, is transfered
from the photons to the pairs. At low fields only a small fraction of
the secondary energy is held by the secondary pairs (e.g., for
$B_{p,12}=1$, $f_E \la 0.05$), but at high fields this fraction is
typically above $50\%$ (e.g,, for $B_{p,12}=1000$, $f_E \ga 0.8$). The
average energy of a secondary electron/positron, $\bar{E}={\cal
E}_{\rm tot}/N_E$, is directly proportional to the total cascade
energy (i.e., $\bar{E} \propto \gamma_0$), but depends only weakly on
$B_p$, $P$, and ${\cal R}_c$ (e.g., $\bar{E}$ is approximately the
same for ${\cal R}_c=R$ and $\gamma_0=2\times 10^7$ as in the dipole
case for $\gamma_0=2\times10^7$).

\begin{table*}
\caption[Cascade energies and multiplicities]
{Energies and multiplicities for a cascade initiated by a single
electron. Listed are the initial and final energy of the primary
electron, $\gamma_0 m_ec^2$ and $\gamma_f m_ec^2$, respectively; the
total energy of the cascade photons, $\varepsilon_{\rm tot}$; and the
multiplicity and total and average energies of the secondary electrons
and positrons, $N_E$, ${\cal E}_{\rm tot}$, $\bar{E}={\cal E}_{\rm
tot}/N_E$, respectively. Each cascade is specified by the magnetic
field strength, field geometry (curvature radius ${\cal R}_c$ or spin
period in the case of dipole fields), and cascade energy ($\gamma_0
m_ec^2$).}
\centering
\begin{tabular}{c c c c c c c c}
\hline\hline
$P$ & $B_p$ & $\gamma_0 m_ec^2$ & $\gamma_f m_ec^2$ & $\varepsilon_{\rm tot}$ & ${\cal E}_{\rm tot}$ & $N_E$ & $\bar{E}$ \\
(1~s) & ($10^{12}$~G) & (MeV) & (MeV) & (MeV) & (MeV) & & (MeV) \\
\hline\hline
10 & 10 & 1.022e7 & 8.1e6 & 2.1e6 & 7.4e3 & 1.3e1 & 5.7e2 \\
\hline
 & 100 & & & 2.0e6 & 8.1e4 & 1.7e2 & 4.8e2 \\
\hline
 & 1000 & & & 1.9e6 & 1.8e5 & 5.1e2 & 3.5e2 \\
\hline\hline
1 & 1 & & 4.8e6 & 5.4e6 & 1.9e4 & 5.3e1 & 3.6e2 \\
\hline
 & 10 & & & 4.9e6 & 5.8e5 & 9.6e2 & 6.0e2 \\
\hline
 & 100 & & & 4.0e6 & 1.4e6 & 3.0e3 & 4.7e2 \\
\hline
 & 1000 & & & 3.4e6 & 2.0e6 & 5.9e3 & 3.4e2 \\
\hline\hline
0.1 & 1 & & 2.4e6 & 7.4e6 & 4.0e5 & 1.5e3 & 2.7e2 \\
\hline
 & 10 & & & 4.4e6 & 3.4e6 & 6.0e3 & 5.7e2 \\
\hline
 & 100 & & & 2.8e6 & 5.0e6 & 1.1e4 & 4.5e2 \\
\hline
 & 1000 & & & 2.1e6 & 5.7e6 & 1.7e4 & 3.4e2 \\
\hline\hline
1 & 1 & 2.044e7 & 4.9e6 & 1.5e7 & 6.3e5 & 8.2e2 & 7.7e2 \\
\hline
 & 10 & & & 9.5e6 & 5.9e6 & 3.8e3 & 1.6e3 \\
\hline
 & 100 & & & 6.3e6 & 9.2e6 & 6.9e3 & 1.3e3 \\
\hline
 & 1000 & & & 4.8e6 & 1.07e7 & 1.1e4 & 1.0e3 \\
\hline\hline
${\cal R}_c=R$ & 1 & 1.022e6 & 5.8e5 & 3.9e5 & 5.2e4 & 6.5e3 & 8.0e0 \\
\hline
 & 10 & & & 8.0e4 & 3.7e5 & 2.0e4 & 1.9e1 \\
\hline
 & 100 & & & 1.3e4 & 4.3e5 & 1.9e4 & 2.3e1 \\
\hline
 & 1000 & & & 1.5e4 & 4.3e5 & 1.8e4 & 2.4e1 \\
\hline\hline
\end{tabular}
\label{tab:energies}
\end{table*}

Figure~\ref{fig:multiplicity} shows the secondary pair multiplicities
$N_E$ as a function of $\gamma_0$, for various field strengths and
periods/field geometries. We find that for dipole geometries $N_E$ is
a strong function of both $\gamma_0$ and the ratio $B_p P^{-2}$ (i.e.,
the polar cap voltage), but depends very weakly on either $P$ or $B_p$
alone. For example, the $N_E$ versus $\gamma_0$ curves for
$B_{p,12}=1,P_0=1$ and $B_{p,12}=100,P_0=10$ are nearly the same, as
is shown in Fig.~\ref{fig:multiplicity}. Assuming that a dense
secondary pair plasma is a necessary ingredient in pulsed radio
emission from neutron stars (see Section~\ref{sec:intro}), we can use
our pair multiplicity results to estimate the conditions for pulsar
``death'' --- the conditions under which a NS is no longer active as a
radio pulsar: For a particular set of cascade parameters $B_p$, $P$,
and ${\cal R}_c$, we define $\gamma_{\rm death}$ to be the value of
$\gamma_0$ at which on average only one electron/positron is created
per primary electron; i.e.,
\be
N_E\left(\gamma_0=\gamma_{\rm death}\right)=1 \,.
\label{eq:Nedeath}
\ee
The value of $\gamma_{\rm death}$ changes very little if we alter the
critical value of $N_E$ in Eq.~(\ref{eq:Nedeath}) by a factor of
$\sim10$, because of the steep dependence of $N_E$ on $\gamma_0$ in
this region. Therefore, although it is unknown exactly what value
$N_E$ must have for a pulsar to be active, $\gamma_0=\gamma_{\rm
death}$ is a good predictor of pulsar death regardless. Using
Fig.~\ref{fig:multiplicity} we find empirically that
\be
\gamma_{\rm death} \simeq 1.5\times10^7 B_{p,12}^{-1/6} {\cal R}_8^{2/3} \,.
\ee
For dipole fields, ${\cal R}_8=0.9P_0^{1/2}$, and we can write
\bal
\gamma_{\rm death} \simeq {}& 1.4\times10^7 B_{p,12}^{-1/6} P_0^{1/3} \nonumber\\
 = {}& \left(\frac{\Phi_{\rm death}}{\Phi_{\rm cap}}\right)^{1/6}
\left(\frac{e\Phi_{\rm death}}{m_ec^2}\right) \,,
\label{eq:gammadeath}
\eal
where $\Phi_{\rm cap}$ is given by Eq.~(\ref{eq:phicap}) and
\be
\Phi_{\rm death} = 7\times10^{12}~{\rm V}.
\ee
Therefore, for dipole fields, the death line is given approximately by
$\Phi_{\rm cap}=\Phi_{\rm death}$, or
\be
P_0 = B_{p,12}^{1/2} \,;
\ee
for $B_{p,12}=1$--$1000$ this is nearly the same as the death line
depicted in Fig.~\ref{fig:death} ($P_0 = 0.6 B_{p,12}^{8/15}$).

An example of how the cascades develop spatially is presented in
Fig.~\ref{fig:dndlnr}, which shows the number of photons and secondary
pairs as a function of the radius at which each particle is created,
for surface fields strengths $B_{p,12}=1$, $44.14$ (i.e., $B_p=B_Q$),
and $1000$. The right two panels show the difference in the behavior
of the cascade when the local magnetic field strength $B$ is above or
below $B_{\rm crit} \simeq 3\times10^{12}$~G [see
Eq.~(\ref{eq:Bcrit})], i.e., below or above $r\simeq 2.5R$ for
$B_p=B_Q$ and $r \simeq 7R$ for $B_{p,12}=1000$ [see
Eq.~(\ref{eq:Blocal})]. For $B \ga B_{\rm crit}$, pair production by
curvature photons is very efficient; for the conditions depicted in
Fig.~\ref{fig:dndlnr} ($P_0=1$ and $\gamma_0=2\times10^7$)
approximately one $e^\pm$ pair is created for every two photons
emitted by the primary electron. This is because about half of all
the photons created at low altitudes lie above the limit
$\epsilon_{\rm min} \sim 200$~MeV of Eq.~(\ref{eq:phminHighB}) (see
Fig.~\ref{fig:BCompR}). Photon splitting has very little effect on the
curvature photons, since the majority ($7/8$) of these photons are
$\parallel$-polarized. There is no synchrotron radiation for
$B>0.5B_Q$, since the $\parallel$ photons will pair produce through
the $(jk)=(00)$ channel and $\perp$ photons will split into two
$\parallel$ photons before pair producing (see
Section~\ref{sub:photon}). For $B<0.5B_Q$, pair production dominates
over photon splitting, such that a few synchrotron photons are emitted
from electrons and positrons in the $n=1$ Landau level. For $B <
B_{\rm crit}$, electrons and positrons are created in higher Landau
levels, such that many synchrotron photons are produced by each
electron or positron ($\sim10$; see Appendix~\ref{sub:photemper}). The
minimum photon energy for pair production $\epsilon_{\rm min} \propto
B^{-1}$ [Eq.~(\ref{eq:phmin})] grows with radius while the energy of
the typical curvature photon $\epsilon_{\rm CR} \propto \gamma^3$
[Eq.~(\ref{eq:ecr})] and synchrotron photon $\epsilon_{\rm SR} \propto
\epsilon_{\rm CR}$ (Appendix~\ref{sub:photemper}) fall with radius;
therefore the number of electron-positron pairs created per photon
decreases rapidly with radius.

Figures~\ref{fig:BComp} and \ref{fig:PComp} show the final spectra
(i.e., the spectra as measured at the light cylinder) of photons and
pairs, as well as the spectra of curvature photons emitted by the
primary electron, for a variety of magnetic field strengths, spin
periods, and cascade energies. Figure~\ref{fig:BCompR} shows
cumulative photon and pair spectra at various magnetosphere
radii. These spectra are generated by recording the energy of any
photon, electron, or positron which reaches various radii (such as
$r/R=1.2,~2,~5,~20$).

Figures~\ref{fig:BComp}--\ref{fig:BCompR} show many of the same trends
as are seen in Table~\ref{tab:energies} and Fig.~\ref{fig:dndlnr}.
For example, the multiplicity $N_E$ increases with increasing $B_p$,
$\gamma_0$, $P^{-1}$, or ${\cal R}_c^{-1}$. But there are also
several new features. First, increasing $B_p$ or $\gamma_0$, or
decreasing $P$ or ${\cal R}_c$, tends to increase the maximum energy
and decrease the minimum energy of the cascade pairs, so that the pair
spectrum becomes broader in energy (see Section~\ref{sub:elecemper}).
Second, the cumulative photon and pair spectra at a given radius $r$
(Fig.~\ref{fig:BCompR}) are nearly independent of $B_p$, as long as
the local field strength at that radius is larger than $B_{\rm crit}$
(e.g., the spectra at $r=2R$ are nearly the same for $B_{p,12}=44.14$
and $1000$). Third, under certain conditions, synchrotron radiation
dominates the cumulative/final photon spectrum at low energies
$\epsilon \la 1$~MeV (e.g., in the top middle and top right panels of
Fig.~\ref{fig:PComp}): For a given altitude $r$, the ratio of the
number of synchrotron photons to the number of curvature photons at
low energies is largest for large $\gamma_0$, small $P$, or small
${\cal R}_c$. The ratio of synchrotron photons to curvature photons is
also largest for $B_p$ close to $B_{\rm crit}$, such that the point of
maximum synchrotron radiation, $B \simeq B_{\rm crit}$ (see
Fig.~\ref{fig:dndlnr}), occurs at a low altitude. Note that for
$\epsilon \la 0.1$~MeV the cumulative and final photon spectra all
have power-law shapes with index $\Gamma=2/3$ (where $dN/d\epsilon =
\epsilon^{-\Gamma}$), regardless of whether synchrotron radiation from
the secondary pairs or curvature radiation from the primary electron
dominates; this is because both processes have spectra that depend on
$F(x)/\epsilon \propto \epsilon^{-2/3}$ at low energies [i.e., for $x
\ll 1$; see Eqs.~(\ref{eq:curvature}) and (\ref{eq:synch})].

Figure~\ref{fig:ICScomp} shows the effect of resonant inverse Compton
scattering on the final photon and pair spectra. Both a ``warm'' hot
spot ($T_6=1$, $\theta_{\rm spot}=0.3$), and a ``cool'' spot
($T_6=0.3$, $\theta_{\rm spot}=\pi/2$) are considered, as well as the
case where ICS is inactive. For the cascade parameters adopted in our
simulations, the results for ``hot'' hot spots ($T_6=3$, $\theta_{\rm
spot}=0.1$) are nearly the same as for warm spots, and so are not
shown. This is because the increased cascade activity due to the
larger $T$ almost exactly cancels the decreased activity due to the
smaller maximum photon-electron intersection angle $\psi_{\rm crit}$).
We find due to resonant ICS, a majority of the electrons and positrons
with energies in the range $E_{\rm RICS} \sim 20$--$2000 B_{p,12}
T_6^{-1}$~MeV [Eq.~(\ref{eq:Eics})] radiatively cool to below that
range. This has a strong effect on the pair spectra if the average
electron/positron energy $\bar{E}$ lies in this range (which occurs,
e.g., for $B_{p,12} \la 44.14$ when $P_0 \ge 0.1$ and $\gamma_0 \le
4$; see Table~\ref{tab:energies}). In general, we find that including
RICS tends to make the cascade pair energy distribution narrower. On
the other hand, ICS has only a minor effect on the photon spectra; its
effect is moderate when $\bar{E}\sim 20$--$2000 B_{p,12} T_6^{-1}$~MeV
and ${\cal E}_{\rm tot} >\varepsilon_{\rm tot}$ (which only occurs
when ${\cal R}_c=R$ and $B_{p,12}=10$; see Table~\ref{tab:energies}).
Regardless of the cascade parameters, resonant ICS by the secondary
particles has very little effect on the multiplicity of photons
$N_\epsilon$ or pairs $N_E$ created in the cascade, so it does not
alter the pulsar death line given by Eq.~(\ref{eq:gammadeath}).

\begin{figure*}
\begin{center}
\begin{tabular}{cc}
\resizebox{0.4\textwidth}{!}{\includegraphics{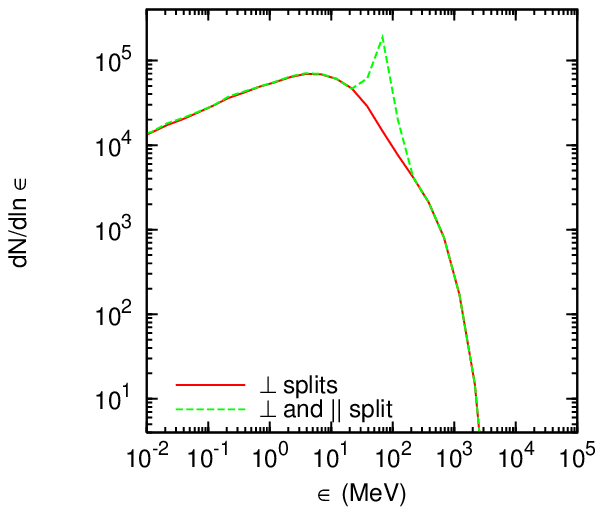}} &
\resizebox{0.4\textwidth}{!}{\includegraphics{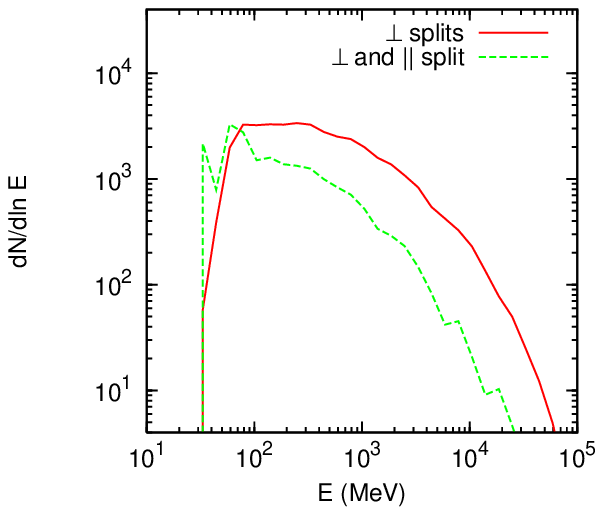}} \\
\end{tabular}
\end{center}
\caption[The effect of photon splitting on the final photon and pair spectra]
{The effect of photon splitting on the final photon and pair spectra.
The solid lines show the case where only photons with perpendicular
polarizations are allowed to split ($\perp \rightarrow
\parallel\,\parallel$), and the dashed lines show the case where
photons of both polarizations are allowed to split. The other cascade
parameters are: $B_{p,12}=1000$, $P_0=1$ (a dipole field geometry is
adopted), and $\gamma_0=4\times10^7$. The photon spectra are shown in
the left panel and the pair spectra are shown in the right panel. For
typical cascade parameters, the spectra when photon splitting is
``turned off'' in the simulation are nearly indistinguishable from the
spectra when only $\perp \rightarrow \parallel\,\parallel$ is allowed;
we therefore do not present the turned-off case here (but see
Fig.~\ref{fig:BComp}).}
\label{fig:split}
\end{figure*}

Figure~\ref{fig:split} shows the effect of photon splitting on the
final photon and pair spectra. We consider both the case where photon
splitting is allowed only in the
$\perp\rightarrow\parallel\,\parallel$ mode (as discussed in
Section~\ref{sub:photon} and in agreement with the selection rule of
\citealt{adler71,usov02}), and the case where both $\perp$ and
$\parallel$ photons are allowed to split (as is suggested in
\citealt{baring01}). For the cascade parameters adopted in
Fig.~\ref{fig:split}, we find that when photon splitting is ``turned
off'' completely, the spectra are nearly indistinguishable from the
case where photon splitting is allowed only in the
$\perp\rightarrow\parallel\,\parallel$ mode. In general, the effect of
photon splitting when both polarizations are allowed to split is
strongest for large $B_p$ and at high energies (e.g., large
$\gamma_0$), such that the quantity $x\beta_Q$ is large [see
Eq.~(\ref{eq:split})]. At altitudes where the local field strength is
$B \ga 0.5B_Q$ (which occurs, e.g., at $r \la 4R$ for
$B_{p,12}=1000$), most photons continue to split until they reach
$\epsilon_{\rm min}$ [Eq.~(\ref{eq:phmin}) or (\ref{eq:phminHighB})],
and very few pairs are created. However, once the cascade reaches an
altitude such that $B < 0.5B_Q$, photon splitting has very little
effect on the cascade. Because a majority of pairs are produced at
altitudes above $B \simeq 0.5B_Q$ (see Figs.~\ref{fig:dndlnr} and
\ref{fig:BCompR}), the pair spectrum is not strongly affected by
photon splitting even when both polarizations are allowed to split and
$B_{p,12}=1000$.

\section{Discussions}
\label{sec:discuss}

We have presented numerical simulations of pair cascades in the
open-field line regions of neutron star magnetospheres, for surface
magnetic fields ranging from $B_p=10^{12}$~G to $10^{15}$~G, rotation
periods $P=0.1$--$10$~s, and dipole and more complex field geometries.
Compared to previous studies (e.g., \citealt{daugherty82,sturner95,
daugherty96,hibschman01b,arendt02}), which were restricted to weaker
magnetic fields ($B\la {\rm a~few}\times 10^{12}$~G) and dipole
geometry, we have incorporated in our simulations additional physical
processes that are potentially important (especially in the high field
regime) but were either neglected or crudely treated before, including
photon splitting with the correct selection rules for photon
polarization modes, one-photon pair production into low Landau levels
for the $e^\pm$, and resonant inverse Compton scattering from polar
cap hot spots [with $T=(0.3$--$3)\times10^6$~K]. Both cascades
initiated by a single electron (representing the entire cascade
process described in Section~\ref{sec:intro}) and by a single photon
(representing one branch of the cascade) are simulated, for a variety
of initial energies of the primary particle ($\gamma_0 m_ec^2$ and
$\epsilon_0$).  We have made an effort to present our numerical
results systematically, including empirical relations for the pair
multiplicity (i.e., the number of electrons and positrons produced per
primary particle).

Locally the cascade behaves very differently above and below the
critical QED field strength $B_Q\equiv4.414\times 10^{13}$~G (e.g.,
pair production is $\sim10$ times more efficient for $B > B_Q$ but
synchrotron emission is highly suppressed; see
Section~\ref{sec:results}). Globally, however, the cascade followed
from the surface to the light cylinder behaves similarly regardless of
surface field strength $B_p$. For example, we find that the total
number of pairs produced in electron-initiated cascades, as well as
their energy spectrum, depends on the polar cap voltage $B_p P^{-2}$
but is weakly dependent on $B_p$ alone.  Additionally, the total
photon spectra for $B > B_Q$ and $B < B_Q$ have similar shapes over
the energy range we consider in our simulation (10~keV-1~TeV); this is
because curvature radiation, which dominates the photon spectrum over
most of this energy range, is independent of magnetic field
strength. Photon splitting, a process that is only active for $B \ga
B_Q$, could potentially distinguish between neutron stars with high
and low surface fields, by lowering the multiplicity of electrons and
positrons $N_E$ produced in the cascade to such a degree that the
radio emission mechanism can not function
(\citealt{baring01}). However, we find that even if both photons
polarized parallel to and perpendicular to the magnetic field are
allowed to split, photon splitting lowers $N_E$ by at most $50\%$ for
$B_{p,12}=1000$ and by $<10\%$ for $B_{p,12} \le 100$.  With the
correct selection rule ($\perp\rightarrow \parallel\,\, \parallel$),
the effect of photon splitting is even smaller.

Our results show that a strongly non-dipole magnetic field with radius
of curvature $\sim10^6$~cm near the stellar surface ($r$ less than a
few stellar radii) can account for pulsars which lie below the
standard death line for dipole fields (see
Section~\ref{sec:accel}). Whether or not such a strongly-curved
geometry could form is another question altogether. It is doubtful
that cascades initiated by resonant inverse Compton scattering in the
open field region of the magnetosphere can account for the ``missing''
pulsars, as the multiplicity of pairs produced in such cascades ($\la
1$ per primary electron) are too low for the current models of the
pulsar radio emission. High-multiplicity cascades due to resonant ICS
may still occur along twisted lines in the closed field region (where
the primary electrons never reach Lorentz factors larger than
$\gamma_0 \sim 10^3$), as they do in magnetars (e.g.,
\citealt{thompson02,beloborodov07,thompson08a,thompson08b,beloborodov09}).

At any altitude in the magnetosphere, the photon spectrum at low
energies ($\la 1$~MeV) has a power-law shape with spectral index
$\Gamma=2/3$ (where $dN/d\epsilon \propto \epsilon^{-\Gamma}$); this
is true regardless of whether synchrotron radiation or curvature
radiation dominates the spectrum, as both processes have the same
low-energy shape. The photon spectrum $dN/d\epsilon\propto
\epsilon^{-2/3}$ is harder than the hard X-ray spectra observed in
several pulsars, which typically have photon indices $\Gamma \sim
1$--$2$ (see, e.g., \citealt{kuiper09,bogdanov09}). One way to
reconcile our results with observations is to include an additional
radiative process, cyclotron resonant absorption, in the
simulation. Such an approach is taken by \citet{harding05} (see also
\citealt{harding08}), who find that in addition to the synchrotron
radiation emitted immediately upon creation, the $e^\pm$ particles
emit an additional component of synchrotron radiation at large
altitudes that dominates the low-energy spectrum. The efficiency of
this high-altitude synchrotron emission is due to the large pitch
angles of the $e^\pm$, which they obtain through resonant absorption
of radio photons that are beamed from some intermediate height in the
magnetosphere. The high-altitude synchrotron emission will have a
hard X-ray part with photon index $\Gamma_{\rm 10-1000~keV}$ generally
different from $2/3$, because the hard X-ray band lies near the peak
of the emission (rather than in the low-energy tail as is the case for
the low-altitude emission). \citet{harding05} find that for typical
millisecond parameters the photon index is in the observed range, $1 <
\Gamma_{\rm 10-1000~keV} < 2$.

To fully incorporate the effects of cyclotron absorption into our
simulation, we would need a model of the radio beam structure and
spectrum, since the evolution in the pitch angle of each $e^\pm$
particle depends on the angle at which radio photons of the resonant
frequency are incident on the particle and their density. Empirical
models of radio beams exist (e.g.,
\citealt*{rankin93,arzoumanian02,kijak03}); however, inclusion of such
a model is beyond the scope of this paper. This means that we can not
say anything in detail about the hard X-ray spectra of
strongly-magnetized neutron stars, such as whether the high-altitude
synchrotron component dominates and what its low-energy cutoff is (but
see \citealt{harding05} for examples of these spectra at several
different cascade parameters). Nevertheless, we can use the $e^\pm$
spectra from our simulation to estimate how the $\Gamma_{\rm
10-1000~keV}$ photon index varies as a function of the cascade
parameters. For synchrotron emission from a distribution of $e^\pm$
particles, the hardness of the $e^\pm$ spectrum and the hardness of
resulting photon spectrum are correlated [for a $dN/dE \propto E^{-p}$
power-law distribution of $e^\pm$ the photon spectrum has an index
given by the familiar expression $\Gamma = (p+1)/2$]. We find that the
$e^\pm$ spectrum is harder, and therefore $\Gamma_{\rm 10-1000~keV}$
is lower, for larger $B_p$ or $\gamma_0$ or shorter $P$ (i.e.,
anything that increases the energy of the cascade); additionally, for
two neutron stars with the same polar cap voltage $B_p P^{-2}$,
$\Gamma_{\rm 10-1000~keV}$ is lower in the neutron star with the
stronger surface field. Therefore, in pulsars where the cyclotron
absorption -- synchrotron emission mechanism dominates the hard X-ray
spectrum, the most-strongly magnetized pulsars will have the hardest
spectra (lowest photon indices), all other parameters being equal.

Luminous hard X-ray (from 20~keV to several hundreds of keV) emission
has also be detected from a number of Anomalous X-ray Pulsars by
INTEGRAL and RXTE (e.g., \citealt{kuiper06}). Possible
mechanisms for this emission were discussed by 
Thompson \& Beloborodov (2005) and Beloborodov \& Thompson (2007).
Since the observed hard X-ray luminosity exceeds the spin-down power by
several orders of magnitude, it must be fed by an alternative source
of energy such as the dissipation of a superstrong magnetic field.
We note, however, that the observed photon indices, $\Gamma\sim
0.8-1.8$, are similar to (but slightly harder than) those of radio
pulsars. Synchrotron radiation by secondary $e^+e^-$ pairs
produced in a cascade similar to those studied in this paper 
could plausibly explain the observations.

In constructing our simulations, we have attempted to rely as little
as possible on the precise nature of the acceleration region. However,
there are several key assumptions that we have made that are only
valid for inner gap accelerators (e.g., that the cascade begins at the
neutron star surface). It has become apparent that models where
particle acceleration occurs only in the inner gap can not account the
observed high-energy gamma ray emission from young pulsars. For
example, the gamma-ray pulse profiles (e.g., the widely separated
double peaks) of the six pulsars detected by EGRET already suggested
that models of high-altitude gamma-ray emission were required.
More recently, observations of the Crab pulsar by the Major
Atmospheric Gamma-ray Imaging Cherenkov Telescope (MAGIC) and around
50 pulsars with ``above average'' spindown powers by the Fermi
Gamma-ray Space Telescope have revealed that the high-energy tails
of the photon spectra fall off exponentially or more slowly than
exponential (\citealt{aliu08,abdo09a,abdo09b,abdo09c,abdo10}),
while inner gap models predict that the tails fall off faster than
exponential (see, e.g., Fig.~\ref{fig:BComp}). The outer gap model and
the high-altitude version of the slot gap model have been successful
in explaining the $\gamma$-ray pulsar light curves (e.g.,
\citealt*{watters09,venter09}; but see \citealt{bai09}).

Although an outer gap or slot gap accelerator model is required to
explain gamma-ray observations, our results based on the inner gap
model still have general applicability. First, even when the
acceleration region is located in the outer magnetosphere a
significant fraction of the pair creation must occur in the inner
magnetosphere (e.g., \citealt{cheng00}). Indeed, pulsar radio emission
is strongly constrained to originate from well inside the light
cylinder radius; the only way to generate highly coherent radio
emission is to have copious $e^\pm$ plasma produced by vigorous pair
cascades (e.g., \citealt{melrose04,lyubarsky08}). Second, both inner
gap and outer gap accelerators may exist in a neutron star
simultaneously, whether as one extended acceleration region (e.g.,
\citealt{hirotani06}) or on different field lines. Radio observations
suggest that radio emission can come from intermediate field lines,
neither along the pole of the star nor at the edge of the open field
region (\citealt{gangadhara04}); if this is true, it provides further
support for an inner gap origin of the $e^\pm$ plasma, since the
plasma generated by an outer gap model is formed on field lines close
to the last open field line. In addition, as discussed above, hard
X-ray (10-1000~keV) observations of both pulsars and magnetars support
a magnetosphere model where the hard X-ray emission is dominated by
radiation from the $e^+e^-$ pairs generated by inner gap
cascades. Third, it is unlikely that the death lines for the inner gap
and outer gap mechanisms overlap completely. There should therefore be
regions of the $P$-$\dot{P}$ diagram where the outer gap mechanism is
excluded but the inner gap mechanism still functions; these regions,
if they exist, will be located near the inner gap death line, where
pulsars have very low spindown powers. Although observations show that
the gamma-ray spectra is dominated by slot or outer gap emission for
pulsars with moderate to large spindown power, they have not yet ruled
out an inner gap origin for pulsars with low spindown power.

\section*{Acknowledgments}

ZM has been supported in part by the Lorne Trottier Chair in
Astrophysics and Cosmology and an NSERC Discovery Grant awarded to
Andrew Cumming. This work forms part of the Ph.D. thesis of ZM at
Cornell University. This work has been supported in part by NASA grant
NNX07AG81G and NSF grants AST 0707628.

\appendix

\section{Resonant inverse Compton scattering}
\label{sec:rics}

\begin{figure}
\includegraphics[width=\columnwidth]{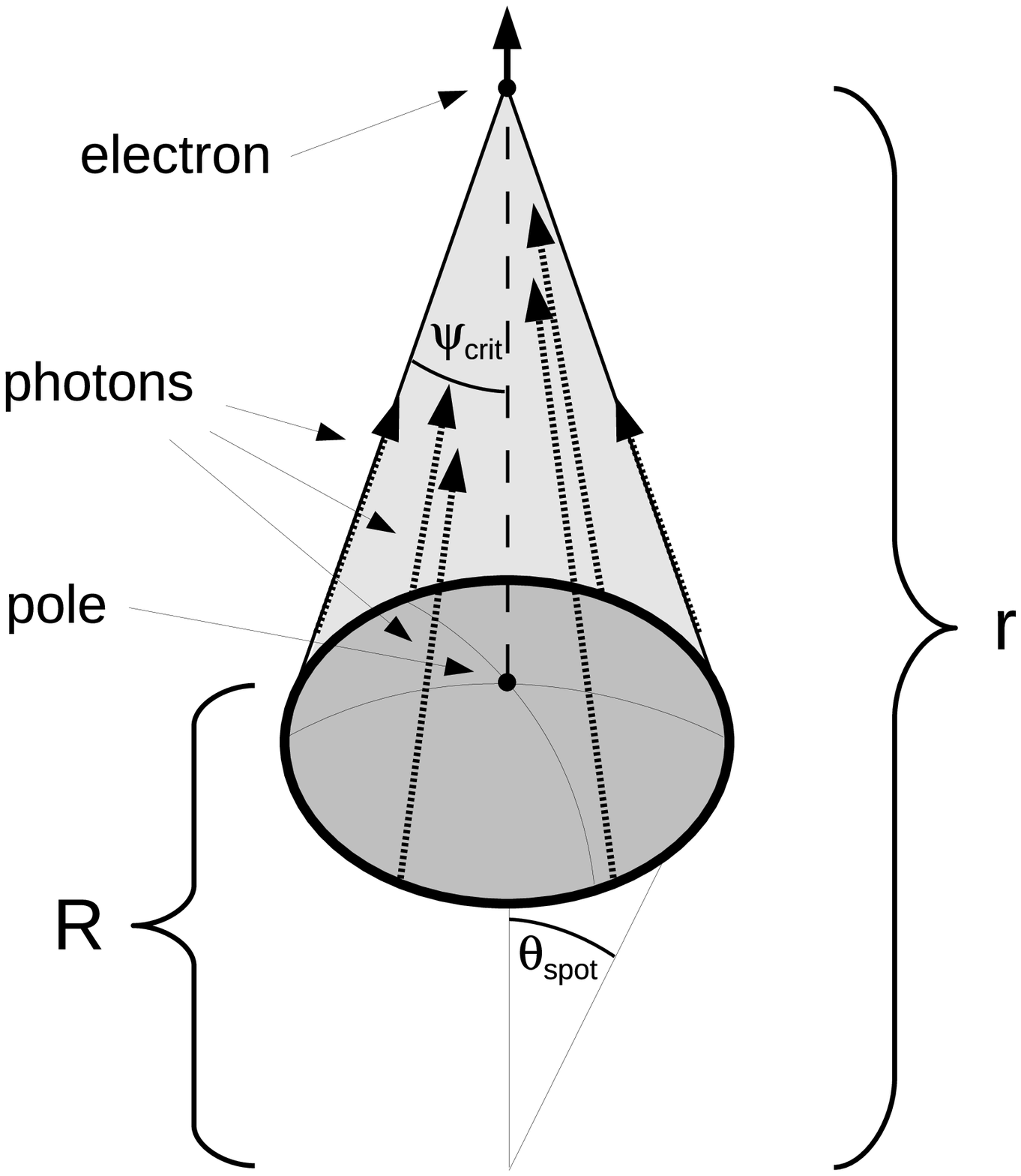}
\caption[A simplified picture of the ICS effect on the electron]
{A simplified picture of the ICS effect on the electron. The
electron is assumed to be directly above the magnetic pole 
and travel radially outward.}
\label{fig:ICS}
\end{figure}

Here we calculate the photon scattering rate for the resonant inverse
Compton process, using the simplified model of an electron positioned
directly above the magnetic pole (hot spot) and traveling radially
outward; see Fig.~\ref{fig:ICS}. The resonant cross section for
inverse Compton scattering, in the rest frame of the electron before
scattering, is
\be
\sigma'_{\rm res} \simeq 2\pi^2 \frac{e^2 \hbar}{m_ec}\,
\delta(\epsilon'-\epsilon_c) \,,
\label{eq:sigma}
\ee
where $\epsilon' = \gamma\epsilon_i(1-\beta\cos\psi)$, $\epsilon_i$ is
the photon energy in the ``lab'' frame, and $\psi$ is the incident
angle of the photon with respect to the electron's trajectory. This
cross section is appropriate even for $B > B_Q$, since the resonant
condition $\epsilon' = \epsilon_c$ holds regardless of field strength
[though at the highest field strengths the prefactor $2\pi^2 e^2
\hbar/m_ec$ in Eq.~(\ref{eq:sigma}) serves only as an upper bound to
the exact expression; see \citealt{gonthier00}]. The polar hot spot
has a angular size of $\theta_{\rm spot}$; this sets a maximum value
for $\psi$ of
\be
\cos\psi_{\rm crit} = \frac{r-R\cos\theta_{\rm spot}}{\sqrt{r^2+R^2-2rR\cos\theta_{\rm spot}}} \,,
\label{eq:cospsi}
\ee
where $r$ is altitude of the electron (from the center of the
star). The intensity of emission from the hot spot is
\be
I_{\epsilon_i}(\psi;r) = \left\{
\begin{array}{ll}
B_{\epsilon_i}(T) = \frac{\epsilon_i^3/(4\pi^3\hbar^3c^2)}{e^{\epsilon_i/kT}-1} \,, & \psi < \psi_{\rm crit} \,; \\
0 \,, & \mbox{otherwise.}
\end{array}
\right.
\ee
Therefore, the photon scattering rate for the resonant ICS process
above a polar hot spot is given by
[see, e.g., Eq.~(B3) of ML07]
\bal
\frac{dN_{\rm ph}}{dt} = {}& \int d\Omega_i \int d\epsilon_i \,(1-\beta\cos\psi) \left(\frac{I_{\epsilon_i}}{\epsilon_i}\right) \sigma'_{\rm res} \\
 = {}& \frac{2\pi^2e^2\hbar}{m_ec\gamma} \int_{\psi<\psi_{\rm crit}} d\Omega_i \left(\frac{B_{\epsilon_i}}{\epsilon_i}\right)_{\epsilon_i=\epsilon_c/[\gamma(1-\beta\cos\psi)]} \\
 = {}& \frac{c}{\gamma^2\beta a_0} \left(\frac{kT}{m_ec^2}\right)\beta_Q \ln\frac{1-e^{-\epsilon_c/[\gamma(1-\beta)kT]}}{1-e^{-\epsilon_c/[\gamma(1-\beta\cos\psi_{\rm crit})kT]}} \,.
\label{eq:dNdt}
\eal

\section{Pair production}
\label{sec:pair}

\subsection{Kinematics}
\label{sub:levels}

Consider the pair production of a photon with energy $\epsilon$ and
angle $\psi$ (the photon -- magnetic field intersection angle). In
the frame where the photon is traveling perpendicular to the local
magnetic field direction (this ``perpendicular'' frame moves at
the velocity $c\cos\psi$ relative to the ``lab'' frame), the photon
has energy $\epsilon'=\epsilon\sin\psi$, and energy conservation demands
\bal
\epsilon' = E'_j+E'_k = {}& \sqrt{p'^{\,2}_zc^2+m_ec^2(1+2\beta_Qj)} \nonumber\\
 & + \sqrt{p'^{\,2}_zc^2+m_ec^2(1+2\beta_Qk)} \,,
\eal
where $E'_j$ and $E'_k$ are the energies of the electron and the
positron and $p'_z$ is the momentum along the magnetic field of either
particle ($p'_{z,j} = -p'_{z,k}$). From this we find
\be
|p'_z| = m_ec\sqrt{x^2-1-(j+k)\beta_Q+(j-k)^2\frac{\beta_Q^2}{4x^2}} \,,
\ee
where $x=\epsilon'/(2m_ec^2)$, and
\be
E'_j = m_ec^2 \sqrt{x^2+(j-k)\beta_Q+(j-k)^2\frac{\beta_Q^2}{4x^2}} \,,
\ee
\be
E'_k = m_ec^2 \sqrt{x^2+(k-j)\beta_Q+(j-k)^2\frac{\beta_Q^2}{4x^2}} \,.
\ee
In the ``lab'' frame, the energies of the electron and
positron at the moment of pair creation are given by
\be
E = \frac{1}{\sin\psi}(E' \pm p'_zc \cos\psi).
\label{eq:Ee}
\ee
In our simulation one particle is randomly assigned the `$+$' energy
and the other the `$-$' energy, with equal probability of either
outcome.

\subsection{Photon attenuation coefficients and optical depth}
\label{sub:atten}

In the perpendicular frame, the first three attenuation coefficients
for $\parallel$ and the first two non-zero attenuation coefficients
for $\perp$ photons are (\citealt{daugherty83})
\be
R'_{\parallel,00} = \frac{1}{2a_0}\frac{\beta_Q}{x^2\sqrt{x^2-1}}e^{-2x^2/\beta_Q} \,, \qquad x > 1 \,;
\label{par00eq}
\ee
\bal
R'_{\parallel,01} = {}& 2\times \frac{1}{2a_0}\frac{2+\beta_Q-\frac{\beta_Q^2}{4x^2}}{\sqrt{x^2-1-\beta_Q+\frac{\beta_Q^2}{4x^2}}}e^{-2x^2/\beta_Q} \,, \nonumber\\
 & \qquad\qquad\qquad\qquad x > \left(1+\sqrt{1+2\beta_Q}\right)/2 \,;
\label{par01eq}
\eal
\bal
R'_{\parallel,02} = {}& 2\times \frac{1}{2a_0}\frac{2x^2}{\beta_Q}\frac{1+\beta_Q-\frac{\beta_Q^2}{2x^2}}{\sqrt{x^2-1-2\beta_Q+\frac{\beta_Q^2}{x^2}}}e^{-2x^2/\beta_Q} \,, \nonumber\\
 & \qquad\qquad\qquad\qquad x > \left(1+\sqrt{1+4\beta_Q}\right)/2 \,;
\label{par02eq}
\eal
\bal
R'_{\perp,01} = {}& 2\times \frac{1}{2a_0}\frac{\beta_Q}{2x^2}\frac{2x^2-\beta_Q}{\sqrt{x^2-1-\beta_Q+\frac{\beta_Q^2}{4x^2}}}e^{-2x^2/\beta_Q} \,, \nonumber\\
 & \qquad\qquad\qquad\qquad x > \left(1+\sqrt{1+2\beta_Q}\right)/2 \,;
\label{perp01eq}
\eal
\bal
R'_{\perp,02} = {}& 2\times \frac{1}{2a_0}\frac{x^2-\beta_Q}{\sqrt{x^2-1-2\beta_Q+\frac{\beta_Q^2}{x^2}}}e^{-2x^2/\beta_Q} \,, \nonumber\\
 & \qquad\qquad\qquad\quad\;\; x > \left(1+\sqrt{1+4\beta_Q}\right)/2 \,.
\label{perp02eq}
\eal
Note that $R'_{\perp,00} = 0$. In the above expressions, the
attenuation coefficients of \citet{daugherty83} are multiplied by a
factor of two for all channels but $(00)$ [i.e., in
Eqs.~(\ref{par01eq})-(\ref{perp02eq})], since we are using the
convention $R'_{jk}=R'^{\,(\rm DH83)}_{jk}+R'^{\,(\rm
DH83)}_{kj}=2R'^{\,(\rm DH83)}_{jk}$ for $j\neq k$.

\begin{figure}
\includegraphics[width=\columnwidth]{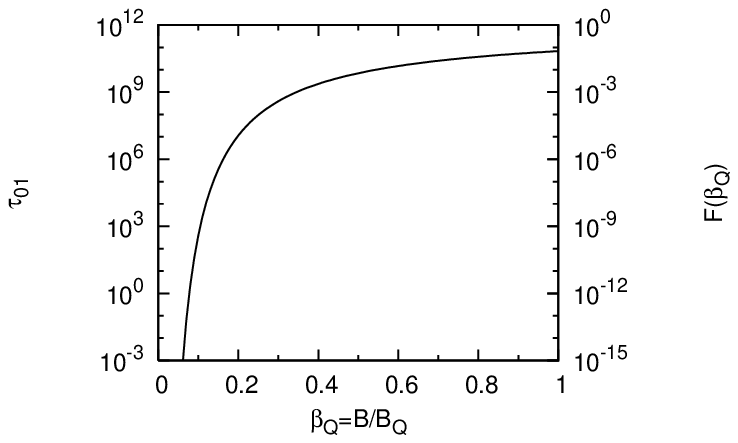}
\caption[The optical depth to reach the second threshold for pair production]
{A plot of the function $T(\beta_Q)$ as given by Eq.~(\ref{fofbeq}) (right axis) 
and the optical depth to reach the second threshold for pair production,
$\tau_{01}$ (left axis), as a function of $\beta_Q$, for $\epsilon=100$~MeV and
${\cal R}_c=10^8$~cm.}
\label{fig:tau}
\end{figure}

We now examine the conditions for pair production by a
$\parallel$-polarized photon; the analysis is similar for a
$\perp$-polarized photon and yields the same result. The optical depth
for pair production is
\be
\tau = \int_0^{s_{\rm ph}} ds \, R(s) = \int_0^{s_{\rm ph}} ds \, R'(s) \sin\psi \,.
\ee
We assume $\psi \ll 1$, which is valid since most photons that can
pair produce will do so long before the angle $\psi$ approaches unity
(only photons with energies $\epsilon \simeq 2m_ec^2$ in the lab frame
must wait until $\psi \sim 1$ to pair produce). In this limit, we have
$\sin\psi \simeq s/{\cal R}_c$, so that $x$ and $s$ are related by
\be
x \simeq \frac{s}{{\cal R}_c}\frac{\epsilon}{2m_ec^2} \,.
\ee
Let $s_{00}$ to be distance traveled by the photon to reach the first
threshold $x=x_{00}\equiv1$, and $s_{01}$ to be the distance traveled
by the photon to reach the second threshold
$x=x_{01}\equiv\left(1+\sqrt{1+2\beta_Q}\right)/2$. The optical depth
to reach the second threshold for pair production is
\bal
\tau_{01} = {}& \int_{s_{00}}^{s_{01}} ds \, R_{\parallel,00}(s) \\
 = {}& \frac{\beta_Q}{2a_0}\left(\frac{2m_ec^2}{\epsilon}\right)^2 R_c \int_{x_{00}}^{x_{01}} \frac{dx}{x\sqrt{x^2-1}}\,e^{-2x^2/\beta_Q} \\
 = {}& 9.87\times10^{11}\left(\frac{{\cal R}_c}{10^8~{\rm cm}}\right)\left(\frac{100~{\rm MeV}}{\epsilon}\right)^2 T(\beta_Q) \,,
\label{eq:tau}
\eal
where
\be
T(\beta_Q) = \beta_Q \int_{x_{00}}^{x_{01}} \frac{dx}{x\sqrt{x^2-1}}\,e^{-2x^2/\beta_Q}\,.
\label{fofbeq}
\ee
We plot $\tau_{01}$ as a function of magnetic field strength in
Fig.~\ref{fig:tau}, for $\epsilon=100$~MeV and $R_c=10^8$~cm.

From Fig.~\ref{fig:tau} we see that pair production occurs in the
$(jk)=(00)$ channel ($\tau_{01} \ge 1$) when
\be
B \ga B_{\rm crit} \simeq 3\times10^{12}~{\rm G}.
\label{eq:critatten}
\ee
Because of the steep dependence of $\tau$ on $B$ for $B \sim
3\times10^{12}$~G, the value of $B_{\rm crit}$ does not change much
for different parameters $\epsilon$ and ${\cal R}_c$. For
example, $B_{\rm crit} = 3\times10^{12}$~G for $\epsilon=100$~MeV and
$R_c=10^8$~cm, and $B_{\rm crit} = 7\times10^{12}$~G for
$\epsilon=10^4$~MeV and $R_c=10^6$~cm. Figure~\ref{fig:tau} also shows
that for $B \la B_{\rm crit}$, the optical depth $\tau_{01}$ is much
less than unity. The same result is found for the optical depth from
the second to the third threshold, and for higher thresholds. The pair
production process can therefore be divided into two regimes: for $B
\la B_{\rm crit}$, photons must travel large distances before pair
producing, at which point the resulting pairs will be in high Landau
levels; for $B \ga B_{\rm crit}$, photons pair produce almost
immediately upon reaching the first threshold, so that the pairs will
be in low Landau levels ($n \la 2$).

\section{Empirical relations for the numerical results}
\label{sec:empirical}

In this section we justify several of the empirical relations given in
Sections~\ref{sub:photi} and \ref{sub:eleci}. In the derivations below
we treat the radius of the emission point for the primary photon,
$r_6=r_{0,\rm ph}/(10^6~\rm cm)$, as a free parameter. This allows the
results to be applicable both to Section~\ref{sub:photi}, where we
assume that $r_6=1$, and to Section~\ref{sub:eleci}, where $r_6 \ge
1$.

\subsection{Photon-initiated cascades}
\label{sub:photemper}

At low fields ($B \la B_{\rm crit}\simeq 3\times10^{12}$~G), we can
use Eq.~(\ref{eq:asymp}) for the attenuation coefficient, which
implies that pair production occurs when $x\beta_Q \simeq
1/15$--$1/10$.  In the following we will use $x\beta_Q=1/10$,
appropriate for photon energy $\epsilon\sim 10^4$~MeV.  To initiate an
effective cascade, a photon must pair produce before traveling a
distance $s_{\rm ph,max}=0.5r_{0,\rm ph}$
[Eq.~(\ref{eq:smax})]. Therefore the minimum photon energy for cascade
is given by
\be
\frac{\epsilon_{\rm min}}{2m_ec^2} \frac{0.5r_{0,\rm ph}}{{\cal R}_c}
\simeq \frac{1}{10\beta_Q} \,,
\ee
where $\beta_Q$ and ${\cal R}_c$ should be evaluated at the pair
creation point $r \simeq s_{\rm ph,max}+r_{0,\rm ph} = 1.5r_{0,\rm
ph}$. Since $\beta_Q = B/B_Q \simeq 0.02 B_{p,12} (r/R)^{-3}$ and
${\cal R}_c \simeq 10^8{\cal R}_8 \sqrt{r/R}$~cm, where $B_{p,12}$
and ${\cal R}_8$ are the magnetic field strength (in units of
$10^{12}$~G) and curvature radius (in units of $10^8$~cm) at the
surface, we find
\be
\epsilon_{\rm min} \sim 3000 B_{p,12}^{-1} {\cal R}_8 r_6^{5/2}~{\rm MeV} \,.
\label{eq:phmin2}
\ee
For a photon injected from the surface ($r_6=1$), this reduces to 
Eq.~(\ref{eq:phmin}).

In each generation of pair production, a photon of energy $\epsilon$
and angle $\psi$ creates an $e^\pm$ pair, each with energy $\gamma
m_ec^2=0.5\epsilon$ and traveling in the same direction as the photon.
The electron and positron radiate synchrotron photons until they reach
an energy of $\gamma_\parallel m_ec^2=m_e c^2/\sin\psi= \gamma
m_ec^2/x \simeq 0.1 B_{12} \epsilon$, since $x=\gamma\sin\psi \simeq
1/(10\beta_Q)$. The synchrotron photons have a characteristic energy
$\epsilon_{\rm SR}=1.5\gamma^2(\sin\psi) \beta_Q m_ec^2 \simeq
0.075\epsilon$. Therefore, in each generation, the energy of the
photons and pairs drops by a factor $\sim 0.075$ while the number of
particles increases by $1/0.075$ (see \citealt{hibschman01a}). When
RICS is inactive (e.g., when $T\le10^5$~K), such that the only
mechanism for energy loss is synchrotron radiation, the final pair
spectrum consists of an $e^\pm$ pair, each with energy
\be
E_{\rm max} \sim 0.1 B_{p,12} r_6^{-3} \epsilon_0 \,,
\label{eq:Emax2}
\ee
created by the primary photon, and a power-law distribution of pairs
with index $p \sim 2$ (where $dN/dE \propto E^{-p}$)
extending from $\sim 0.075E_{\rm max}$ down to
$\sim0.1B_{12}\epsilon_{\rm min}$ (see
Fig.~\ref{fig:photiLowB}). Hence the total energy of the cascade pairs
produced by the primary photon (of energy $\epsilon_0$) is
\be
{\cal E}_{\rm tot} \sim 2E_{\rm max} + 0.1B_{p,12}r_6^{-3}\epsilon_{\rm min} n_E \ln \left(\frac{0.075\epsilon_0}{\epsilon_{\rm min}}\right) \,.
\label{eq:Etot2}
\ee
which reduces to Eq.~(\ref{eq:Etot}) for $r_6=1$.

For the range of high fields considered in this work ($B_{\rm crit}
\la B \le 1000$), pair production occurs when $x=1$ (for $x=x_{00}$)
or $x \simeq 1$--$4$ (for $x=x_{01}$; see
Section~\ref{sub:photon}). For a photon emitted from the point
$(r_{0,\rm ph},\theta_{0,\rm ph})$, the maximum intersection angle
between the photon and the magnetic field is given by
[Eqs.~(\ref{eq:chiangle}), (\ref{eq:thph}), and (\ref{eq:psi})]
\bal
\sin\psi_{\rm max} \simeq {}& \chi\left[\chi\left(\theta_{0,\rm ph}\right)\right]-\chi\left(\theta_{0,\rm ph}\right) \\
 = {}& \arctan\left\{\frac{1}{2}\tan\left[x+\arctan\left(\frac{\tan\theta_{0,\rm ph}}{2}\right)\right]\right\} \,;
\eal
for ``small'' angles $\theta_{0,\rm ph} \le 0.6$,
\be
\sin\psi_{\rm max} \la \sin\theta_{0,\rm ph} \,.
\ee
For $\theta < 0.6$ the local radius of
curvature is given by ${\cal R}_c \simeq 1.3 r/\sin\theta$; therefore,
using $x=\epsilon\sin\psi/(2m_ec^2) \sim 2$ we have
\be
\epsilon_{\rm min} \sim 200 {\cal R}_8 r_6^{-1/2}~{\rm MeV} \,,
\label{eq:phminHighB2}
\ee
which at $r_6=1$ is Eq.~(\ref{eq:phminHighB}).

At high fields, each electron-positron pair is created with energies
$\gamma m_ec^2 \simeq 0.5\epsilon$ (for $\parallel$ photons) or
$\simeq 0.5\epsilon/x_{01}$ and $\simeq 0.5\epsilon (2-1/x_{01})$ (for
$\perp$ photons). After synchrotron radiation the electron and
positron have energies $0.5\epsilon$ or
$0.5\epsilon/x_{01}$. Therefore, the final energies of the electron
and positron created by the primary photon are given by
\be
E_{\rm max} \simeq \frac{\epsilon_0}{2x_{01}}~{\rm or}~\frac{\epsilon_0}{2x_{00}} 
\sim 0.1\epsilon_0~{\rm or}~0.5\epsilon_0 \,.
\label{eq:EmaxHighB}
\ee

In order for resonant ICS to modify the cascade spectra, the electron
(positron) must have an energy (after synchrotron radiation) larger
than the minimum energy at which RICS is effective, i.e., $E_{\rm max}
> 0.3\gamma_{\rm crit} m_ec^2$ [where $\gamma_{\rm
crit}=\epsilon_c/kT$; see Eq.~(\ref{eq:Eics})], which implies
\be
\epsilon_0 \ga 70 B_{p,12} T_6^{-1}~{\rm MeV} \,.
\label{eq:eminICS2}
\ee

\subsection{Electron-initiated cascades}
\label{sub:elecemper}

The spectrum of curvature photons extends from
approximately one photon at
\be
\epsilon_{\rm max} \sim 10\epsilon_{\rm CR}\left(\gamma_0\right) = 3\times10^3 \gamma_7^3 {\cal R}_8^{-1}~{\rm MeV} \,,
\ee
where $\gamma_7$ is the primary electron's initial Lorentz factor
$\gamma_0$ in units of $10^7$, up to a maximum of $\sim 6\times10^4
{\cal R}_8^{1/2}$ photons at $\epsilon_{\rm peak} \sim 6 {\cal
R}_8^{1/2}$~MeV. For photon energies below $\epsilon_{\rm peak}$ the
spectrum is a power law with spectral index $\Gamma = 2/3$ (where
$dN/dE \propto E^{-\Gamma}$), characteristic of curvature and
synchrotron radiation at low energies (e.g., \citealt{erber66}).

At low fields $B<B_{\rm crit}$ the pair spectrum extends from
\be
E_{\rm max} \simeq 0.1 B_{p,12} \epsilon_{\rm max} \sim 300 \gamma_7^3 B_{p,12} {\cal R}_8^{-1}~{\rm MeV}
\label{eq:Emax4}
\ee
down to
\be
E_{\rm min} \simeq 0.1 B_{12} \epsilon_{\rm min} \sim 300 {\cal R}_8 r_6^{-1/2}~{\rm MeV} \,;
\label{eq:Emin2}
\ee
at high fields
\be
E_{\rm max} \simeq 0.5\epsilon_{\rm max} \sim 6\times10^3 \gamma_7^3 {\cal R}_8^{-1}~{\rm MeV}
\label{eq:EmaxHighB4}
\ee
and
\be
E_{\rm min} \simeq 0.5\epsilon_{\rm min}/x_{01} \sim 20 {\cal R}_8 r_6^{-1/2}~{\rm MeV} \,,
\label{eq:EminHighB2}
\ee
(see Section~\ref{sub:photemper}).

\label{lastpage}


\begin{thebibliography}{}


\bibitem[Abdo et al.(2009a)]{abdo09a}
Abdo A. A. et al. (Fermi Collab.), 2009,
ApJ, 696, 1084.


\bibitem[Abdo et al.(2009b)]{abdo09b}
Abdo A. A. et al. (Fermi Collab.), 2009,
Science, 325, 848.


\bibitem[Abdo et al.(2009c)]{abdo09c}
Abdo A. A. et al. (Fermi Collab.), 2009,
ApJ, 706, 1331.


\bibitem[Abdo et al.(2010)]{abdo10}
Abdo A. A. et al. (Fermi Collab.), 2010,
ApJ, 708, 1254.


\bibitem[Aliu et al.(2008)]{aliu08}
Aliu et al. (MAGIC Collab.), 2008,
Science, 322, 1221. 


\bibitem[Adler(1971)]{adler71}
Adler S. L., 1971,
Ann. Phys., 67, 599.


\bibitem[Arendt \& Eilek(2002)]{arendt02}
Arendt P. N., Eilek J. A., 2002,
ApJ, 581, 451.


\bibitem[Arons(1983)]{arons83}
Arons J., 1996, ApJ, 266, 215.


\bibitem[Arons(1996)]{arons96}
Arons J., 1996, A\&A, 120, 49.


\bibitem[Arons(1998)]{arons98}
Arons J., 1998, in Shibazaki et al., eds, Proc. Intl. Conf. on Neutron
Stars and Pulsars, Neutron Stars and Pulsars: Thirty Years after the
Discovery. UAP, Tokyo, p. 339.


\bibitem[Arons(2007)]{arons07}
Arons J., 2008, in Becker W., ed, Neutron Stars and Pulsars, 40 Years
After the Discovery. [arXiv:0708.1050]


\bibitem[Arons \& Scharlemann(1979)]{arons79}
Arons J., Scharlemann E. T., 1979,
ApJ, 231, 854.


\bibitem[Arzoumanian et al.(2002)Arzoumanian, Chernoff, \& Cordes]{arzoumanian02}
Arzoumanian Z., Chernoff D. F., Cordes J. M., 2002,
ApJ, 568, 289.


\bibitem[Bai \& Spitkovsky(2009)]{bai09}
Bai X.-N., Spitkovsky A., 2009,
ApJ, submitted. [arXiv:0910.5741]


\bibitem[Baring \& Harding(1997)]{baring97}
Baring M. G., Harding A. K., 1997,
ApJ, 482, 372.


\bibitem[Baring \& Harding(2001)]{baring01}
Baring M. G., Harding A. K., 2001,
ApJ, 547, 929.


\bibitem[Baring \& Harding(2007)]{baring07}
Baring M. G., Harding A. K., 2007,
Ap\&SS, 308, 109.


\bibitem[Beloborodov(2008)]{beloborodov08}
Beloborodov A. M., 2008, ApJ, 683, L41.


\bibitem[Beloborodov(2009)]{beloborodov09}
Beloborodov A. M., 2009, ApJ, 703, 1044.


\bibitem[Beloborodov \& Thompson(2007)]{beloborodov07}
Beloborodov A. M., Thompson C., 2007, ApJ, 657, 967.


\bibitem[Beskin(1999)]{beskin99}
Beskin V. S., 1999, Physics-Uspekhi, 42, 1071.


\bibitem[Bogdanov \& Grindlay(2009)]{bogdanov09}
Bogdanov S., Grindlay J. E., 2009,
ApJ, 703, 1557.


\bibitem[Camilo et al.(2007)]{camilo07}
Camilo F. et al., 2007, ApJ, 669, 561.


\bibitem[Camilo et al.(2008)]{camilo08}
Camilo F. et al., 2008, ApJ, 679, 681.


\bibitem[Chen \& Ruderman(1993)]{chen93}
Chen K., Ruderman M., 1993, ApJ, 402, 264.


\bibitem[Cheng et al.(1986a)Cheng, Ho, \& Ruderman]{cheng86a}
Cheng K. S., Ho C., Ruderman, M., 1986, ApJ, 300, 500.


\bibitem[Cheng et al.(1986b)Cheng, Ho, \& Ruderman]{cheng86b}
Cheng K. S., Ho C., Ruderman, M., 1986, ApJ, 300, 522.


\bibitem[Cheng et al.(2000)Cheng, Ruderman, \& Zhang]{cheng00}
Cheng K. S., Ruderman, M., Zhang L., 2000,
ApJ, 537, 964.


\bibitem[Contopoulos(2005)]{contopoulos05}
Contopoulos I., 2005, A\&A, 442, 579.


\bibitem[Contopoulos et al.(1999)Contopoulos, Kazanas, \& Fendt]{contopoulos99}
Contopoulos I., Kazanas D., Fendt C., 1999, ApJ, 511, 351.


\bibitem[Contopoulos \& Spitkovsky(2006)]{contopoulos06}
Contopoulos I., Spitkovsky A., 2006, ApJ, 643, 1139.


\bibitem[Daugherty \& Harding(1982)]{daugherty82}
Daugherty J. K., Harding A. K., 1982,
ApJ, 252, 337.


\bibitem[Daugherty \& Harding(1983)]{daugherty83}
Daugherty J. K., Harding A. K., 1983, ApJ, 273, 761.


\bibitem[Daugherty \& Harding(1989)]{daugherty89}
Daugherty J. K., Harding A. K., 1989, ApJ, 336, 861.


\bibitem[Daugherty \& Harding(1996)]{daugherty96}
Daugherty J. K., Harding A. K., 1996,
ApJ, 458, 278.


\bibitem[Dermer(1990)]{dermer90}
Dermer C. D. 1990, ApJ, 360, 197.


\bibitem[Dyks \& Rudak(2003)]{dyks03}
Dyks J., Rudak B., 2003, ApJ, 598, 1201.


\bibitem[Dyks et al.(2004)Dyks, Rudak, \& Harding]{dyks04}
Dyks J., Rudak B., Harding A. K., 2004, ApJ, 607, 939.


\bibitem[Erber(1966)]{erber66}
Erber T., 1966, Rev. Mod. Phys., 38, 626.


\bibitem[Gangadhara(2004)]{gangadhara04}
Gangadhara R. T., 2004, ApJ, 609, 335.


\bibitem[Gangadhara \& Gupta(2001)]{gangadhara01}
Gangadhara R. T., Gupta Y., 2001, ApJ, 555, 31.


\bibitem[Gonthier et al.(2000)]{gonthier00}
Gonthier P. L., Harding A. K., Baring M. G., Costello R. M., Mercer C. L.,
2000, ApJ, 540, 907.


\bibitem[Gruzinov(2005)]{gruzinov05}
Gruzinov A., 2005, PRL, 94, 021101.


\bibitem[Harding et al.(1997)Harding, Baring, \& Gonthier]{harding97}
Harding A. K., Baring M. G., Gonthier P. L., 1997, ApJ, 476, 246.


\bibitem[Harding \& Muslimov(2002)]{harding02}
Harding A. K., Muslimov A. G., 2002,
ApJ, 568, 862.


\bibitem[Harding et al.(2002)Harding, Muslimov, \& Zhang]{hardingetal02}
Harding A. K., Muslimov A. G., Zhang, B., 2002, ApJ, 576, 366.


\bibitem[Harding \& Preece(1987)]{harding87}
Harding A. K., Preece R., 1987, ApJ, 319, 939.


\bibitem[Harding et al.(2008)]{harding08}
Harding A. K., Stern J. V., Dyks J., Frackowiak M., 2008,
ApJ, 680, 1378.


\bibitem[Harding et al.(1978)Harding, Tademaru, \& Esposito]{harding78}
Harding A. K., Tademaru E., Esposito L. W., 1978,
ApJ, 225, 226.


\bibitem[Harding et al.(2005)Harding, Usov, \& Muslimov]{harding05}
Harding A. K., Usov V. V., Muslimov A. G., 2005,
ApJ, 622, 531.


\bibitem[Herold et al.(1982)Herold, Ruder, \& Wunner]{herold82}
Herold H., Ruder H., Wunner G., 1982, A\&A, 115, 90.


\bibitem[Hibschman \& Arons(2001a)]{hibschman01a}
Hibschman J. A., Arons. J., 2001, ApJ, 554, 624.


\bibitem[Hibschman \& Arons(2001b)]{hibschman01b}
Hibschman J. A., Arons. J., 2001, ApJ, 560, 871.


\bibitem[Hirotani(2006)]{hirotani06}
Hirotani K., 2006, ApJ, 652, 1475.


\bibitem[Kalapotharakos \& Contopoulos(2009)]{kalapotharakos09}
Kalapotharakos C., Contopoulos I., 2009, A\&A, 496, 495.


\bibitem[Kaspi \& McLaughlin(2005)]{kaspi05}
Kaspi V. M., McLaughlin M. A., 2005, ApJ, 618, 41.


\bibitem[Kijak \& Gil(2003)]{kijak03}
Kijak J., Gil, J., 2003,
A\&A, 397, 969.


\bibitem[Jackson(1998)]{jackson98}
Jackson J. D., 1998,
Classical Electrodynamics, 3rd edition. Wiley, New York.


\bibitem[Komissarov(2006)]{komissarov06}
Komissarov S. S., 2006, MNRAS, 367, 19.


\bibitem[Kuiper et al.(2006)]{kuiper06}
Kuiper L., Hermsen W., den Hartog P. R., Collmar W., 2009, ApJ, 645, 556.


\bibitem[Kuiper \& Hermsen (2009)]{kuiper09}
Kuiper L., Hermsen W., 2009, A\&A, 501, 1031.


\bibitem[Levinson et al.(2005)]{levinson05}
Levinson A. et al., 2005, ApJ, 631, 456.


\bibitem[Luo \& Melrose(2008)]{luo08}
Luo Q., Melrose D., 2008, MNRAS, 387, 1291.


\bibitem[Lyubarsky(2002)]{lyubarsky02}
Lyubarsky Y. E., 2002, in Becker W., Lesch H., Tr\"umper J., eds,
Proc. WE-Heraeus Seminar 270, Neutron Stars, Pulsars, and
Supernova Remnants. MPE Rep. 278, p. 230.


\bibitem[Lyubarsky(2008)]{lyubarsky08}
Lyubarsky Y. E., 2008, in Bassa et al., eds, AIP Conf. Proc. 983, 40
Years of Pulsars: Milliscond Pulsars, Magnetars and More. AIP, New
York, p. 29.


\bibitem[Lyutikov(2007)]{lyutikov07}
Lyutikov M., 2007, MNRAS, 381, 1190.


\bibitem[Medin(2008)]{thesis}
Medin Z., 2008, Ph.D. Thesis, Cornell University, New York.


\bibitem[Medin \& Lai(2007)]{medin07b}
Medin Z., Lai D., 2007,
MNRAS, 382, 1833. [ML07]


\bibitem[Melikidze et al.(2000) Melikidze, Gil, \& Pataraya]{melikidze00}
Melikidze G. I., Gil J. A., Pataraya A., 2000, ApJ, 544, 1081.


\bibitem[Melrose(1995)]{melrose95}
Melrose D. B., 1995, J. Astrophys. Astr., 16, 137.


\bibitem[Melrose(2004)]{melrose04}
Melrose D. B., 2004, in Camilo F., Gaensler B. M., eds, Proc. IAU
Symp. 218, Young Neutron Stars and Their Environments. ASP, San
Francisco, p.349.


\bibitem[Muslimov \& Harding(2003)]{muslimov03}
Muslimov A. G., Harding A. K., 2003,
ApJ, 588, 430.


\bibitem[Muslimov \& Harding(2004)]{muslimov04}
Muslimov A. G., Harding A. K., 2004,
ApJ, 606, 1143.


\bibitem[Muslimov \& Tsygan(1992)]{muslimov92}
Muslimov A. G., Tsygan A. I., 1992,
MNRAS, 255, 61.


\bibitem[Pavan et al.(2009)]{pavan09}
Pavan L., Turolla R., Zane S., Nobili L., 2009,
MNRAS, 395, 753.


\bibitem[Pons et al.(2007)]{pons07}
Pons J. A., Link B., Miralles J. A., Geppert U., 2007,
PhRvL, 98, 1101.


\bibitem[Rankin(1993)]{rankin93}
Rankin J. M., 1993,
ApJ, 405, 285.


\bibitem[Rankin \& Wright(2003)]{rankin03}
Rankin J. M., Wright G. A. E., 2003, A\&A Rev., 12, 43.


\bibitem[Romani(1996)]{romani96}
Romani R. W., 1996, ApJ, 470, 469.


\bibitem[Romani \& Yadigaroglu(1995)]{romani95}
Romani R. W., Yadigaroglu I. A., 1995, ApJ, 438, 314.


\bibitem[Ruderman \& Sutherland(1975)]{ruderman75}
Ruderman M. A., Sutherland P. G., 1975, ApJ, 196, 51.


\bibitem[Rybicki \& Lightman(1979)]{rybicki79}
Rybicki G. B., Lightman A. P., 1979, Radiative Processes in
Astrophysics. Wiley-Interscience, New York.


\bibitem[Sakai \& Shibata(2003)]{sakai03}
Sakai N., Shibata S., 2003, ApJ, 484, 427.


\bibitem[Sokolov \& Ternov(1968)]{sokolov68}
Sokolov A. A., Ternov I. M., 1968,
Synchrotron Radiation, Pergamon, New York.


\bibitem[Spitkovsky(2006)]{spitkovsky06}
Spitkovsky A., 2006, ApJ, 648, 51.


\bibitem[Sturner et al.(1995)Sturner, Dermer, \& Michel]{sturner95}
Sturner S. J., Dermer C. D., Michel F. C., 1995, ApJ, 445, 736.


\bibitem[Sturrock(1971)]{sturrock71}
Sturrock P. A., 1971,
ApJ, 164, 529.


\bibitem[Takata et al.(2006)]{takata06}
Takata J., Shibata S., Hirotani K., Chang H. K., 2006,
MNRAS, 366, 1310.


\bibitem[Thompson(2008a)]{thompson08a}
Thompson C., 2008, ApJ, 688, 1258.


\bibitem[Thompson(2008b)]{thompson08b}
Thompson C., 2008, ApJ, 688, 499.


\bibitem[Thompson \& Beloborodov(2005)]{thompson2005}
Thompson C., Beloborodov A. M., 2005, ApJ, 634, 565.


\bibitem[Thompson et al.(2002)Thompson, Lyutikov, \& Kulkarni]{thompson02}
Thompson C., Lyutikov M., Kulkarni S. R., 2002,
ApJ, 574, 332.


\bibitem[Thompson et al.(1999)]{thompson99}
Thompson D. J. et al., 1999, ApJ, 516, 297.


\bibitem[Thompson(2004)]{thompson04}
Thompson D. J., 2004, in Cheng K. S., Romero G. E., eds, Cosmic
Gamma-Ray Sources. Kluwer, Dordrecht, p. 149.


\bibitem[Thompson(2008)]{thompson08}
Thompson D. J., 2008, in 40 Years of Pulsars: Millisecond Pulsars,
Magnetars, and More. AIP Conf. Proc. No. 983. AIP, New York, p. 56.


\bibitem[Timokhin(2006)]{timokhin06}
Timokhin A. N., 2006, MNRAS, 368, 1055


\bibitem[Usov(2002)]{usov02}
Usov V. V., 2002, ApJ, 572, L87.


\bibitem[von Hoensbroech et al.(1998)von Hoensbroech, Lesch, \& Kunzl]{hoensbroech98}
von Hoensbroech A., Lesch H., Kunzl T., 1998,
A\&A, 336, 209.


\bibitem[Vranevsevic et al.(2007)Vranevsevic, Manchester, \& Melrose]{vranevsevic07}
Vranevsevic N., Manchester R. N., Melrose D. B., 2007,
in Becker W., Huang H. H., eds, Proc. WE-Heraeus Seminar 363,
Neutron Stars and Pulsars. MPE Rep. 291, p. 88.


\bibitem[Venter et al.(2009)Venter, Harding, \& Guillemot]{venter09}
Venter C., Harding A. K., Guillemot L., 2009,
ApJ, 707, 800.


\bibitem[Watters et al.(2009)]{watters09}
Watters K. P., Romani R. W., Weltevrede P., Johnston, S.,
2009, ApJ, 695, 1289.


\bibitem[Woods \& Thompson(2006)]{woods06}
Woods P. M., Thompson, C., 2006, in Lewin W., van der Klis M., eds,
Compact Stellar X-ray Sources. Cambridge Univ. Press, p. 547.


\bibitem[Zhang \& Harding(2000)]{zhang00}
Zhang, B., Harding, A. K., 2000, ApJ, 532, 1150.


\end{thebibliography}
\end{document}